\documentclass[12pt]{iopart}

\usepackage[jphysicsB]{harvard}
\usepackage{graphicx,setstack,rotating,threeparttablex,pdflscape,longtable}
\usepackage{bm}
\usepackage{pifont}

\newcommand{\cov}{\rm cov}
\newcommand{\pad}{p$_\mathrm{A\hbox{-}D}$}
\newcommand{\plb}{p$_\mathrm{L\hbox{-}B}$}
\newcommand{\totalsamplesize}{39}
\newcommand{\nulag}{\mbox{$\nu$-lag}}
\newcommand{\chg}[1]{#1}
\newcommand{\chgg}[1]{#1}

\def\jnl@style{\rm}
\def\aaref@jnl#1{{\jnl@style#1}}

\def\aaref@jnl#1{{\jnl@style#1}}

\def\aj{\aaref@jnl{AJ}}                   
\def\araa{\aaref@jnl{ARA\&A}}             
\def\apj{\aaref@jnl{ApJ}}                 
\def\apjl{\aaref@jnl{ApJ}}                
\def\apjs{\aaref@jnl{ApJS}}               
\def\ao{\aaref@jnl{Appl.~Opt.}}           
\def\apss{\aaref@jnl{Ap\&SS}}             
\def\aap{\aaref@jnl{A\&A}}                
\def\aapr{\aaref@jnl{A\&A~Rev.}}          
\def\aaps{\aaref@jnl{A\&AS}}              
\def\azh{\aaref@jnl{AZh}}                 
\def\baas{\aaref@jnl{BAAS}}               
\def\jrasc{\aaref@jnl{JRASC}}             
\def\memras{\aaref@jnl{MmRAS}}            
\def\mnras{\aaref@jnl{MNRAS}}             
\def\pra{\aaref@jnl{Phys.~Rev.~A}}        
\def\prb{\aaref@jnl{Phys.~Rev.~B}}        
\def\prc{\aaref@jnl{Phys.~Rev.~C}}        
\def\prd{\aaref@jnl{Phys.~Rev.~D}}        
\def\pre{\aaref@jnl{Phys.~Rev.~E}}        
\def\prl{\aaref@jnl{Phys.~Rev.~Lett.}}    
\def\pasp{\aaref@jnl{PASP}}               
\def\pasj{\aaref@jnl{PASJ}}               
\def\qjras{\aaref@jnl{QJRAS}}             
\def\skytel{\aaref@jnl{S\&T}}             
\def\solphys{\aaref@jnl{Sol.~Phys.}}      
\def\sovast{\aaref@jnl{Soviet~Ast.}}      
\def\ssr{\aaref@jnl{Space~Sci.~Rev.}}     
\def\zap{\aaref@jnl{ZAp}}                 
\def\nat{\aaref@jnl{Nature}}              
\def\iaucirc{\aaref@jnl{IAU~Circ.}}       
\def\aplett{\aaref@jnl{Astrophys.~Lett.}} 
\def\apspr{\aaref@jnl{Astrophys.~Space~Phys.~Res.}}
\def\bain{\aaref@jnl{Bull.~Astron.~Inst.~Netherlands}} 
\def\fcp{\aaref@jnl{Fund.~Cosmic~Phys.}}  
\def\gca{\aaref@jnl{Geochim.~Cosmochim.~Acta}}   
\def\grl{\aaref@jnl{Geophys.~Res.~Lett.}} 
\def\jcp{\aaref@jnl{J.~Chem.~Phys.}}      
\def\jgr{\aaref@jnl{J.~Geophys.~Res.}}    
\def\jqsrt{\aaref@jnl{J.~Quant.~Spec.~Radiat.~Transf.}}
\def\memsai{\aaref@jnl{Mem.~Soc.~Astron.~Italiana}}
\def\nphysa{\aaref@jnl{Nucl.~Phys.~A}}   
\def\pasa{\aaref@jnl{PASA}}               
\def\physrep{\aaref@jnl{Phys.~Rep.}}   
\def\physscr{\aaref@jnl{Phys.~Scr}}   
\def\planss{\aaref@jnl{Planet.~Space~Sci.}}   
\def\procspie{\aaref@jnl{Proc.~SPIE}}   
\def\rmxaa{\aaref@jnl{Rev.~Mex. Astronom{\'i}a \& Astrof{\'i}sica}} 
\def\na{\aaref@jnl{New~Ast.}}

\newcounter{magicrownumbers}
\newcommand\rownumber{\stepcounter{magicrownumbers}\arabic{magicrownumbers}}

\begin{document}
\title[Stochastic variability of ALMA calibrators]{Stochastic modeling of the time variability of ALMA calibrators }
\author{A.\ E.\ Guzm\'an,$^1$ C.\ Verdugo,$^2$ H.\ Nagai,$^1$ Y.\ Contreras,$^3$ G.\ Marinello,$^2$ R.\ Kneissl,$^2$ K.\ Nakanishi,$^1$ J.\ Ueda.$^1$}
\address{$^1$ National Astronomical Observatory of Japan, 2-21-1 Osawa, Mitaka, Tokyo 181-8588, Japan}
\address{$^2$ Joint Alma Observatory (JAO), Alonso de C\'ordova 3107, Vitacura, Santiago, Chile}
\address{$^3$ Leiden Observatory, Leiden University, PO Box 9513, NL-2300 RA Leiden, the Netherlands}
\begin{indented}
\item[]\today
\end{indented}
\ead{andres.guzman@nao.ac.jp}
\begin{abstract}
Characterizing the variability of the extragalactic sources used for calibration in the \emph{Atacama Large Millimeter/Sub-millimeter Array} (ALMA) is key to assess the flux scale uncertainty of science observations. To this end, we model the variability of  \totalsamplesize\ quasars which have been used by ALMA as secondary flux calibrators using continuous time stochastic processes. This formalism is specially adapted to the multi-frequency, quasi-periodic sampling which characterizes the calibration monitoring of ALMA. We find that simple mixtures of Ornstein-Uhlenbeck processes can describe well the flux and spectral index variability of these sources \chg{for  Bands 3 and 7 (91.5 and 103.5, and 343.5 GHz, respectively)}. The spectral shape of the calibrators are characterized by negative spectral indices, mostly between $-0.35$ and $-0.80$, and with additional concavity. The model provides forecasts, interpolations, and uncertainty estimations for the observed fluxes \chg{that depend on the} intrinsic variability \chg{of the source}. These can be of practical use for the ALMA data calibrator survey and data quality assurance.
 \end{abstract}

\submitto{Publications of the Astronomical Society of the Pacific}
\maketitle
\section{Introduction}

The flux density scale and calibration of the science observations performed by the \emph{Atacama Large
Millimeter/Sub-millimeter Array} (ALMA) relies heavily on contemporaneous comparison with calibrator sources.  
Properties of these calibrators need to be predictable to an
acceptable uncertainty level.  The flux scaling, in particular, depends 
ultimately on comparison with solar system objects (SSO) \chg{which are used as primary flux calibrators. For these, 
(sub-)millimeter emission models} are accurate within a 3--5\% level \cite{almamemo594}.

\chg{In the ideal case, each observation should include one measurement of a primary flux calibrator, 
whose flux is then bootstrapped to an amplitude calibrator 
\cite{almamemo372}.
The scarcity of SSO calibrators and its restricted location near the ecliptic forces us to rely on bright quasars for flux calibration in case
\chgg{SSO  observations are inconvenient or inaccessible} \cite{Fomalont2014Msngr}.} Quasars have some advantages as  calibrators for ALMA compared with
SSO: they are in most cases unresolved point-like sources with little spectral features like emission or absorption lines, which makes
them ideal for phase and bandpass calibration. 

However, quasars are also known to be highly variable. Moreover, most ALMA calibrators
are bright mm/sub-mm blazars of various types \cite{Bonato2018MNRAS}. Blazars 
are known to be amongst the most variable type of quasar \cite{Falomo2014A&ARv,Ulrich1997ARA&A,Wagner1995ARA&A}. 
\chg{In contrast to the \chgg{more} predictable variability observed toward SSOs, the complexity of the
quasar phenomena makes it  very difficult to accurately predict its flux variations. 
Therefore, their use as secondary flux calibrators requires  constant monitoring and 
cross calibration against primary flux calibrators. 
To this end, the ALMA calibration procedure relies on frequent observations of a subset of $\sim40$ 
  bright and \chgg{homogeneously} distributed in the sky extragalactic sources known as the ``grid'' calibrators.  These are monitored and compared
with near-to-simultaneous observations of SSO  objects every 10--14 days \cite{2012ALMAN...9....8V,ATM6}.}
\chg{The typical calibration procedure for ALMA 
therefore includes an observation of a \chgg{grid source 
whose flux} is assumed to be 
 known. This assumed flux  is calculated by interpolating 
 the flux and spectral index from  the closest measurements calibrated against primary flux calibrators.
   This same grid  source is commonly used to determine the bandpass response as well. 
The amplitude and phase calibrator is usually another, fainter quasar, located close \chgg{($\le10^\circ$ for mm wavelength and compact array configuration;  and $\le2^\circ$ if possible for sub-mm wavelength and extended configurations)} to the science target, whose flux is determined from
comparison with the flux calibrator \chgg{and transferred to the science target}.}

\chg{In consequence, the flux density of the grid quasar used as secondary flux calibrator 
has an additional uncertainty due to its intrinsic variability, uncertainty that is transmitted to any  source which uses said quasar for  flux scaling. 
This uncertainty depends on how much time has elapsed between the closest 
comparison performed against a primary flux calibrator  and the time of observation:
the larger this time interval, the larger the  uncertainty.
Thus, it is \chgg{key to 
adopt strategies} to \emph{minimize and quantify} 
the uncertainties produced by the time variability of the secondary flux calibrator. Intrinsic, non-predictable variability 
of the flux calibrator has not been traditionally \chgg{included explicitly} as a source of uncertainty before 
\cite{almamemo211,almamemo372,almamemo599}.}


\chg{In order to estimate quantitatively the uncertainty of the flux of the quasars due to variability, in this work we model the 
 calibrator's fluxes using time-series stochastic processes.} Stochastic time series 
 \citeaffixed{Scargle1981ApJS,1982ApJ...263..835S,2018FrP.....6...80F}{e.g.,} 
 have been used in astronomy to model the light-curves of time-varying phenomena which have a random or unpredictable component. This randomness can arise in a variety of forms, either due to an intrinsic unpredictability of the phenomena, or due to uncertainties introduced in the measurement process. 

The main objective of this work  is to characterize the variability of the ALMA grid calibrators statistically. 
The formalism is largely based on the Ornstein-Uhlenbeck (OU) mixture models described in \citeasnoun{Kelly2009ApJ} and
 \citeasnoun{Kelly2011ApJ} \citeaffixed{Kelly2014ApJ}{and generalized in}, which allows us to derive flux predictions, interpolations, and time dependent uncertainties. 
 Albeit the method may also provide constrains on the actual, physical emission mechanisms of these blazars, this is not the main driving goal of the present study.  Quasar variability is thought to originate from, for example, variable accretion onto the central black hole, variable jet ejection, relativistic beaming and amplification, development of instabilities, and from line-of-sight effects like occultation events and stellar microlensing from intervening galaxies. Despite the physical interpretation of the optical variability proposed by \citeasnoun{Kelly2009ApJ}
 is unlikely to be applicable to the physics of blazar emission, the techniques and stochastic modeling are  well adapted (as we also show in the present work) to observed variability at mm/sub-mm wavelengths. 

\Sref{sec-data} describes briefly the ALMA calibrator public dataset. In \Sref{sec-met} we present the model and its hypotheses. We find best-fit  models for the grid calibrators and describe these results in \Sref{sec-res}. We discuss some applications of the modeling to the calibration and quality assurance process of ALMA in section \Sref{sec-dis}. \Sref{sec-con} briefly summarizes the mains results of this work.

\section{\chg{ALMA grid calibrator data}}\label{sec-data}

\begin{table}
\caption{\label{tab-data} Observational characteristics of the core grid calibrators list. \chg{Columns (1) to (3) show the name of the source, its
Right Ascension (J2000), and its Declination (J2000), respectively. Columns (4) and (5) show the first and last day of observation of the light-curves presented in this study. 
Column (6) shows the total number of measurements. Columns (7) and (8) show the mean flux at the fiducial frequency of 100 GHz and the mean spectral index 
fitted to the data according to \eref{eq-si}, respectively. Column \chgg{(9)} shows the variability index (\Eref{eq-vix}).}}\lineup
\begin{indented}
\item[]\hspace*{-7.5em}\begin{tabular}{rlllllrrrr}
\br
&Source & RA & DEC & Start & End & \#  & $\langle F_{\rm100}\rangle$ &$\langle\alpha\rangle$ & \textit{vix}\\
&            & (J2000) & (J2000) &  & & meas. & (Jy)& & (\%) \\
\mr
\rownumber &  J0006$-$0623 & 00:06:13.893 & $-$06:23:35.34 & 2013-07-23 & 2018-07-04 & 387 & $3.85$ & $-0.58$ & $26.1$ \\ 
\rownumber &  J0237$+$2848 & 02:37:52.406 & $+$28:48:08.99 & 2012-06-30 & 2018-07-02 & 366 & $2.13$ & $-0.62$ & $12.8$ \\ 
\rownumber &  J0238$+$1636 & 02:38:38.930 & $+$16:36:59.27 & 2012-06-30 & 2018-07-02 & 446 & $1.56$ & $-0.49$ & $33.0$ \\ 
\rownumber &  J0319$+$4130 & 03:19:48.160 & $+$41:30:42.11 & 2012-08-15 & 2018-06-30 & 286 & $18.3$ & $-0.73$ & $16.5$ \\ 
\rownumber &  J0334$-$4008 & 03:34:13.654 & $-$40:08:25.40 & 2012-06-30 & 2018-07-07 & 506 & $0.762$ & $-0.67$ & $43.0$ \\ 
\rownumber &  J0423$-$0120 & 04:23:15.801 & $-$01:20:33.07 & 2012-06-30 & 2018-07-07 & 599 & $1.63$ & $-0.51$ & $60.4$ \\ 
\rownumber &  J0510$+$1800 & 05:10:02.369 & $+$18:00:41.58 & 2012-07-18 & 2018-07-07 & 448 & $2.54$ & $-0.45$ & $28.6$ \\ 
\rownumber &  J0519$-$4546 & 05:19:49.723 & $-$45:46:43.85 & 2012-07-18 & 2018-07-07 & 495 & $1.2$ & $-0.39$ & $11.9$ \\ 
\rownumber &  J0522$-$3627 & 05:22:57.985 & $-$36:27:30.85 & 2012-07-18 & 2018-07-07 & 463 & $5.57$ & $-0.24$ & $19.9$ \\ 
\rownumber &  J0538$-$4405 & 05:38:50.362 & $-$44:05:08.94 & 2012-07-31 & 2018-07-07 & 563 & $2.04$ & $-0.65$ & $34.6$ \\ 
\rownumber &  J0635$-$7516 & 06:35:46.508 & $-$75:16:16.82 & 2012-06-30 & 2018-07-07 & 758 & $1.25$ & $-0.92$ & $12.1$ \\ 
\rownumber &  J0750$+$1231 & 07:50:52.046 & $+$12:31:04.83 & 2012-10-06 & 2018-07-07 & 373 & $1.32$ & $-0.67$ & $13.9$ \\ 
\rownumber &  J0854$+$2006 & 08:54:48.875 & $+$20:06:30.64 & 2012-08-26 & 2018-07-07 & 451 & $4.54$ & $-0.46$ & $29.1$ \\ 
\rownumber &  J0904$-$5735 & 09:04:53.179 & $-$57:35:05.78 & 2015-05-31 & 2018-07-07 & 95 & $1.16$ & $-0.42$ & $18.3$ \\ 
\rownumber &  J1037$-$2934 & 10:37:16.080 & $-$29:34:02.81 & 2012-08-26 & 2018-07-07 & 481 & $1.22$ & $-0.52$ & $26.2$ \\ 
\rownumber &  J1058$+$0133 & 10:58:29.605 & $+$01:33:58.82 & 2012-10-06 & 2018-07-05 & 389 & $4.26$ & $-0.48$ & $20.1$ \\ 
\rownumber &  J1107$-$4449 & 11:07:08.694 & $-$44:49:07.62 & 2012-10-06 & 2018-07-07 & 439 & $1.16$ & $-0.76$ & $11.3$ \\ 
\rownumber &  J1127$-$1857 & 11:27:04.392 & $-$18:57:17.44 & 2013-12-29 & 2018-07-07 & 84 & $0.93$ & $-0.62$ & $19.7$ \\ 
\rownumber &  J1146$+$3958 & 11:46:58.298 & $+$39:58:34.30 & 2012-08-26 & 2018-07-05 & 266 & $1.2$ & $-0.56$ & $47.5$ \\ 
\rownumber &  J1229$+$0203 & 12:29:06.700 & $+$02:03:08.60 & 2012-06-30 & 2018-07-05 & 493 & $9.42$ & $-0.79$ & $32.7$ \\ 
\rownumber &  J1256$-$0547 & 12:56:11.167 & $-$05:47:21.53 & 2012-06-30 & 2018-07-05 & 612 & $13$ & $-0.58$ & $19.0$ \\ 
\rownumber &  J1331$+$3030 & 13:31:08.288 & $+$30:30:32.96 & 2012-08-02 & 2018-07-05 & 35 & $0.742$ & $-1.1$ & $4.4$ \\ 
\rownumber &  J1337$-$1257 & 13:37:39.783 & $-$12:57:24.69 & 2012-06-30 & 2018-07-05 & 437 & $3.55$ & $-0.6$ & $24.8$ \\ 
\rownumber &  J1427$-$4206 & 14:27:56.298 & $-$42:06:19.44 & 2012-06-30 & 2018-07-05 & 554 & $3.4$ & $-0.58$ & $27.3$ \\ 
\rownumber &  J1517$-$2422 & 15:17:41.813 & $-$24:22:19.48 & 2012-06-30 & 2018-07-05 & 470 & $2.39$ & $-0.29$ & $24.2$ \\ 
\rownumber &  J1550$+$0527 & 15:50:35.269 & $+$05:27:10.45 & 2012-06-30 & 2018-07-05 & 538 & $1.01$ & $-0.72$ & $5.0$ \\ 
\rownumber &  J1617$-$5848 & 16:17:17.891 & $-$58:48:07.86 & 2012-06-30 & 2018-07-05 & 554 & $1.07$ & $-0.91$ & $15.6$ \\ 
\rownumber &  J1642$+$3948 & 16:42:58.810 & $+$39:48:36.99 & 2012-07-29 & 2018-07-04 & 330 & $3.19$ & $-0.74$ & $17.6$ \\ 
\rownumber &  J1733$-$1304 & 17:33:02.706 & $-$13:04:49.55 & 2012-06-30 & 2018-07-05 & 527 & $2.84$ & $-0.67$ & $10.9$ \\ 
\rownumber &  J1751$+$0939 & 17:51:32.819 & $+$09:39:00.73 & 2012-06-30 & 2018-07-04 & 454 & $2.9$ & $-0.47$ & $30.1$ \\ 
\rownumber &  J1924$-$2914 & 19:24:51.056 & $-$29:14:30.12 & 2012-06-30 & 2018-07-04 & 745 & $5.56$ & $-0.65$ & $10.4$ \\ 
\rownumber &  J2000$-$1748 & 20:00:57.090 & $-$17:48:57.67 & 2013-12-30 & 2018-07-04 & 126 & $1.03$ & $-0.44$ & $41.2$ \\ 
\rownumber &  J2025$+$3343 & 20:25:10.842 & $+$33:43:00.21 & 2012-06-30 & 2018-07-02 & 483 & $1.46$ & $-0.8$ & $30.2$ \\ 
\rownumber &  J2056$-$4714 & 20:56:16.360 & $-$47:14:47.63 & 2012-06-30 & 2018-07-04 & 568 & $1.22$ & $-0.63$ & $21.7$ \\ 
\rownumber &  J2148$+$0657 & 21:48:05.459 & $+$06:57:38.60 & 2012-06-30 & 2018-07-04 & 601 & $2.09$ & $-0.96$ & $23.1$ \\ 
\rownumber &  J2232$+$1143 & 22:32:36.409 & $+$11:43:50.90 & 2012-06-30 & 2017-09-29 & 357 & $3.76$ & $-0.48$ & $40.5$ \\ 
\rownumber &  J2253$+$1608 & 22:53:57.748 & $+$16:08:53.56 & 2013-07-23 & 2018-07-04 & 350 & $14.8$ & $-0.53$ & $17.4$ \\ 
\rownumber &  J2258$-$2758 & 22:58:05.963 & $-$27:58:21.26 & 2012-06-30 & 2018-07-04 & 737 & $1.93$ & $-0.68$ & $56.3$ \\ 
\rownumber &  J2357$-$5311 & 23:57:53.266 & $-$53:11:13.69 & 2012-06-30 & 2018-07-04 & 755 & $0.954$ & $-0.77$ & $15.0$ \\ 
\br
\end{tabular}
\end{indented}
\end{table}

\begin{figure}
\includegraphics[width=\textwidth]{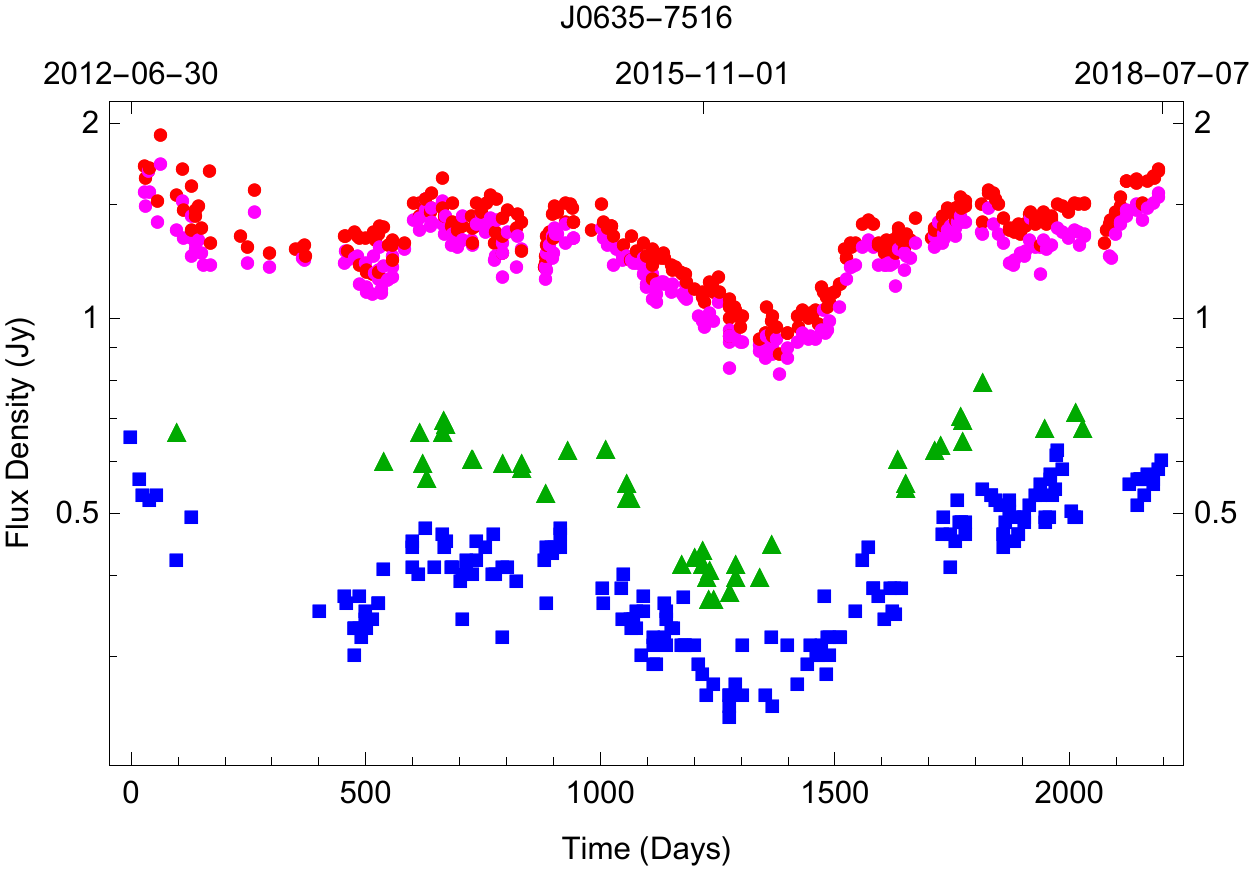} 
\caption{\chg{Flux density light-curve  for the source with the largest number of measurements, J0635$-$7516. 
Band 6, and 7 data are marked with green triangles and blue squares, respectively. Red and magenta circles indicate LSB and USB Band 3 data, respectively.\label{fig-data}}}
\end{figure}

\begin{figure}
\centering\includegraphics[width=0.5\textwidth]{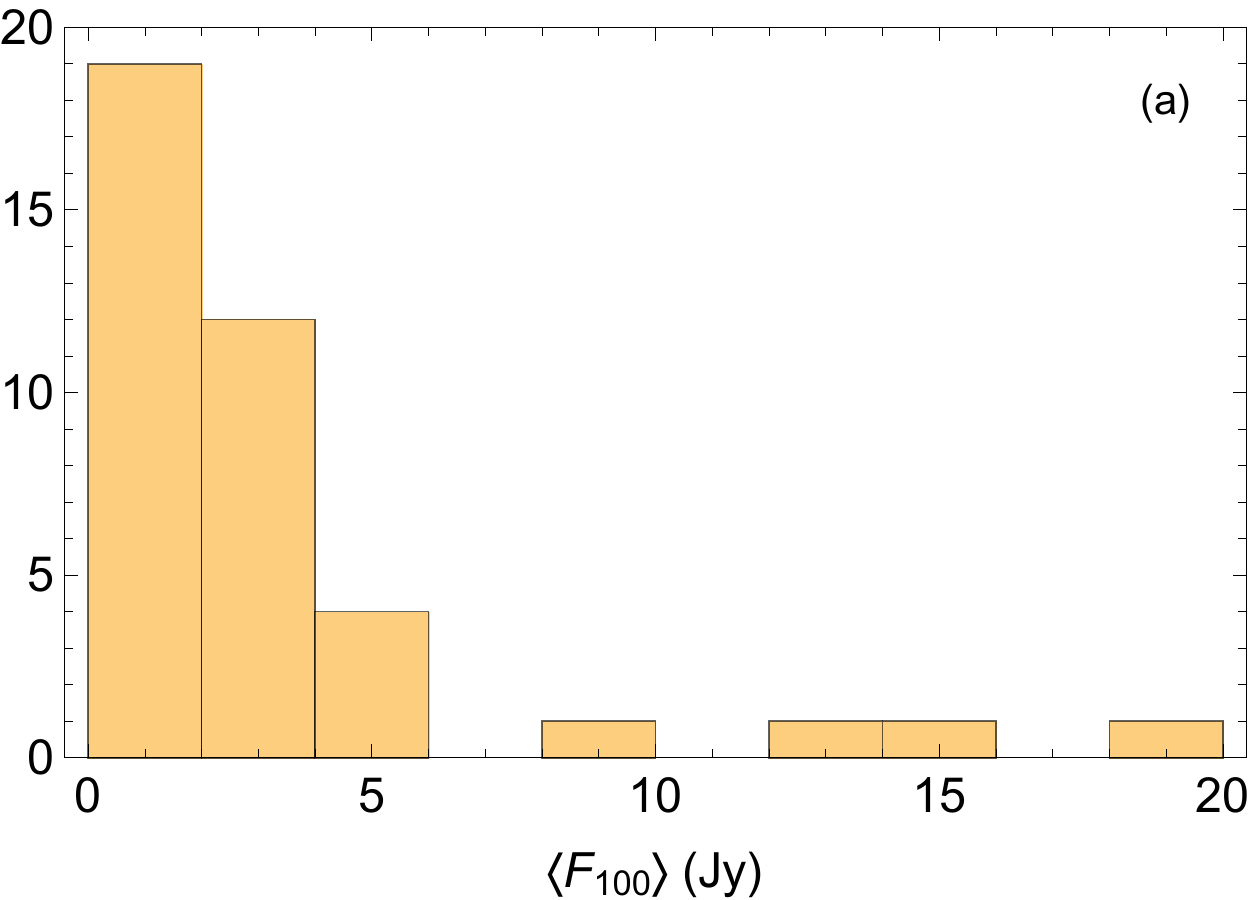}\\
\includegraphics[width=0.5\textwidth]{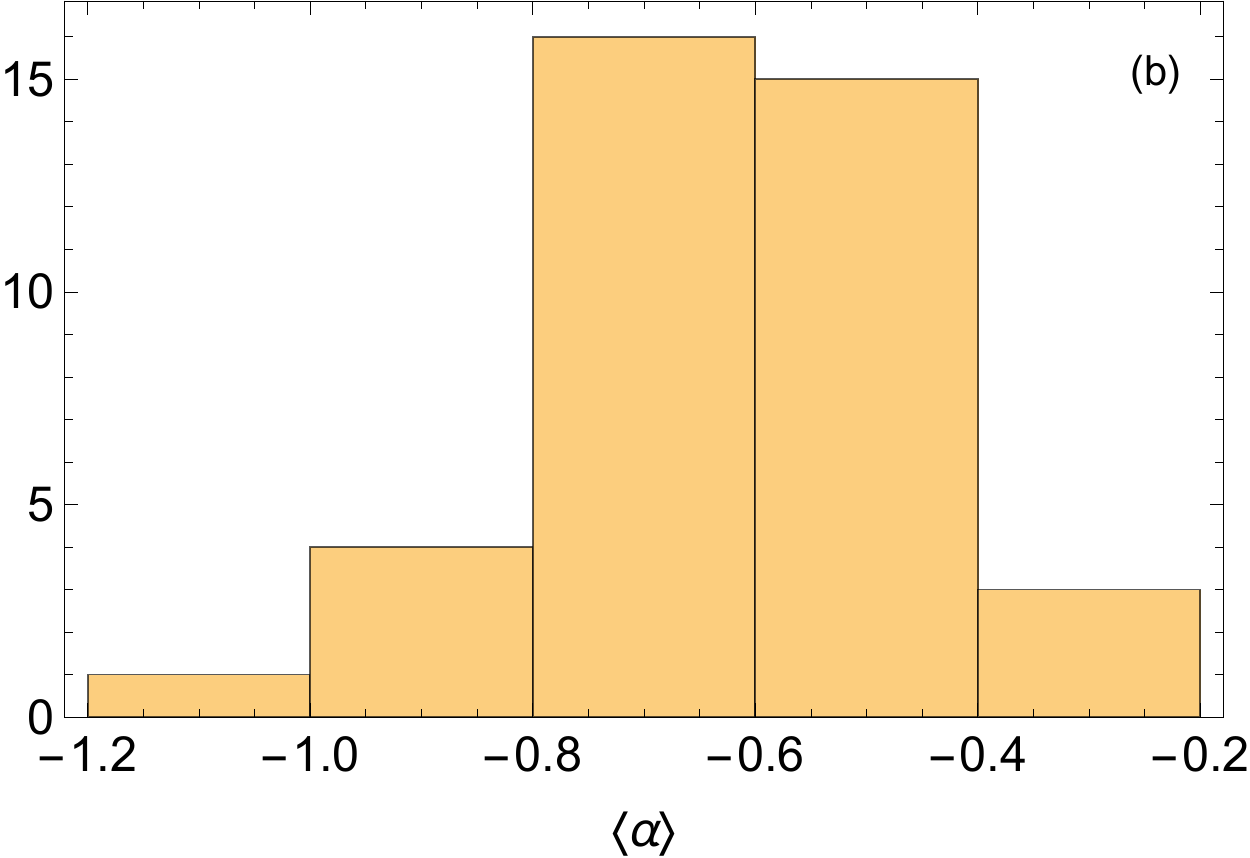}\\
\includegraphics[width=0.5\textwidth]{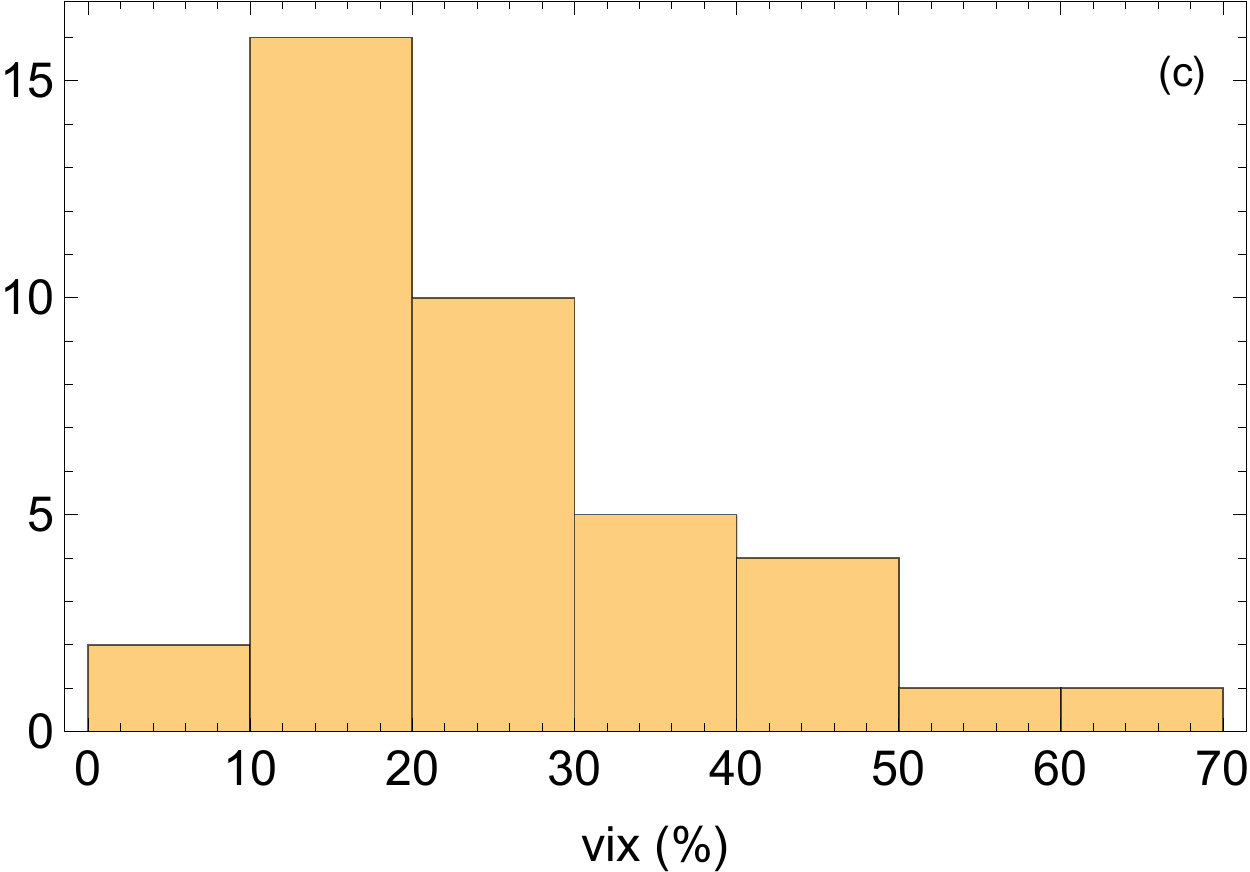}%
\caption{\chg{Panels (a) to (c) show, respectively,  the histograms of $\langle F_{100}\rangle$, $\langle\alpha\rangle$, and \textit{vix} from \Tref{tab-data}}. \label{fig-histos}}
\end{figure}

\chg{The list of secondary flux calibrators known  as the grid sources is given by Table 10.1 from \citeasnoun{ATM6}. 
Flux densities, positions, and \chgg{calibrator type}  can be  obtained from the online source catalogue.\footnote{\texttt{https://almascience.org/sc/}}}
\chg{The goals of the \chgg{grid} monitoring calibration survey are to quantify the variability of calibrators, provide up to date flux densities, and 
 identify the best  bandpass candidates and  investigate their suitability as secondary flux calibrators. The observational aim for the grid sample is
  to obtain a minimum of one Band 3 and one Band 6 or 7 measurement every two weeks. Band 3 was selected as the lower frequency 
because of the smaller weather effects, higher sensitivity, and better instrumental stability. Band 7 was preferred as the second frequency because it 
combines a large frequency lever arm to estimate the calibrator's spectral index with a relatively large fraction ($>40\%$) of observing time  with compatible 
weather conditions. Band 6 is the fallback option in case Band 7 observations are not possible, and there are instances in which Band 3 observations are the 
only option. In some cases, Band 6 data are also taken in addition to the Band 3 and 7 observations.}
Data reduction and treatment  of the calibration data roughly follows  the same  procedures as normal science observations, but the  
\chg{specific characteristics}  have been adapted through the years to accommodate  the needs and constrains of the observatory.
Currently and from Cycle 5 on, observations for the calibrator survey are performed using the Morita ALMA compact array (ACA) and the ACA correlators. 
The grid calibrators are observed in partially overlapping  LST range groups together (ideally) with one of the
 main primary \chg{flux} calibrators:  Uranus, Neptune, Callisto, Ganymede \chgg{and Mars}. 
After a pointing calibration, system temperature measurement helps  correcting  for atmospheric opacity 
including the most conspicuous atmospheric absorption lines.
The bandpass solution is determined from the brightest source of \chg{each} LST group. Phase self-calibration is performed directly onto the targets.
On-source time spent on each grid calibrator varies between 2--4 minutes.   \chg{More details about the design,
observational strategy, and data reduction procedure  of the grid calibrator survey can be found in \citeasnoun{almamemo599}.}

\Tref{tab-data} shows some of the observational characteristics of \totalsamplesize\ \chg{grid} calibrator sources. 
The  \emph{Joint ALMA Observatory} (JAO) revises and updates the list of sources used \chg{for secondary flux calibration,
which makes the full set of sources being used   to slightly vary over time.}
The sample in \Tref{tab-data} can be considered the ``core"  set of \chg{sources}  used by ALMA at least during \chg{Cycle 6.}
Columns (1) to (6) show, respectively, the source name, right ascension, declination, the starting day of observations (as it appears in the publicly available catalog), the last day of observations considered in this study, and the number of measured flux densities considered in this study. These sources have been observed approximately every two weeks, but in some cases with significant irregularity. There are four \chg{grid} calibrators with more sporadic observations which have  less than 150 measurements in total,  J0904$-$5735, J1127$-$1857,  J2000$-$1748, and J1331+3030. 
\chg{These  were included as grid calibrators only during the last year. This explains  why they have less measurements: the 
calibrator survey for non-grid sources monitors a sample of approximately 800 quasars \cite{almamemo599}, with necessarily a much lower cadence compared to the grid survey.}

The bulk ($\ge90\%$) of the \chg{grid source} observations consist of measurements of the flux densities at Bands 3 and 7, 
mostly at  91.5 and 103.5 GHz, and at 343.5 GHz, respectively. \chg{Because the relative separation of the LSB and USB spectral windows in Band 3 
is more than 10\% of the band's typical frequency, and  because it is possible to obtain sufficient SNR in each sideband of the Band 3 observations; it was decided by the calibrator group 
to report flux densities separated for both sidebands. The rest of the data consist mostly of observations at Band 6 and measurements taken at different frequencies within Bands 3 and 7. 
In addition, very sporadic observations at Band 4 were taken for sources J0334$-$4008, J1733$-$1304, and J2056$-$4714.} 

Using these data we can fit a simple spectral model defined by 
\begin{equation}
\ln\left(\frac{F_\nu}{{\rm Jy}}\right)=
\ln\left(\frac{F_{\nu_0}}{{\rm Jy}}\right)+\alpha\ln\left(\frac{\nu}{\nu_0}\right)~~,\label{eq-si}
\end{equation}
 where $\alpha$ is the spectral index and $\nu_0$ is a fiducial frequency \chg{arbitrarily} chosen to be 100 GHz. 
 Using  model \eref{eq-si}, we obtain least squared fitted parameters  to all the data available for each calibrator,  
 disregarding its variability. These least-squares best-fit parameters 
 represent time-averaged values, and we denote them $\langle F_{100}\rangle$ and
  $\langle\alpha\rangle$, given in columns (7) and (8) of \Tref{tab-data}.  Finally, column (9) shows the variability index 
  \citeaffixed{Bonato2018MNRAS}{\textit{vix,}} defined as 
  \begin{equation}{\it vix}=\frac{100}{\langle S \rangle\sqrt{N}}\sqrt{\sum\left(S_i-\langle S\rangle\right)^2-\sum\sigma_i^2}~~,\label{eq-vix}\end{equation}
  where $\langle S\rangle$ is the time-averaged flux density of the source, 
  calculated at 100 GHz using the  fitted law defined by \eref{eq-si}, and $\sigma_i$ is the observational uncertainty \chgg{taken from the source catalog}.
 The \textit{vix} provides a quick quantitative estimation of the percentage of relative  time-variability of the source.

\Fref{fig-data} shows the data of the source J0635$-$7516, which will be used  as a representative example of the data obtained for the grid calibrators. Similar plots for the rest of the sources in \Tref{tab-data} can be found as supplementary material  (\ref{sec-supp}). 
In this work, we adopt the approach followed by previous optical/IR studies of quasar variability  
which have used the magnitude (or logarithmic) scale as it is better adapted for Gaussian stochastic modeling \citeaffixed{Liodakis2017MNRAS,Kushwaha2016ApJ,Edelson2013ApJ,Kelly2009ApJ}{e.g.,}.  In the case of the mm/sub-mm data of the grid calibrators, there are two additional reasons: as shown in \Fref{fig-data},  the light-curve of Bands 3 and 7 data  looks  similar but displaced by a constant amount in the log-lin plot, 
which  suggests a relative flux change  with a stable spectral index. This type of relative variability is  more simply modeled as additive quantities in the logarithmic scale. The second reason is that the main source of flux uncertainty likely arises from multiplicative (calibration-type) errors.  Typically,  for these bright sources the additive,  thermal noise uncertainty does not rise above 1\% \cite{2012ALMAN...9....8V}.
 
According to \Tref{tab-data}, most  sources (35 from \totalsamplesize) have fluxes projected at 100 GHz below 6.5 Jy, with four sources above 9 Jy. 
The spectral indices range between $-0.2$ and $-1.1$, with 32 out of \totalsamplesize\ spectral indices between $-0.35$ and $-0.80$. 
The 100 GHz \textit{vix} spans  between 6.7 and 60.7\%. \chg{The first, second, and third quantiles of \textit{vix} of the sources 
with monitoring times within 30\% of 800 days  in the source frame given by  Bonato et al.\ (2018, Table 2) are 13.9, 23.9, and 32.8. These values are comparable 
 with the same quantiles of the \textit{vix} of those sources in Table 1, with monitoring times compatible with the 800 days criterion, which are 13.5, 19.0, and 37.3, respectively.
 Since the flux and spectral indices of these sources are variable, the mean spectral index \eref{eq-si} is highly dependent on the time-frequency sampling, which is typically very different between the calibrator  survey and the study by \citeasnoun{Bonato2018MNRAS}. 
 In any case, the absolute differences in mean spectral index between Table 1 and those calculated for the data presented in \citeasnoun{Bonato2018MNRAS} have a median of $0.04$, and are in all cases $<0.24$. 
 \Fref{fig-histos} shows histograms of $\langle F_{100}\rangle$, $\langle \alpha\rangle$, and \textit{vix} for the grid sources.}

%
%


\begin{figure}
\centering\includegraphics[width=0.7\textwidth]{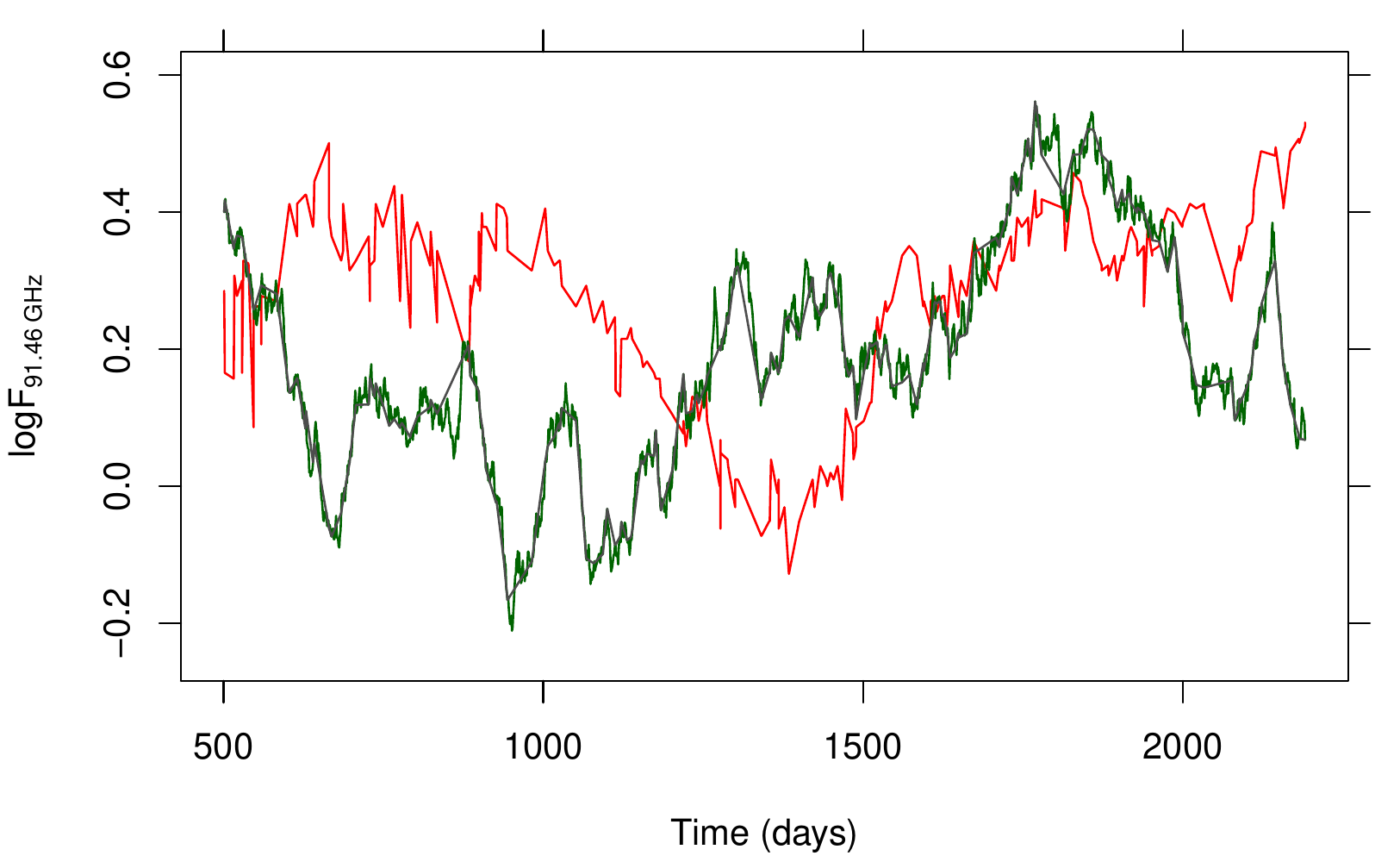}\\
\includegraphics[width=0.7\textwidth]{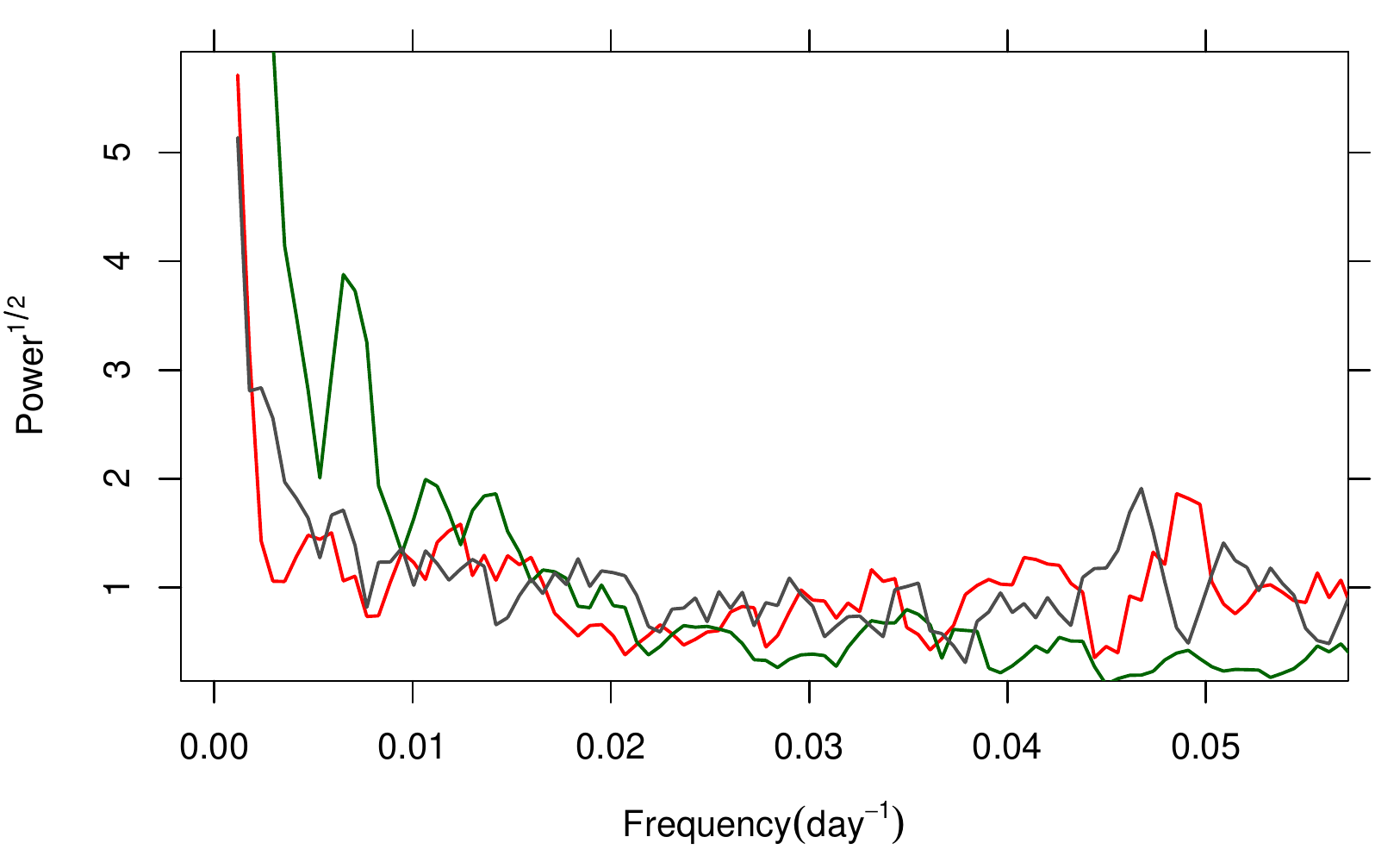}%
\caption{\emph{Top panel.} Red curve shows the  log-flux time-curve of J0635$-$7516 at 91.46 GHz. \chg{Days in the abscissa are taken with respect to the start day of observations (Table 1)}. Green curve shows  a random realization of an  OU process with decorrelation time 500 d and variance rate 0.015 (see \sref{sec-met}), sampled every day.  Gray curve shows a random realization of the same  OU-process sampled at the same times as the red curve. \emph{Bottom panel.} Lomb-Scargle periodograms of the data on top panel (same color coding).\label{fig-lsp}}
\end{figure}

Another important  characteristic of the quasar variability is the apparent lack of periodic components \citeaffixed{Goyal2018ApJ}{e.g.}. A method to assess the presence of (quasi-) periodicity in the data applicable to  irregular time sampling  is to examine its Lomb-Scargle periodogram \cite{2018ApJS..236...16V}.  This type of periodogram  is analogous to a Fourier transform, and is sensitive to periodic behavior of  the auto-covariance function.
The red lines in the top and bottom  panels of  \Fref{fig-lsp} show the 91.46 GHz flux densities of  J0635$-$7516  and its periodogram, respectively.  We calculate 
the periodogram using the \texttt{lsp} task within the \texttt{R} software \cite{R}. 
It is patent that the only conspicuous peak occurs at \chgg{zero} time-frequency, after which the periodogram power decreases toward higher time-frequencies.
This same behavior can be reproduced with stochastic models with autocorrelation, but without any periodicity. For example,  the green line in \Fref{fig-lsp}  shows a random realization
of an OU-process (see \sref{sec-met}) with decorrelation time 400 d and variance rate 0.015, sampled evenly every 7 days.  The parameters of the process were selected to nearly match  the variability index of the data shown in \Tref{tab-data}.

We stress on a note of caution: one noticeable difference between the periodograms of the green model and the data 
is that the  power of the latter seems to plateau  at time-frequencies above 0.01 d$^{-1}$ at $\approx1$~($\ln$~Jy)$^{2}$, whereas the 
periodogram  power of the model decreases toward zero more steadily. This may be in part due to the presence of nearly 
uncorrelated white noise in the data of the quasar (e.g., instrumental errors), but we find that the main explanation is probably due the inhomogeneous  
sampling. This is illustrated by the gray curves in \Fref{fig-lsp}, which  show another realization of the same OU-process as the green model, 
but sampled at the times of the red data curve.  Similar to the periodogram power of the data,  the  gray curve 
decreases much more slowly toward zero compared with the green curve, and both display an excess of power 
around 0.05 d$^{-1}$ --- which naively  may hint to quasi-periodicity --- but it is an effect attributable only to the inhomogeneous sampling.


\section{Methods}\label{sec-met}

To analyze the time-series data of the calibrator, we follow  an additive ``classical'' decomposition approach \cite{Brockwell2002ITS&F}. 
According to this notion, we can express the log-flux  (or any other adequate function of the flux) 
at time $t$ and frequency $\nu$ as the sum
\begin{equation}
\ln F_\nu(t)=P_\nu(t)+y_\nu(t)~~,\label{eq-cd}
\end{equation}
where $P_\nu(t)$ and $y_\nu $ are  the `deterministic' and `stochastic' models, respectively. The flux density $F_\nu$ unit is Jy, unless stated otherwise.
$P_\nu$ and $y_\nu$ are characterized by several parameters, which  need to be estimated from the data (e.g., through maximum likelihood estimation).  

The frequency dependence of the deterministic models consist  of (possibly slightly modified) power laws in frequency.
For this study, we assume that the stochastic model is stationary and centered (i.e., zero expectation). Non-stationary features like
 trends (that is, a long-term increase/decrease of flux) or  seasonal (periodic) components can be included in the deterministic modeling. 
 
\chg{ The stochastic model is based on the OU-process, used to describe the light-curves of quasars by \citeasnoun{Kelly2009ApJ}. 
Subsequent studies have confirmed  this is a reasonably adequate and simple model for the stochastic nature of the light-curves of many quasars 
 \cite{Kozlowsky2010ApJ,MacLeod2010ApJ,Ruan2012ApJ,Andrae2013AA,Wang2019APSS}. It has been found, however, that  sometimes this model is too 
 simple  to describe the variability on very high cadency data \citeaffixed{Kasliwal2015MNRAS}{$\sim10^3$ s,}. Extensions to the simple OU-process 
   have been proposed by, for example, \citeasnoun{Kelly2011ApJ} (finite mixtures), \citeasnoun{Kelly2014ApJ} (continuous auto-regressive mean-average, CARMA),
  and  \citeasnoun{Takata2018ApJ} (infinite mixtures).}
 
 In its most general form, the  stochastic models proposed in this work consist of \chg{finite} mixtures  of OU-processes \cite{Kelly2011ApJ}
for the log-flux of the quasar at a fiducial frequency and for its spectral index, plus uncorrelated noise. 
\chg{An important difference between  the modeling performed in previous studies and this one is the inclusion of a spectral deterministic model for the source, which 
allow us to fit together the data at different frequencies within a single model. In contrast, the stochastic analysis of multi-frequency data is done usually  \citeaffixed{Goyal2018ApJ}{e.g.,} 
on monochromatic  light-curves.}
We will denote the process including a mixture of $r$ OU-processes for the fiducial log-flux and 
$s$ OU-processes for the spectral index as OU-$r$F$s$a.  In the literature, one of the  extensions of a single  OU-process 
is  a CARMA process. An OU-$r$F$s$a process is equivalent to
a \mbox{CARMA($r+s, r+s-1$)} process  \cite{Kelly2014ApJ,10.2307/2345178} associated with a characteristic equation with only real solutions. 

We can gain insight of the time-series methods applied in this paper analyzing one of \chg{its simplest models}:
 the Gaussian autoregressive (AR) model \cite{Scargle1981ApJS}.
In  \ref{sec-ar1} we present a short description of the AR(1) process, which is an evenly sampled, 
discrete time, and noiseless version of the  stochastic processes  we describe  in the following sections. 
The main characteristics, advantages,  and demerits of the modeling we propose in this work can be generally understood
 from the AR(1) process. 

\subsection{Ornstein-Uhlenbeck mixture models.}

\subsubsection{Constant spectral index: OU-1F0a process.}\label{sec-OU}%
We start describing the simplest mixture model:  one OU-process characterizing the log-flux at a fiducial frequency, and using 
\eref{eq-si} for the fluxes at other frequencies. This OU-1F0a model is defined by  
\begin{eqnarray}
\fl P_{n}=\mu+\alpha\ln\left(\nu_n/{\nu_0}\right)~~,\label{eq-csi}\\
\fl y_n=f_n+\sigma_N(\nu_n) Z_n~~,\label{eq-ynOU10}\\
\fl f_n=e^{-\delta t_n/\tau}f_{n-1}+ \sqrt{\frac{\varsigma^2\tau}{2} \left(1-e^{-2\delta t_n/\tau}\right)}Z'_n~~.\label{eq-ou10}
\end{eqnarray}
\Eref{eq-csi} defines the  deterministic model $P_\nu$ measured at time $t_n$ ($n\ge1$). The observed frequency at time $t_n$ is $\nu_n$. The deterministic model depends on two parameters:  the spectral index   $\alpha$  and the long-term mean log-flux at the fiducial frequency ($\nu_0=100$~GHz), $\mu$.
In \Eref{eq-ynOU10}, the stochastic component $y_n$ (see \eref{eq-cd}) consist of the sum of a Gaussian noise of variance $\sigma_N^2(\nu_n)$ and 
 a time variable term $f_n$. 
We introduce  heteroscedasticity in the model assuming that the noise term $\sigma_N(\nu)$ depends on frequency as  $\nu^{1/2}$. 
This dependence  roughly matches  the \chg{estimated ratio of the flux density uncertainty at \chgg{Band 7 and Band 3}, which is $\approx2$  \cite[\S 10]{ATM6}. 
The  ratio between the average flux uncertainties at \chgg{Band 7 and Band 3} given in the source catalog  is $2.1\pm0.4$, where the error represents the dispersion among the grid sources.} 
  
  The $f_n$ term follows an OU process, defined explicitly in \eref{eq-ou10}, 
  where the  time between observations is  $\delta t_n=t_n-t_{n-1}$.  Both $(Z_n, Z'_n)$ 
are a series of uncorrelated normal standard random variables.
 The stochastic parameters of the model are three:
\begin{itemize}
\item[]{$\varsigma^2$, the variance rate (time$^{-1}$ units),}
\item[]{$\tau$, the decorrelation time (time units),}
\item[]{$\sigma_N^2(\nu_0)$, the uncorrelated \chg{(``measurement'')} noise \chg{at the fiducial frequency}.}
\end{itemize}
Equations \eref{eq-csi} to \eref{eq-ou10} also form a state space representation (see \ref{sec-ssr}) 
of the stochastic process $\ln F_{\nu}(t)$, in which the ``state" role  is taken 
by $f_n$. 
The variance rate is \chg{called} this way because for $\delta t_n\ll\tau$,
 the variance of $\ln{F_0(t_n)}$ assuming  $\ln{F_0(t_{n-1})}$  is given by 
 $\varsigma^2\delta t_n$.  Note that it is not rare to have 
 simultaneous measurements (taken during the same day, in our case) of the source at different frequencies. 
 These observations can be included as part of the time series  in any order. 
  They  cannot give us information about the time variability, 
 but they are  useful to estimate the uncorrelated noise level  $\sigma_N$.
 

\subsubsection{OU-$r$F$s$a plus noise process.}\label{sec-OUp}%
This is a more generic model than the one presented in the previous section. It consists of  a mixture of OU-processes for the flux  and for 
the spectral index. They are defined by
\begin{eqnarray}
P_n=\mu+\alpha\ln\left(\nu_n/{\nu_0}\right)~~,\label{eq-vsi1}\\
 y_n=\sum_{i=1}^r f^i_n+\sum_{j=1}^s\beta^j_n\ln\left(\nu_n/{\nu_0}\right)+\sigma_N(\nu_n) Z_n~~,\label{eq-vsi}\\
\left(\begin{array}{c}\mathrm{\bf f}_n\\ \bm{\beta}_n\\\end{array}\right)%
= e^{-\delta t_n \bm{\tau}^{-1}} \left(\begin{array}{c}\mathrm{\bf f}_{n-1}\\ \bm{\beta}_{n-1}\\\end{array}\right) +\eta_n ~~,\label{eq-srgral}\\
\bm{\tau}=\mathrm{diag}\left(\tau_1,\ldots, \tau_r,T_1,\ldots,T_s\right)~~,\label{eq-taumatrix}\\[.5ex]
\bm{\Sigma}=\mathrm{diag}\left(\varsigma_1\ldots,\varsigma_r,\zeta_1,\ldots,\zeta_s\right)~~,\label{eq-sigmamatrix}\\
\eta_n \sim\mathrm{GWN}\left(\bm{0}, \int_{t_{n-1}}^{t_n} e^{(t_n-s)\bm{\tau}^{-1}}\bm{\Sigma}\,\Gamma\,\bm{\Sigma}e^{(t_n-s)\bm{\tau}^{-1}}ds\right)~~.\label{eq-aleamat}
\end{eqnarray}
In \eref{eq-vsi1}, the deterministic model is the same as in the previous section, but $\alpha$ in this context represent the long-term mean spectral index.
In \eref{eq-vsi}, $f^i$ and $\beta^j$ are centered  OU-processes and the components of the vectors $\mathrm{\bf f}$ and $\bm{\beta}$, respectively. 
The definition of the OU-process mixture is given by Equations \eref{eq-srgral} to \eref{eq-aleamat}. 
 In Equations \eref{eq-taumatrix} and \eref{eq-sigmamatrix} 
$\mathrm{diag}(\ldots)$ represents a diagonal matrix, with the listed values as the diagonal elements.
Finally, in \eref{eq-aleamat} $\Gamma$ is a symmetric $(r+s)\times(r+s)$ matrix  representing  cross-correlations between the terms, 
with $|\Gamma_{k,l}|<\Gamma_{k,k}=1$ for all $k,l=1,\ldots,(r+s)$, $k\neq l$.
Decorrelation times and variance rates are given by $\tau_i$ ($T_j$) and $\varsigma_i$ ($\zeta_j$) for each $f^i$ ($\beta^j$), 
respectively. In \eref{eq-aleamat}, GWN stands for Gaussian white noise, meaning that $\eta_n$ forms a series of uncorrelated
--- in time --- multivariate normal variables. 
We emphasize that the model includes normal variables in two instances: as a ``noise" term  in \eref{eq-vsi}, and as part of the OU-process in \eref{eq-aleamat}.
Similarly as in \sref{sec-OU}, Equations \eref{eq-vsi} to \eref{eq-aleamat} define a state space representation of the measurements with a state  
  vector $(f^1,\ldots,f^r,\beta^1,\ldots,\beta^s)^\mathrm{T}$.

In practice, for this work we use  $r\le2$ and $s\le1$ models. Note that the matrix $\Gamma$ defines a set of additional $(r+s)(r+s-1)/2$ 
parameters which need to be adjusted or assumed.  While choosing $\Gamma$ equal to the identity matrix --- that is, no cross-correlations ---
might look like a natural choice, we find that it has the undesirable consequence of making the model too dependent on 
 the fiducial frequency. Indeed, at the fiducial frequency the model would be characterized by 
only $f$ terms, and no $\beta$ terms, effectively reducing the order of the process. 
Since the fiducial frequency (100 GHz) is arbitrarily chosen, this dependence is unjustified.
 Including cross-correlation terms allows us to moderate this  artificial and unbalanced role given to the fiducial frequency.
 Ultimately, the  reason  behind this complication lies in the difficulty of describing a varying spectral index independently 
 of flux variations, \chg{since these are correlated.}
An alternative to include cross-correlations may be to leave the fiducial frequency 
 $\nu_0$ as a free parameter, possibly even time-variable. 
Exploring all these alternatives, however, is beyond the scope of this work.

For a general OU-1F1a process (dropping the $i=j=1$ indices), the cross-correlation between $f_n$ and $\beta_n$
is defined by a single parameter $\rho$, with  $\Gamma_{1,2}=\Gamma_{2,1}=\rho$. Hence, $\eta_n$ is a 
centered bi-normal variable with covariance matrix
\begin{equation}
\left[\begin{array}{cc}
\frac{\varsigma^2 \tau}{2}\left(1-e^{-2\delta t_n/\tau}\right) & \rho\frac{\varsigma \zeta \tau T}{\tau+ T}\left(1-e^{-\left(\frac{\delta t_n}{\tau}+\frac{\delta t_n}{T}\right)}\right) \\ 
 \rho\frac{\varsigma \zeta \tau T}{\tau+ T}\left(1-e^{-\left(\frac{\delta t_n}{\tau}+\frac{\delta t_n}{T}\right)}\right) & \frac{\zeta T^2}{2}\left(1-e^{-2\delta t_n/T}\right)\\
\end{array}\right]~~.\label{eq-11gral}
\end{equation} 

\subsubsection{Additional features.\label{sec-soph}}

According to the prevalent physical model of blazars, their   emission  is mainly   synchrotron 
radiation arising from a relativistic jet, directed nearly in our line of sight, and
powered by accretion onto a supermassive black-hole 
\cite{Blandford1979ApJ,Begelman1984RvMP}.  Shocks within the jet may increase the luminosity of the jet, and they may be the root cause of some of the observed variability \cite{Ulrich1997ARA&A}. 
Based more or less on this physical picture, we study two more additions to the model which help to fit  better the quasars light-curves \citeaffixed{Trippe2011AA}{e.g.,}. 
These are:

\noindent \emph{Spectral curvature.}

Synchrotron cooling  may cause a decrease in the spectral index of the emission starting at high frequencies, 
an effect also known as ``ageing" of the spectrum \cite{1988gera.book.....K}.  We  can generalize the spectral 
model for $P_\nu$ in the following way \chg{\citeaffixed{Xue2016MNRAS}{e.g.,}}
\begin{equation}
P_\nu=\mu+\left(\alpha+\kappa\ln(\nu/\nu_0)\right)\ln(\nu/\nu_0)~~,\label{eq-conc}
\end{equation}
which amounts to a spectral index varying with frequency. The sign of  $\kappa$ being negative or 
positive defines \chg{what we call} concave and convex spectral models, respectively. 
The local spectral index  at frequency $\nu$ is 
\begin{equation}
 \frac{d\ln(F_\nu)}{d\ln(\nu/\nu_0)}=\alpha+2\kappa\ln(\nu/\nu_0)+\sum_{j=1}^s\beta^j(t_n)~~.\label{eq-lsi}
 \end{equation}

\noindent\emph{Frequency depending lag (\nulag).}

A common interpretation of the mm/sub-mm blazar variability is provided by the  shock-in-jet model 
\chg{\cite{2004ApJ...613..725S,Joshi2011ApJ,2011Natur.477..185H}}. According to 
it, flares of various intensities are generated by the development of shock waves traveling downstream through the jet body.   
This type of model produces 
 naturally time-variable emission features with frequency depending  light-curves 
\citeaffixed{2015A&A...580A..94F}{e.g.,}. Light-curves with comparable frequencies display qualitatively similar morphology, 
but with a time displacement or delay dependent on frequency. 
 Therefore,   we  add another potential feature to the  blazar light-curves
 by  mixing frequency and temporal coordinates, that is, 
\begin{equation}
y_n(t,\nu)=y_n(t+\Delta t(\nu))~~,\label{eq-lag}
\end{equation}
where $\Delta t(\nu)$ is a frequency dependent time lag, chosen to be $\Delta t(\nu)=\epsilon \ln(\nu/\nu_0)$. 
A positive $\epsilon$ is associated  with  corresponding 
variations at   frequencies higher than $\nu_0$  occurring at later times than those observed at $\nu_0$, and vice-versa.

In the shock-in-jet model, the mm/sub-mm emission arises mainly from synchrotron. Due to the synchrotron opacity
increasing upstream in the jet and decreasing with frequency, we expect higher-frequency emission tracing an \emph{earlier} development
of the shock wave. That is, the most common physical interpretation predicts only negative $\epsilon$, as it is 
commonly observed \cite{2014MNRAS.441.1899F}. 
We must note also that there is the possibility of degeneracy between the \nulag\ and a variable spectral index, specially 
affecting observations at only two frequencies. 
For example, a short flare led by the high-frequency flux can be equally well described by a sudden increase and later  
decrease of the spectral index. Alternatively, a decrease and later increase in the spectral index may be interpreted as a 
positive lag. We test the ability of the model to resolve  this  degeneracy with data at only two bands 
 in \Sref{sec-sim}.
For the current modeling, we leave the sign of $\epsilon$  free and evaluate its reliability
later.

\subsection{Determination of the best-fit stochastic model.}

Estimation of the best-fit   model parameters and their uncertainties  is done through likelihood 
maximization. To calculate the likelihood function we use the state space representation of each 
process together with the Kalman recursions, as described in \ref{sec-ssr}.  The Kalman recursions \cite{Hamilton1994TSA}
are a group of recursive equations which allow us to obtain best estimators (in the least squares sense) 
of forecasts and interpolations of stochastic time series. For Gaussian time series,  Kalman recursions provide the
conditional expectations of forecasts and interpolations given the observed data.

Briefly,  the procedure to obtain the maximum likelihood estimators (MLE) has the following steps
\begin{enumerate}
\item[1.]{For each set of parameters, subtract from the log-fluxes the deterministic model.}
\item[2.]{Calculate the log-likelihood as a function of the  stochastic parameters and the Kalman predictions  \eref{eq-ll} (\ref{sec-ssr}).}
\item[3.]{Re-iterate 1--2 in order to maximize the log-likelihood with respect to the free parameters. We use the \texttt{Minuit} \cite{JAMES1975343} 
package implemented within \texttt{PDL}\footnote{\texttt{http://pdl.perl.org/}} to obtain the  MLE and their uncertainties. }
\end{enumerate}

\subsubsection{Quality of the best-fit model.} \label{sec-gof} %
Testing the hypothesis that the data originated from a certain stochastic model entails evaluating
 the fit according to the following criteria \citeaffixed{2005MNRAS.361..887K}{e.g.,}:
\begin{enumerate}
\item{The Akaike information criterion \citeaffixed{Feigelson2012msma}{AIC,}.  Defined as $2k-2\ln\mathcal{L}_{\rm max}$, where $k$ 
is the number of free parameters of the model and $\mathcal{L}_{\rm max}$ is the maximum attained likelihood. 
The AIC allows us to compare between models and make a first evaluation about the merit of including more free parameters in the model.
Better-fit models are associated with a lower AIC.}
\item Normalized residual plot. We define the normalized residuals of the observed stochastic components $y_n$ with
respect to the best-fit predictions $y_{n|n-1}$ by
\begin{equation}
\chi_n=\frac{y_n-y_{n|n-1}}{\sqrt{r_n}}~~,\label{eq-res}
\end{equation} 
where the notations follows from \eref{eq-predY} (\ref{sec-ssr}), that is,  $y_{n|n-1}$ is the best forecast of the n-th measurement 
and $r_n$ is the   uncertainty of this forecast.
We inspect the plot of residuals vs.\ time coloring them by  band in order to identify  any systematic 
effect depending on frequency or time, for example, a change in spectral index. 
\item According to the hypotheses, $\chi_n$ in \eref{eq-res} should resemble a standard GWN series for large $n$ \cite{Kelly2011ApJ}. 
We test the normality of the residuals running the Anderson-Darling test \cite{Feigelson2012msma} using the \texttt{ad.test} 
task from package \texttt{goftest} \cite{goftest}. We also examine the histogram of $\chi_n$ and compare it with a standardized 
Gaussian probability density.
\item We test the hypotheses that $\chi_n$ forms an uncorrelated white noise series by calculating its autocorrelation function 
(ACF) using the task \texttt{acf} from  \texttt{R}. For a white noise sequence, the only significant peak of the
 ACF should be located at zero. 
\end{enumerate}

The AIC is more useful as a relative criterion to compare between different models, while the other three  evaluate the  goodness-of-fit.
\chg{These goodness of fit  criteria test how well  the stochastic and deterministic models can describe the calibrators light-curves, 
leaving final residuals consistent with Gaussian white noise.}
Additional features  may provide a slightly better fit to the data (model more adequate) but at the expense of
 introducing  excessive sophistication and degeneracy between the parameters (less parsimony). Some heuristic assessment of the
 merits of the optimized models is necessary to finally decide which is the best model for each source.
From the best-fit model, using the Kalman recursions  (\ref{sec-ssr}) we derive predictions and interpolations for the 
process at arbitrary time, together with its uncertainties. 

\subsection{Uncertainty of the predictions}\label{sec-uncs}

There are three types of uncertainty associated with the predictions or forecasts made by the model. The first two are 
determined by $\sigma_N(\nu)$ (what we call \chg{measurement noise}), and the stochastic variability, determined by the variance rates and decorrelation times 
\chgg{through the Kalman recursions (\Eref{eq-predY} in \ref{sec-ssr}).
 The modeling assumes these two types of uncertainty are uncorrelated, and its sum in quadrature is denoted by $r_n$ (\ref{sec-ssr}).} 
This would be the only source of uncertainty for a process with perfectly determined parameters.
The third type of uncertainty \chg{is  derived} from the uncertainty associated with the best-fit parameters.

\chg{The} actual flux prediction is not given by $y_{n|n-1}$, but by $P_{\nu_n}+y_{n|n+1}$ \eref{eq-cd}.  In both the deterministic model $P_\nu$
 and in the stochastic model, \chg{best-fit} parameters have been estimated from the data itself. Therefore, there is an error level associated with 
 \chgg{the ambiguity about which model best fit the data.} We refer hereafter to this forecast (or interpolation) error level  as the ``parametric"  uncertainty.
Uncertainties of the MLE parameters can be estimated using their covariance matrix calculated by \texttt{Minuit} using the
 second derivative matrix of the likelihood function at maximum.
When the fit performs poorly and the likelihood is very non-Gaussian --- which happens, for example, when the decorrelation times are long --- 
the covariance matrix is still  useful as an order of magnitude estimator.

To estimate the parametric uncertainty of the forecast, we consider the estimators as random variables  
around  the MLE solution with  the  covariance matrix describing its  uncertainty. 
This random variable is taken as 
independent of any single measurement (particularly, the last one), a hypothesis 
valid for large number of samples.
As a simplified example, \ref{sec-ar1} discusses  the  parametric uncertainty associated with an  unknown long-term mean $\mu$
for an AR(1) process. An estimation of this error term is given in  \eref{eq-parUncAR1}.
These formulae quantify the intuitive idea  that  the  most relevant information in order to make a 
forecast is the last (that is, the most recent) measurement. However, if this last sampling was taken  too long ago (several decorrelation times), 
it is no longer useful as a predictor, and the best forecast becomes the long-term mean  with an
error  comparable to the source's observed variability. 
 
 In order to obtain an explicit parametric uncertainty estimation, we construct our forecast using the  Kalman filter (\ref{sec-ssr})
 using only the last  measurement.  We also make the  assumption that there is no covariance between the $f$ and $\beta$ terms.
 The  forecast at time $t$ and frequency $\nu$, based on a single  flux $F_1$ measured at time $t=0$ and at frequency $\nu_1$, is given by
\begin{equation}
P_\nu+\left(\ln F_1-P_{\nu_1}\right)\frac{H e^{-t\bm{\tau}^{-1}}C H_1^\mathrm{T}}{\sigma_N^2(\nu)+H_1 C H_1^\mathrm{T}}~~,\label{eq-1forecast}
\end{equation}
where 
$C=\frac{1}{2}\mathrm{diag}\left(\varsigma_1\tau_1^2,\ldots,\varsigma_r^2\tau_r,\zeta_1^2T_1,\ldots,\zeta_s^2T_s\right)$ and 
 $H$ is given by \eref{eq-H} in \ref{sec-ssr}. 
Let the  gradient of \eref{eq-1forecast} respect to the parameters be $\bm{J}$. 
Then,  the uncertainty of the forecast of  $\ln F_\nu(t)$  is 
\begin{equation}\bm{J}\Sigma_\mathrm{MLE}\bm{J}^\mathrm{T}~~,\label{eq-pu}\end{equation}
where $\Sigma_\mathrm{MLE}$ is the  covariance matrix at the maximum likelihood  \citeaffixed{Bevington2003DRDP}{e.g.,}. 
Finally, for simplicity, we calculate the total uncertainty on the forecast as the sum in quadrature of the parametric and
 stochastic uncertainties.

\subsection{Performance in simulated data}\label{sec-sim}
We test the ability of the MLE procedure to recover the parameters of simulated, irregularly sampled data, under the 
hypotheses described in previous sections. 
In order to do this,
we generate 100 OU-1F1a simulated light-curves at 91.5, 103.5, and 343.5 GHz with  parameters
$\mu=0$, $\alpha=-0.7$, $\epsilon=-2$ (d), $\kappa=-0.1$, $\sigma_N=0.04\sqrt{\nu/\nu_0}$, $\tau=360$ d, $T=50$ d,  
$\varsigma=0.05~{\rm d}^{-1/2}$, $\zeta=0.01~{\rm d}^{-1/2}$, and monitoring time 1080 d. The curves have 400 measurements each with a 
30\% of the fluxes in Band 7, and the rest in Band 3, which is comparable to the frequency breakdown of the real data. 
We intentionally take the monitoring time covering only three  decorrelation times for the $f$ component in order to explore
the effects of a poorly constrained $\tau$  \cite{2017A&A...597A.128K}. 
We analyze the consequences of this particular limitation in more detail in \Sref{sec-dis} and \ref{sec-ar1}.
The simulated parameters produce light-curves with a similar appearance compared with those of  the grid calibrators. 
One example of these curves (dubbed J2600+0010) is shown in the \chgg{top} panel of 
\Fref{fig-recovery}, and a \texttt{R} script  is provided in the supplementary material to produce this one and the rest of the simulated light-curves.
\begin{figure}
\centering\hspace*{1em}\includegraphics[width=0.52\textwidth]{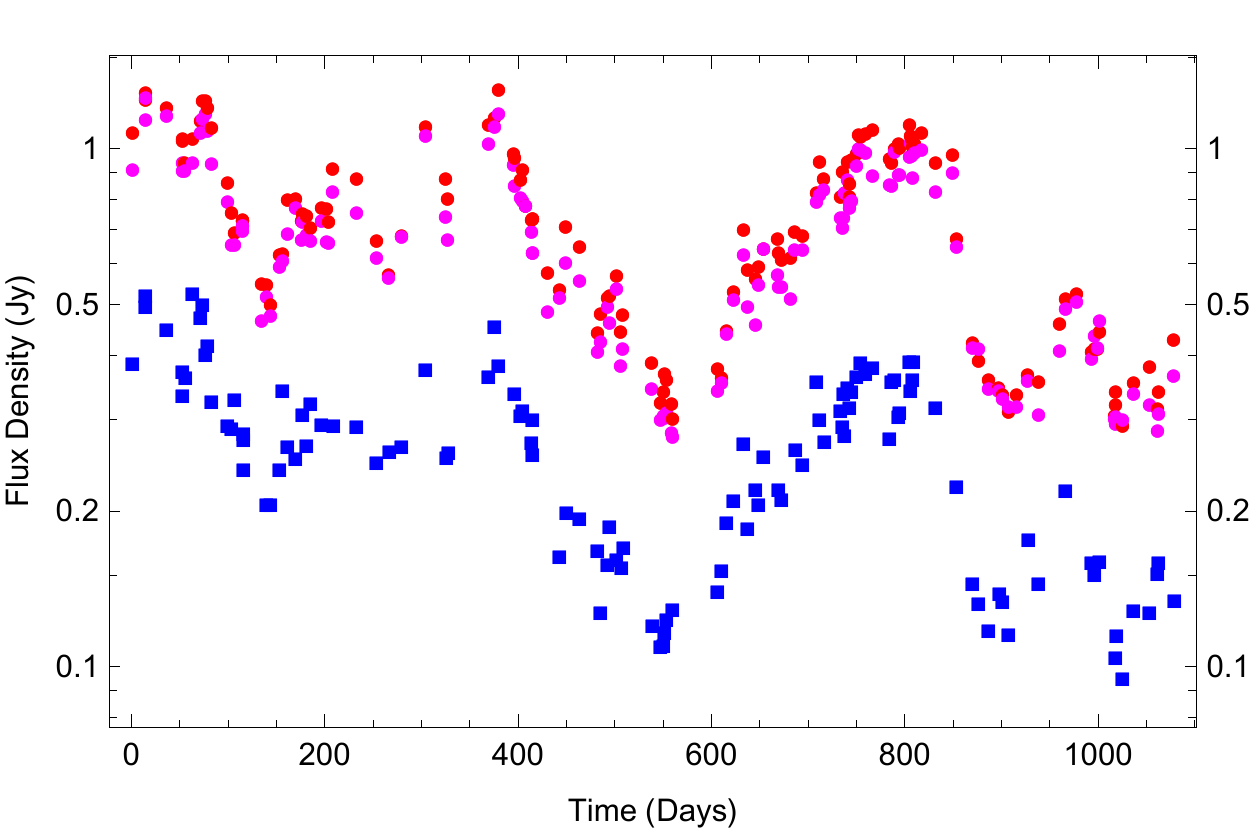}\\
\includegraphics[width=0.5\textwidth]{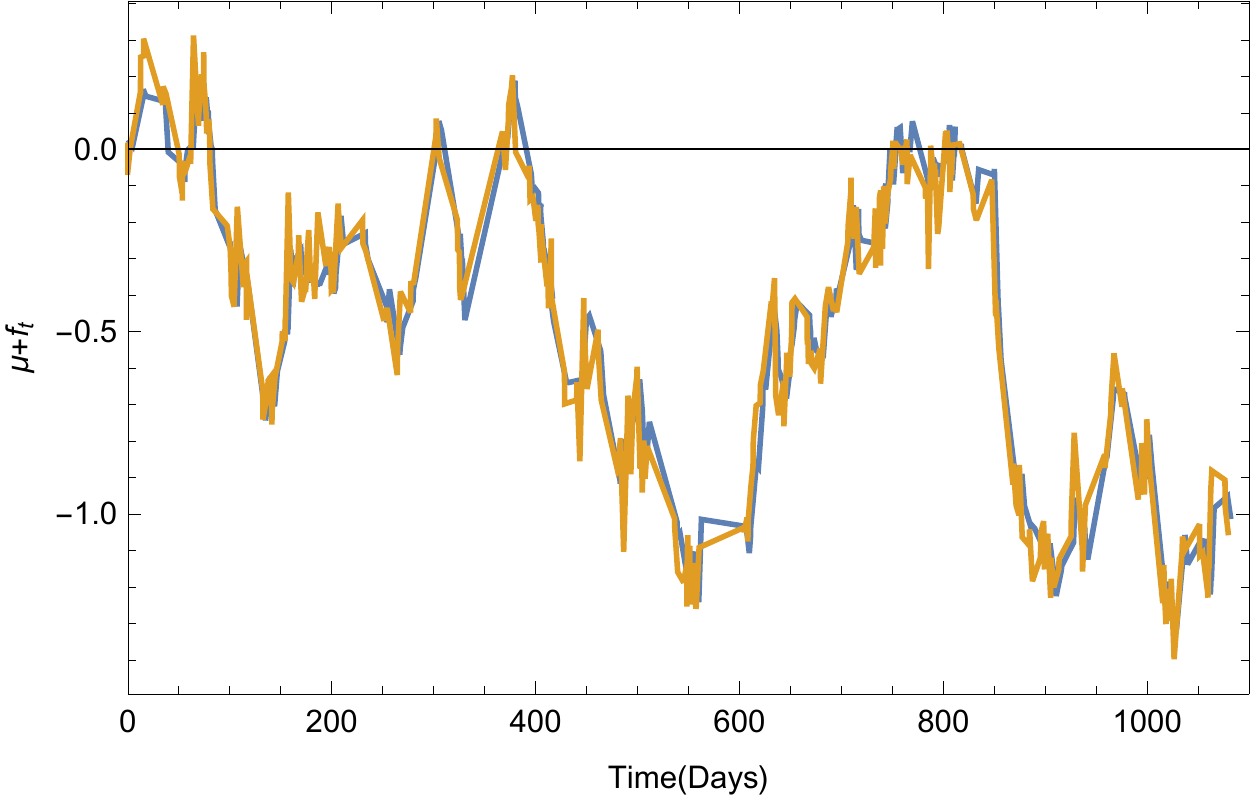}\\
\includegraphics[width=0.5\textwidth]{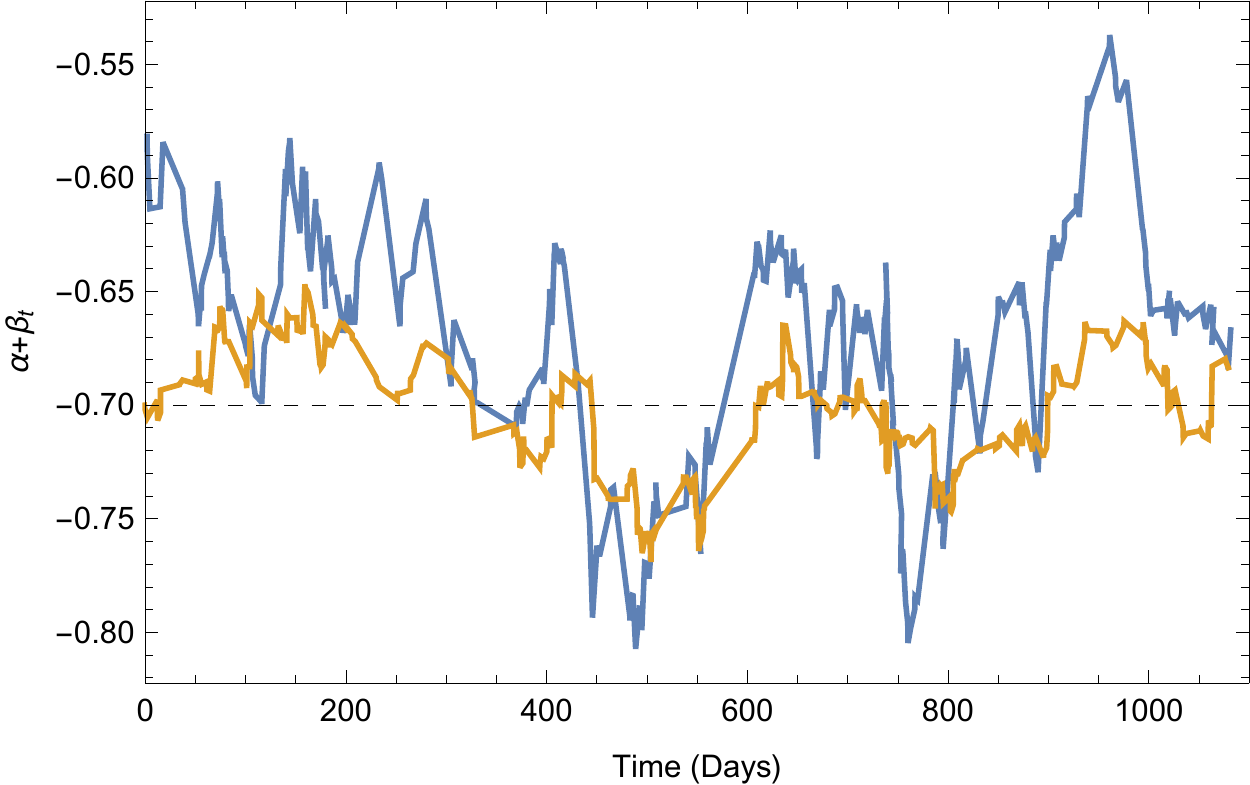}
\caption{\emph{\chgg{Top} panel.} One example (J2600+0010) of the 100 simulated OU-1F1a light-curves 
at \chgg{91.5 and  103.5 GHz (in red and magenta circles, respectively)}, and 343.5 GHz \chg{(blue squares)}.  
\emph{\chgg{Middle} panel.} Blue and orange lines show the simulated and MLE $\mu+f_n$ curves, respectively. 
\emph{\chgg{Bottom} panel.} Same as the previous panel but showing the $\alpha+\beta_n$curves. \label{fig-recovery}}
\end{figure}

\begin{figure}
\centering\includegraphics[width=0.55\textwidth]{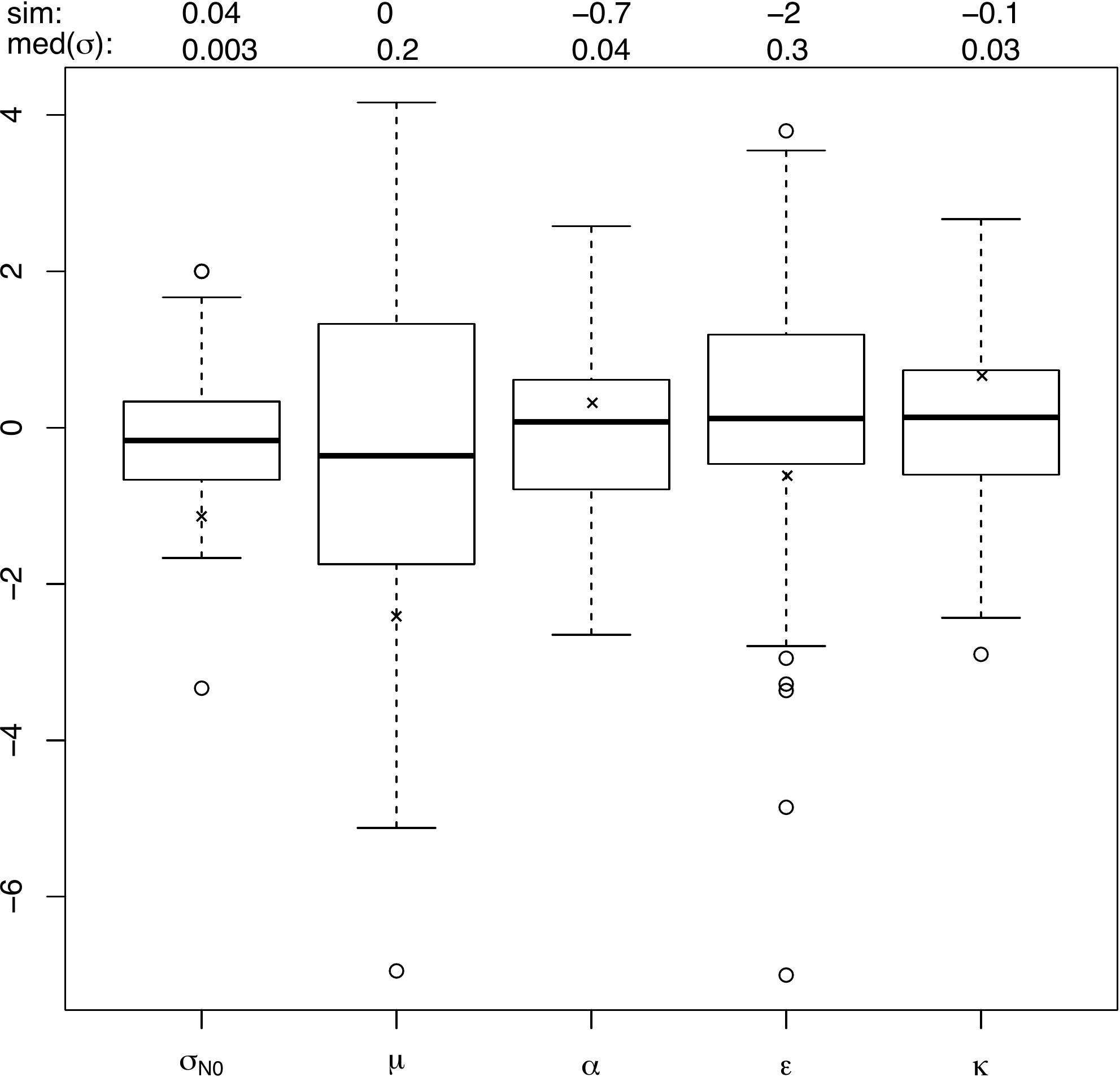}%
\caption{Box-plot \cite[Section 5.9]{Feigelson2012msma} of the scaled differences between the given simulated parameters $\sigma_{N,0}$, $\mu$, $\alpha$, $\epsilon$, and $\kappa$ (see \Sref{sec-OUp} and \ref{sec-soph}) and their MLE best fit, for the 100 synthetic light-curves. The differences have been scaled by the median MLE uncertainty  of each parameter. The values adopted for the simulations and the median MLE uncertainties  are given on top of the plot. The box-plot shows the median (thick black line) and   the interquartile range as boxes encompassing half (50) of the points. The error bars encompass the data within 1.5 times the inter-quartile distance from the box limits.  Points outside these limits are marked individually with circles. The crosses mark the recovered MLE parameters for the simulated light-curve J2600+0010, shown in \Fref{fig-recovery}.
   \label{fig-box}}
\end{figure}

\Fref{fig-box} shows box-plots of the differences between MLE recovered values and  simulated ones for the 100 synthetic light-curves. 
 We find that the  parameters describing the spectral shape of the source (like the spectral index and curvature parameter)
 are in general well recovered by MLE, with little bias. 
As expected, $\tau$ (not shown in \Fref{fig-box}) is not  recovered accurately. The median value of the MLE $\tau$ is 170 d, not close the value of the simulation. 
 For $T$, the MLE  median is 45 d, but with a very large dispersion. Large uncertainties in the decorrelation times also imply large uncertainties 
in the long-term mean estimations. Indeed, as shown in \Fref{fig-box}, the MLE of $\mu$ has a  relatively large dispersion respect to the simulated value, although not obviously biased.
The rest of the parameters seems to be reasonably well recovered, with the MLE uncertainty providing a sensible scale for its dispersion.
 
The simulations also allow us to estimate the behavior of the model in an idealized scenario.   Based on these results, we do not expect the 
fitting to recover the spectral index and its curvature parameter 
 with an uncertainty better than 0.05, and the \nulag\ parameter $\epsilon$ with an uncertainty better than 0.5 d, despite the MLE uncertainty of these
 sometimes suggesting even smaller error bars. Considering also that the time resolution of the  light-curves is 1 d, a minimum uncertainty of 
 0.5 d in the $\epsilon$ parameter seems sensible.
 
\Fref{fig-recovery} also shows the simulated and recovered $\mu+f_n$ and $\alpha+\beta_n$  for J2600+0010 in the \chgg{middle and bottom} panel, respectively; 
the latter  showing the variable spectral index of the simulated quasar. While $\mu+f_n$ is obtained very well by MLE, the spectral index seems less well
recovered. The MLE spectral index curve follows only qualitatively the simulated variable spectral index, with typical differences within 0.05. Note, however, that differences of this same order of magnitude are seen in the \chgg{middle} panel, but they are less noticeable due to the larger dynamic range. 

Part of the difficulties to recover the spectral index history may also be related to the need of observing at more than one frequency to get basic spectral information. 
\chg{Band 3 observations are more frequent than those at band Band 7. The monitoring is somewhat inhomogeneous \chgg{due to observational circumstances}, 
but we can define a mean cadency given by the average separation in days 
between consecutive observations. Among the grid sources, the median separation between consecutive 
 Band 7 observations is 17 days, while for Band 3 observations is 12 days.}
Henceforth, we do not expect to recover spectral index features on  \chg{timescales shorter than} the mean cadency of the Band 7 monitoring.

\section{Results}\label{sec-res}
\subsection{Best fit models.}
As a first step we  fit the OU-1F0a model to the grid calibrators. 
Table \ref{tab-OU-1F0a} lists the best fit parameters for each calibrator. Columns (2) to (6) give the MLE $\mu$, $\alpha$, $\sigma_{N,0}$ (at $\nu_0$),
$\tau$, and $\varsigma$ (see \eref{eq-csi} and \eref{eq-ou10}), respectively. 
We find that the heteroscedasticity prescription of $\sigma_N(\nu)$ depending as $\nu^{1/2}$ is  reasonable, providing in all cases 
a lower AIC compared either with a single $\sigma_N$ (homoscedasticity) 
or with a  heteroscedasticity prescription $\propto\nu$. Because data other than those  at Band 3 and 7 are very 
limited, we do not test extensively other dependences of $\sigma_N$ with frequency. 

The (1-sigma) parabolic uncertainties in \Tref{tab-OU-1F0a} have been calculated from the likelihood function around the minimum. 
Columns (7) and (8) show the AIC and the p-value of the Anderson-Darling test of normality of the residuals (\pad), respectively. 
Column (9) shows the ratio between the total time span of the monitoring $\Delta T$ 
versus $\tau$. 
This quantity is crucial to determine the confidence on the best-fit $\tau$ values \cite{2017A&A...597A.128K}. 
Values of  $\Delta T/\tau\ge10$ indicate reliable MLE $\tau$, while for $\Delta T/\tau<5$  the estimation is not reliable at all. 
Altogether, the  formal uncertainties of $\tau$ and $\mu$ are higher for  lower values of $\Delta T/\tau$. Also, 
 in this case,  the log-likelihood function around the maximum is not  symmetric and therefore the 
parabolic uncertainties only give a rough idea of the parameter error, which may be significantly skewed.

\begin{figure}
\centering\includegraphics[width=0.65\textwidth]{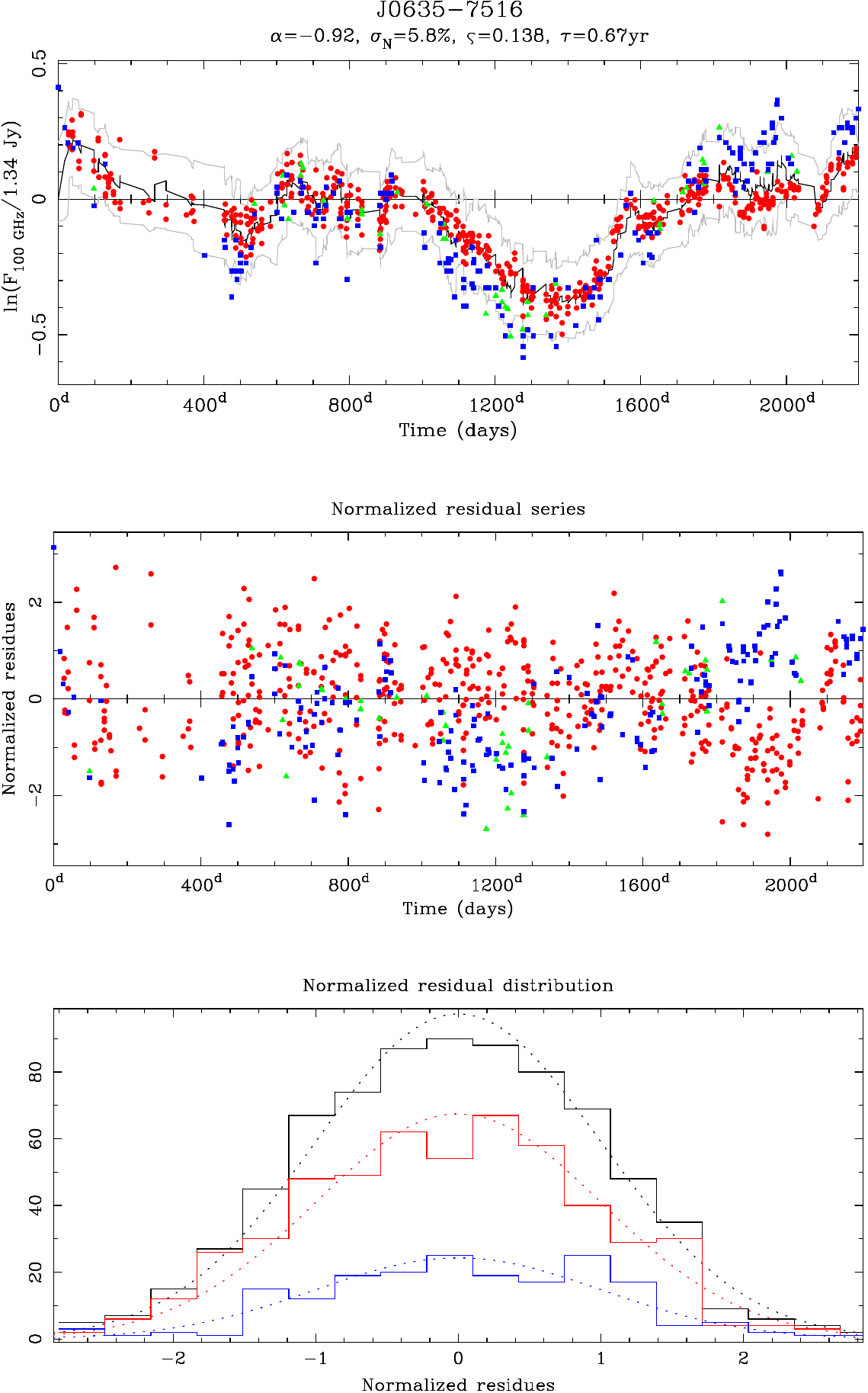}%
\caption{Best-fit OU-1F0a model applied to J0635$-$7516.  
\emph{Top panel.}  Centered log-flux curve at 100 GHz ($\ln(F_{100})-\mu$).  Red points and blue squares indicate Band  3 and 7 
data, respectively. Green triangles  represent data between  116--275 GHz (mostly Band 6, with very sporadic band 4 data).  Black line shows the predictions  based on previous data ($y_{n|n-1}$, see \eref{eq-predY} in \ref{sec-ssr} for notation). Grey lines encloses the 95\% confidence  interval around the predictions. 
\emph{Middle panel.} Normalized residuals $\chi_n$ vs.\ time  \eref{eq-res}.  \emph{Bottom panel}. Histograms of $\chi_n$. Black is all data, red and blue represent the Band 3 and Band 7 residuals, respectively. Dashed line shows  scaled standard Gaussian probability densities.
 \label{fig-OU10ex}}
\end{figure}

Despite the very simple hypotheses of the OU-1F0a model, it performs well as first approximation 
in optical data \cite{Kelly2009ApJ} and in the mm/sub-mm data of the grid calibrators. 
As a representative example, Figure \ref{fig-OU10ex} shows the results of the OU-1F0a fitting to J0635$-$7516.
The middle and bottom panels of Figure \ref{fig-OU10ex} show  the time-series of normalized residuals 
and its histogram, respectively.  The histogram and the \pad$ =0.28$ 
indicates  that the aggregated residuals are consistent 
with the normal hypothesis.  In fact, for 27 of the sources the Anderson-Darling test cannot 
reject the hypothesis of normality at the 0.05 confidence level. 
Similar plots as \Fref{fig-OU10ex}  for the \totalsamplesize\ grid calibrators are given
in the supplementary material.

The simple OU-1F0a model is also useful to identify possible  outliers in the data. For example, data points located at more than 
$5\sigma$ from the best prediction may be considered suspicious, specially if they do not seem to be associated with 
a more consistent and relatively longer ``flaring" type of activity. In any case, censoring data of these  varying 
sources should be performed conservatively. In the case of J0635$-$7516, we removed 8 data points, corresponding to a 1.06\% of the 
total amount of data: those  measurements  taken in 2013-12-29, and 
the data taken at the non-standard frequency 104.2 GHz at the end of January 2015. 
The remotion of these points does improve the general quality 
of the light-curve fittings.

\begin{table}
\caption{\label{tab-OU-1F0a} Best-fit OU-1F0a model parameter and results.}
\begin{indented}
\item[]\hspace*{-10em}\begin{tabular}{@{}crllllrlr}
\br
Source &$\mu$ & \multicolumn{1}{c}{$\alpha$} & $\sigma_{N,0}$ & $\tau$ & $\varsigma$ & AIC & \pad & $\frac{\Delta T}{\tau}$\\
            & (ln Jy)$^{\rm a}$           &                &         		 &  (yr)    	&  ($\sqrt{\rm yr}^{-1}$)& & &\\
\mr
J0006$-$0623 &  $1.161\pm0.08$ & $-0.565\pm0.006$ & $0.046\pm0.002$ & $0.96\pm0.3$ & $0.202\pm0.02$  & $-1017.4$ &   6.5E-01 & 6 \\ 
J0237+2848 &  $0.758\pm0.05$ & $-0.614\pm0.005$ & $0.044\pm0.002$ & $0.49\pm0.2$ & $0.260\pm0.03$  & $-942.0$ &   1.1E-01 & 12 \\ 
J0238+1636 &  $0.34\pm0.1$ & $-0.478\pm0.005$ & $0.043\pm0.002$ & $0.58\pm0.2$ & $0.509\pm0.04$  & $-1047.3$ &   3.4E-03 & 10 \\ 
J0319+4130 &  $2.777\pm0.09$ & $-0.711\pm0.006$ & $0.043\pm0.002$ & $1.37\pm0.7$ & $0.158\pm0.02$  & $-780.3$ &   8.9E-04 & 4 \\ 
J0334$-$4008 &  $-0.09\pm0.8$ & $-0.618\pm0.005$ & $0.043\pm0.001$ & $20.8\pm10$ & $0.270\pm0.02$  & $-1335.7$ &   2.1E-03 & 0 \\ 
J0423$-$0120 &  $1.12\pm0.3$ & $-0.588\pm0.007$ & $0.071\pm0.002$ & $4.2\pm2$ & $0.251\pm0.02$  & $-1070.6$ &   4.1E-01 & 1 \\ 
J0510+1800 &  $0.87\pm0.1$ & $-0.411\pm0.006$ & $0.053\pm0.002$ & $0.65\pm0.2$ & $0.408\pm0.03$  & $-943.4$ &   1.0E+00 & 9 \\ 
J0519$-$4546 &  $0.318\pm0.08$ & $-0.381\pm0.006$ & $0.052\pm0.002$ & $0.89\pm0.4$ & $0.206\pm0.02$  & $-1184.6$ &   4.2E-02 & 6 \\ 
J0522$-$3627 &  $1.719\pm0.05$ & $-0.239\pm0.005$ & $0.048\pm0.002$ & $0.259\pm0.07$ & $0.458\pm0.03$  & $-1044.7$ &   5.8E-01 & 23 \\ 
J0538$-$4405 &  $0.86\pm0.2$ & $-0.605\pm0.005$ & $0.052\pm0.002$ & $2.5\pm1$ & $0.213\pm0.02$  & $-1360.3$ &   7.3E-03 & 2 \\ 
J0635$-$7516 &  $0.288\pm0.04$ & $-0.918\pm0.005$ & $0.057\pm0.002$ & $0.68\pm0.2$ & $0.138\pm0.01$  & $-1813.2$ &   3.0E-01 & 9 \\ 
J0750+1231 &  $0.314\pm0.1$ & $-0.657\pm0.005$ & $0.039\pm0.002$ & $1.38\pm0.7$ & $0.195\pm0.02$  & $-1055.7$ &   1.3E-02 & 4 \\ 
J0854+2006 &  $1.524\pm0.08$ & $-0.424\pm0.007$ & $0.062\pm0.002$ & $0.35\pm0.1$ & $0.585\pm0.04$  & $-754.3$ &   1.2E-01 & 17 \\ 
J0904$-$5735 &  $0.089\pm0.1$ & $-0.407\pm0.01$ & $0.049\pm0.005$ & $0.28\pm0.1$ & $0.65\pm0.1$  & $-163.7$ &   8.1E-01 & 20 \\ 
J1037$-$2934 &  $0.21\pm0.1$ & $-0.524\pm0.006$ & $0.053\pm0.002$ & $0.81\pm0.4$ & $0.349\pm0.03$  & $-1057.1$ &   4.9E-01 & 7 \\ 
J1058+0133 &  $1.19\pm0.2$ & $-0.472\pm0.005$ & $0.046\pm0.002$ & $2.2\pm1$ & $0.248\pm0.03$  & $-953.7$ &   3.5E-01 & 3 \\ 
J1107$-$4449 &  $0.163\pm0.04$ & $-0.748\pm0.006$ & $0.054\pm0.002$ & $0.53\pm0.2$ & $0.170\pm0.02$  & $-1042.1$ &   1.0E-01 & 11 \\ 
J1127$-$1857 &  $0.09\pm0.2$ & $-0.583\pm0.007$ & $0.024\pm0.002$ & $1.64\pm0.8$ & $0.280\pm0.05$  & $-283.3$ &   2.7E-01 & 3 \\ 
J1146+3958 &  $0.16\pm0.1$ & $-0.587\pm0.01$ & $0.084\pm0.004$ & $0.96\pm0.3$ & $0.361\pm0.03$  & $-342.4$ &   1.1E-01 & 6 \\ 
J1229+0203 &  $2.165\pm0.08$ & $-0.775\pm0.007$ & $0.066\pm0.002$ & $0.75\pm0.2$ & $0.262\pm0.02$  & $-942.1$ &   6.5E-01 & 8 \\ 
J1256$-$0547 &  $2.96\pm0.3$ & $-0.579\pm0.005$ & $0.051\pm0.002$ & $5.7\pm3$ & $0.186\pm0.02$  & $-1508.8$ &   2.2E-03 & 1 \\ 
J1331+3030 &  $-0.275\pm0.03$ & $-1.091\pm0.03$ & $0.074\pm0.009$ & $0.25\pm0.1$ & $0.17\pm0.1$  & $-50.3$ &   5.4E-01 & 20 \\ 
J1337$-$1257 &  $1.42\pm0.1$ & $-0.621\pm0.006$ & $0.053\pm0.002$ & $1.55\pm0.6$ & $0.190\pm0.02$  & $-1025.3$ &   1.0E-02 & 4 \\ 
J1427$-$4206 &  $1.37\pm0.1$ & $-0.558\pm0.004$ & $0.044\pm0.002$ & $0.81\pm0.3$ & $0.411\pm0.03$  & $-1356.1$ &   7.3E-02 & 7 \\ 
J1517$-$2422 &  $0.741\pm0.08$ & $-0.271\pm0.005$ & $0.046\pm0.002$ & $0.67\pm0.2$ & $0.320\pm0.02$  & $-1150.6$ &   9.7E-01 & 9 \\ 
J1550+0527 &  $0.029\pm0.01$ & $-0.701\pm0.005$ & $0.045\pm0.001$ & $0.243\pm0.06$ & $0.115\pm0.01$  & $-1523.7$ &   6.3E-02 & 24 \\ 
J1617$-$5848 &  $0.24\pm0.2$ & $-0.891\pm0.004$ & $0.044\pm0.001$ & $3.8\pm2$ & $0.136\pm0.01$  & $-1571.7$ &   1.1E-02 & 1 \\ 
J1642+3948 &  $1.214\pm0.05$ & $-0.749\pm0.008$ & $0.059\pm0.002$ & $0.79\pm0.2$ & $0.166\pm0.02$  & $-719.0$ &   7.3E-01 & 8 \\ 
J1733$-$1304 &  $0.969\pm0.04$ & $-0.670\pm0.005$ & $0.049\pm0.002$ & $0.56\pm0.2$ & $0.180\pm0.02$  & $-1356.2$ &   1.1E-01 & 11 \\ 
J1751+0939 &  $1.034\pm0.08$ & $-0.476\pm0.005$ & $0.049\pm0.002$ & $0.33\pm0.1$ & $0.589\pm0.04$  & $-957.7$ &   1.4E-02 & 16 \\ 
J1924$-$2914 &  $1.724\pm0.04$ & $-0.643\pm0.004$ & $0.044\pm0.001$ & $0.36\pm0.1$ & $0.210\pm0.02$  & $-2063.6$ &   4.5E-03 & 13 \\ 
J2000$-$1748 &  $0.63\pm0.2$ & $-0.472\pm0.009$ & $0.043\pm0.004$ & $0.87\pm0.4$ & $0.71\pm0.1$  & $-260.0$ &   1.6E-01 & 7 \\ 
J2025+3343 &  $0.49\pm0.1$ & $-0.770\pm0.005$ & $0.046\pm0.002$ & $0.84\pm0.3$ & $0.336\pm0.03$  & $-1224.1$ &   1.7E-01 & 5 \\ 
J2056$-$4714 &  $0.223\pm0.06$ & $-0.634\pm0.004$ & $0.043\pm0.001$ & $0.56\pm0.1$ & $0.253\pm0.02$  & $-1519.7$ &   9.8E-02 & 11 \\ 
J2148+0657 &  $0.722\pm0.07$ & $-0.971\pm0.005$ & $0.052\pm0.002$ & $1.45\pm0.5$ & $0.139\pm0.01$  & $-1511.4$ &   1.8E-01 & 4 \\ 
J2232+1143 &  $1.302\pm0.08$ & $-0.411\pm0.01$ & $0.102\pm0.004$ & $0.346\pm0.08$ & $0.404\pm0.04$  & $-397.2$ &   8.9E-01 & 9 \\ 
J2253+1608 &  $2.647\pm0.05$ & $-0.533\pm0.009$ & $0.072\pm0.003$ & $0.48\pm0.1$ & $0.217\pm0.02$  & $-628.8$ &   7.4E-01 & 12 \\ 
J2258$-$2758 &  $0.83\pm0.1$ & $-0.665\pm0.006$ & $0.063\pm0.002$ & $1.05\pm0.2$ & $0.243\pm0.02$  & $-1541.3$ &   4.8E-02 & 6 \\ 
J2357$-$5311 &  $-0.002\pm0.03$ & $-0.794\pm0.004$ & $0.041\pm0.001$ & $0.539\pm0.1$ & $0.136\pm0.01$  & $-2259.7$ &   1.7E-01 & 11 \\ 
\br
\end{tabular}
\item[]\hspace*{-9em}{$^{\rm a}$ ``ln Jy" is not a  unit, but is indicated to emphasize the definition given in \eref{eq-csi}.}
\end{indented}
\end{table}

\begin{figure}
\centering%
\includegraphics[width=0.8\textwidth]{{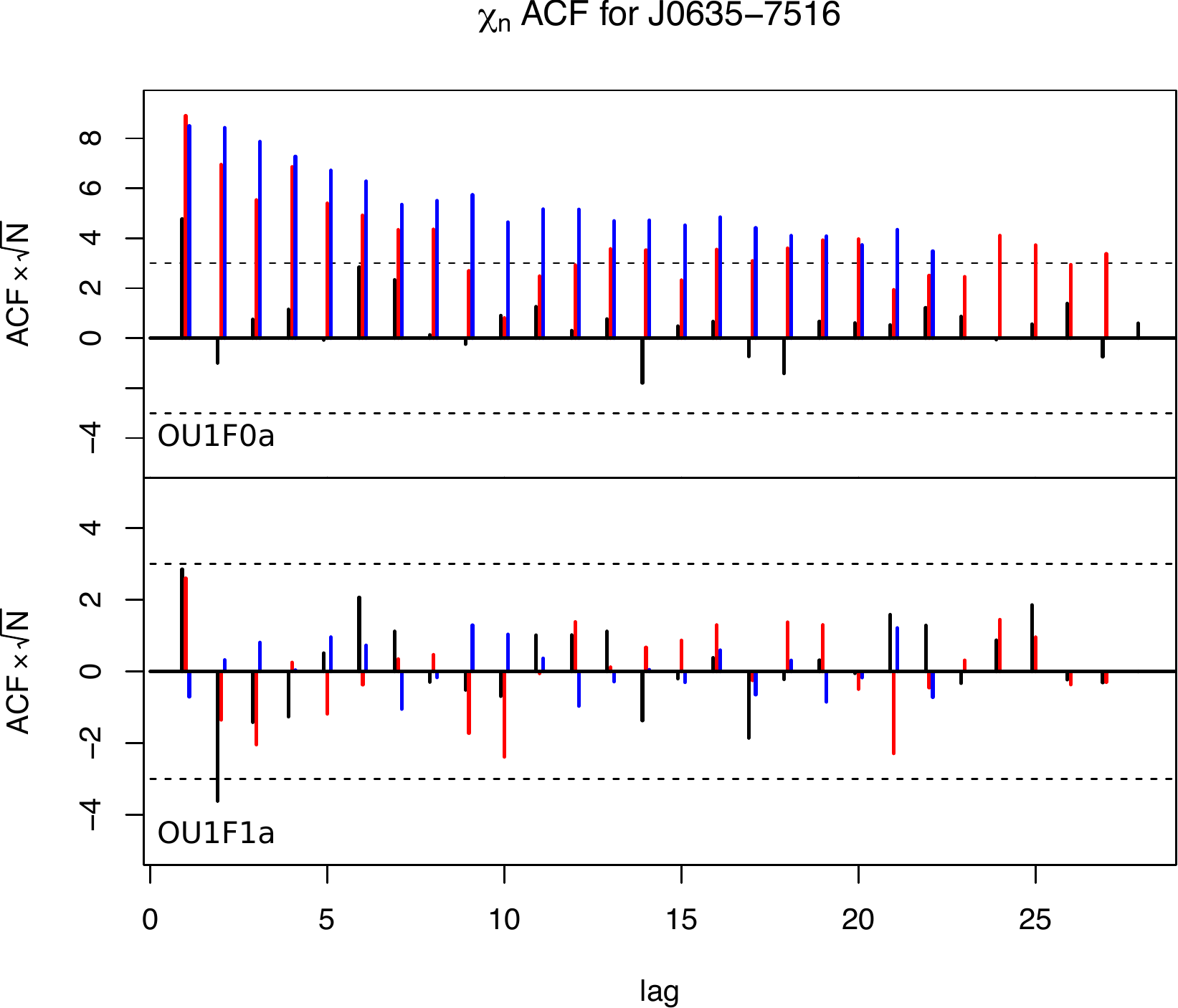}}
\caption{Red, blue, and black bars show the standardized ($\times\sqrt{N}$) ACF of the normalized residuals 
at Band 3, 7, and all data, respectively. Dashed lines mark the $\pm$3-$\sigma$ level assuming 
no cross-correlation above zero lag.
 \emph{Top panel.} Standardized ACF of the OU-1F0a model, parameters in \Tref{tab-OU-1F0a}. 
\emph{Bottom panel.} The same for the OU-1F1a model with curvature and $\nu$-lag (\Tref{tab-sel}).\label{fig-acfComp}}
\end{figure}

Despite the apparent adequacy of the OU-1F0a model,
 a more careful analysis of the residuals reveals its shortcomings. 
For example, the second panel of Figure \ref{fig-OU10ex} shows that constant spectral index hypothesis is 
not the most adequate: there are systematic patterns respect to frequency  in the range between 
900--1300 d, and specially after day 1800. These systematics are somewhat confused when aggregating 
all the residuals as in the third panel. 
Furthermore, the ACF of the $\chi_n$ displayed in \Fref{fig-acfComp} 
shows that the residuals separated by Band have significant cross-correlation not
consistent with white noise.

\begin{table}
\caption{Comparison of different models applied to J0635$-$7516.\label{tab-comp}}
\begin{indented}
\item[]\begin{tabular}{@{}lccrcc}
\br
\multicolumn{3}{c}{Model} & $\Delta$ AIC$^a$ & \pad & \plb$^b$ \\ %
OU-mix 	& Curvature& \nulag  			& & & \\
\mr
1F0a		&		&			&$0$		&  2.8E-01&  2.2E-04 \\
2F0a		&		&			&$-58.4$	&  5.8E-01 & 2.3E-08 \\
1F1a	 	&		&			&$-335.4$	&  6.5E-02 &  1.4E-02 \\
1F1a		&\ding{51}&			&$-421.8$	&   4.0E-02 &  6.4E-06\\
1F1a		&		&\ding{51}		&$-333.5$	&   2.8E-02 &  8.2E-02\\
%
1F1a		&\ding{51}&\ding{51}		&$-419.8$	&  1.1E-01 & 4.0E-03 \\
2F1a 	&		&			&$-331.4$	&  6.5E-02 & 1.4E-02 \\
\br
\end{tabular}
\item[]{$^a$ Difference with respect to the  OU-1F0a AIC given in \Tref{tab-OU-1F0a}. }
\item[]{$^b$ We use lag $=20$ except for J1331+3030, for which we use lag $=16$. }
\end{indented}
\end{table}

\Tref{tab-comp} shows a comparison of different  models applied to J0635$-$7516. The first column shows the 
OU-\chg{mixture} model. Curvature and \nulag\ refers to  the additional features described 
in Section \ref{sec-soph}. 
Because the AIC is useful as comparison criteria between models, 
 column (4) shows the difference between the AIC attained by the specified model and the OU-1F0a.
Columns (5) and (6) shows  the \pad\ and the p-value of the Ljung-Box test (\plb). The null hypotheses of the Ljung-Box test
assumes that the residuals arise from GWN.

\begin{figure}
\centering\includegraphics[width=1.15\textwidth]{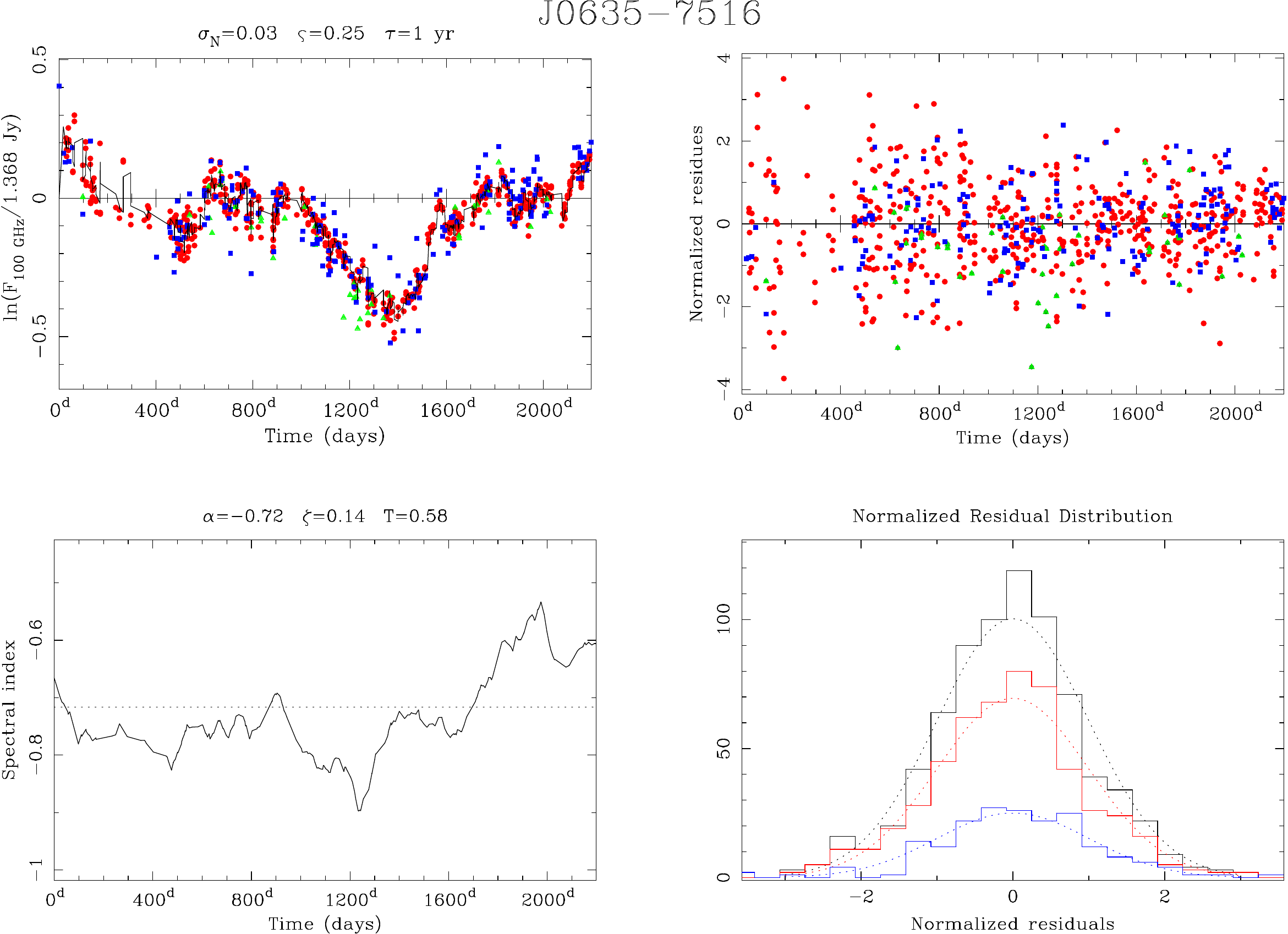}%
\caption{Best-fit OU-1F1a model applied to J0635$-$7516 with curvature \chgg{and lag}. MLE parameters are given in \Tref{tab-sel}.  
Symbol definition and colors as in \Fref{fig-OU10ex}. \emph{Top-left.} Centered data and predictions.  
\emph{Top-right.} $\chi_n$ time series. \emph{Bottom-left.} Local spectral index time series at 100 GHz \eref{eq-lsi}.
 \emph{Bottom-right.} Residuals histogram.\label{fig-sel}}
\end{figure}

We see that the addition of stochastic time-variability  to the spectral index (a $\beta$ term)  accounts for a significant diminishing in the AIC 
($\sim420$).  The second sophistication in importance, judged from the AIC variation,  
is  the addition of curvature --- specifically, \emph{concavity} --- which reduces the AIC by $\sim90$.
As shown in  \Tref{tab-comp}, the inclusion of \nulag\ in the model does not seem to improve the AIC, but it does 
increase \pad\ and \plb, indicating that the residuals seem to be better described by GWN. 
Based on the criteria defined in \sref{sec-gof}, we conclude that the model which 
 is simultaneously simpler and more adequate for J0635$-$7516 is the OU-1F1a model with 
 lag and additional concavity. 
 \Fref{fig-acfComp} shows in the second panel the  ACF of the  total residuals 
and those segregated by band of this model applied to J0635$-$7516, which are much smaller 
and similar to a GWN compared with those of the OU-1F0a model.
\Fref{fig-sel} show the best-fit flux and spectral index predictions of the model in the left panels, while the right panels show the normalized
 residuals ($\chi_n$) time series and distribution histogram.
Less band-systematics in  the residuals time series are evident when comparing them with those of   \Fref{fig-OU10ex}.

This same analysis --- that is, examination of the residuals quality, systematics, biases, and the final values of the 
AIC and p-values of statistical tests --- is applied  to the rest of the grid calibrators, allowing us to pick the
optimal model for each  light-curve. Hereafter, we refer to these as the ``selected models." 
\Tref{tab-sel} shows the MLE parameters for all selected models. More detailed MLE results
 including plots as those of Figures 7 and  \ref{fig-sel} for the rest of the grid calibrators are presented in  the supplementary material (\ref{sec-supp}). 

We find that, by a similar evaluation as the one performed on J0635$-$7516, 
the selected model  includes a spectral index variation term ($\beta$-term) for all grid calibrators, except J1127$-$1857. 
Additionally, a negative curvature parameter $\kappa$  is included in all selected models except that of J1127$-$1857 and
J1331+3030. These sources have the least number of measurements, which impedes us to fit reliably more sophisticated models.
The inclusion of \nulag\ ($\epsilon$) in the fitting does improve in 15 cases the quality of the residuals, that is, their ACF and statistical tests indicate 
they are more consistent with GWN than the residuals of the model without \nulag. However, as it can be deduced from the commonly 
 large error  bars  of $\epsilon$ in \Tref{tab-sel}, the \nulag\ inclusion does not affect much the  likelihood value. 
 As described in \Sref{sec-sim}, we take a more conservative approach and consider that the actual uncertainty of 
 $\epsilon$ is bounded from below by 0.5 d. Considering this, 
there are only five sources with estimated  $\epsilon$ magnitudes larger than 1.5 d, four of them with $\epsilon<0$. 

Only for  J2148+0657 we find an $\epsilon$ MLE which appears  significantly positive. It is difficult to evaluate in detail why 
this source displays such a feature, and even more difficult to explain it physically. 
We note, however,  that Band 6 data for this source does appear to be overestimated by the model, and that the remotion of these data 
produces a MLE for $\epsilon$ which is positive but smaller ($\sim0.5$ d) and only marginally significant. This example indicates us that, 
while in the ideal case (\Sref{sec-sim}) and under all the hypotheses (e.g., normality, independence, spectral shape, etc\ldots) we can recover a solid \nulag\ 
as small of 1--2 d, divergences from the ideal case can bias the MLE estimation towards an artificial lag. 

\begin{landscape}
\scriptsize
\begin{ThreePartTable}
  \begin{longtable}{lcccccccccc}
    \caption{MLE parameters of the selected OU model for each source.\label{tab-sel}}\\
    \br\\
   Source  &$\mu$ & $\alpha$ & $\kappa$& $\epsilon$& $\sigma_{N,0}$ & $\varsigma_1$ & $\zeta$ & $\tau_{1}$ & $T$ & $\rho$\\
    &&  & & (d)& & $\left({\rm yr}^{-1/2}\right)$  & $\left({\rm yr}^{-1/2}\right)$  & (yr) & (yr) & \\ 
   \hline\\
J0006$-$0623 &  $0.97\pm0.3$ & $-0.326\pm0.04$ & $-0.191\pm0.01$ & --- & $0.017\pm0.0008$ & $0.221\pm0.02$ & $0.152\pm0.02$ & $5.4\pm5$ & $0.61\pm0.2$ & $0.66\pm0.1$    \\ 
J0237+2848 &  $0.781\pm0.06$ & $-0.449\pm0.02$ & $-0.147\pm0.01$ & --- & $0.017\pm0.001$ & $0.316\pm0.02$ & $0.324\pm0.04$ & $0.52\pm0.2$ & $0.091\pm0.03$ & $0.44\pm0.1$    \\ 
J0238+1636 &  $0.39\pm0.2$ & $-0.307\pm0.02$ & $-0.145\pm0.01$ & $-0.15\pm0.2$ & $0.020\pm0.001$ & $0.540\pm0.03$ & $0.419\pm0.08$ & $0.79\pm0.3$ & $0.054\pm0.01$ & $0.56\pm0.1$    \\ 
J0319+4130 &  $2.80\pm0.1$ & $-0.561\pm0.02$ & $-0.131\pm0.02$ & --- & $0.017\pm0.001$ & $0.166\pm0.02$ & $0.227\pm0.08$ & $2.1\pm1$ & $0.065\pm0.04$ & $0.03\pm0.2$    \\ 
J0334$-$4008 &  $-0.2\pm1$ & $-0.496\pm0.02$ & $-0.106\pm0.01$ & $0.67\pm0.2$ & $0.019\pm0.001$ & $0.355\pm0.02$ & $0.263\pm0.04$ & $19.8\pm20$ & $0.128\pm0.05$ & $0.56\pm0.1$    \\ 
J0423$-$0120 &  $1.16\pm0.7$ & $-0.328\pm0.03$ & $-0.218\pm0.02$ & --- & $0.025\pm0.001$ & $0.410\pm0.03$ & $0.286\pm0.04$ & $7.2\pm8$ & $0.248\pm0.07$ & $0.60\pm0.1$    \\
J0510+1800 &  $0.91\pm0.2$ & $-0.193\pm0.02$ & $-0.185\pm0.01$ & --- & $0.020\pm0.001$ & $0.520\pm0.03$ & $0.292\pm0.04$ & $0.82\pm0.4$ & $0.144\pm0.06$ & $0.43\pm0.1$    \\ 
J0519$-$4546 &  $0.279\pm0.07$ & $-0.197\pm0.02$ & $-0.167\pm0.01$ & --- & $0.018\pm0.0009$ & $0.376\pm0.03$ & $0.382\pm0.04$ & $0.46\pm0.2$ & $0.093\pm0.03$ & $0.09\pm0.1$    \\ 
J0522$-$3627 &  $1.737\pm0.06$ & $-0.043\pm0.02$ & $-0.167\pm0.02$ & $-1.79\pm0.2$ & $0.017\pm0.001$ & $0.567\pm0.03$ & $0.342\pm0.08$ & $0.242\pm0.07$ & $0.059\pm0.02$ & $0.51\pm0.1$    \\ 
J0538$-$4405 &  $0.92\pm0.5$ & $-0.437\pm0.02$ & $-0.115\pm0.01$ & --- & $0.021\pm0.0009$ & $0.309\pm0.02$ & $0.165\pm0.02$ & $7.8\pm8$ & $0.303\pm0.08$ & $0.736\pm0.09$    \\ 
J0635$-$7516 &  $0.31\pm0.1$ & $-0.716\pm0.04$ & $-0.175\pm0.02$ & $0.1\pm1$ & $0.035\pm0.001$ & $0.245\pm0.02$ & $0.143\pm0.03$ & $1.02\pm0.5$ & $0.58\pm0.3$ & $0.25\pm0.2$    \\ 
J0750+1231 &  $0.31\pm0.1$ & $-0.497\pm0.03$ & $-0.146\pm0.02$ & --- & $0.027\pm0.001$ & $0.234\pm0.02$ & $0.113\pm0.03$ & $1.7\pm1$ & $0.40\pm0.2$ & $0.39\pm0.2$    \\ 
J0854+2006 &  $1.51\pm0.1$ & $-0.217\pm0.03$ & $-0.181\pm0.02$ & $-0.853\pm0.04$ & $0.026\pm0.002$ & $0.703\pm0.04$ & $0.217\pm0.04$ & $0.43\pm0.2$ & $0.204\pm0.09$ & $0.17\pm0.2$    \\ 
J0904$-$5735 &  $0.07\pm0.1$ & $-0.325\pm0.05$ & $-0.051\pm0.03$ & --- & $0.018\pm0.003$ & $0.735\pm0.08$ & $0.45\pm0.1$ & $0.30\pm0.2$ & $0.131\pm0.09$ & $0.14\pm0.3$    \\ 
J1037$-$2934 &  $0.25\pm0.2$ & $-0.322\pm0.03$ & $-0.181\pm0.02$ & $0.22\pm0.4$ & $0.031\pm0.002$ & $0.422\pm0.03$ & $0.267\pm0.04$ & $1.3\pm1$ & $0.125\pm0.05$ & $0.44\pm0.1$    \\ 
J1058+0133 &  $1.30\pm0.2$ & $-0.368\pm0.04$ & $-0.121\pm0.02$ & $-0.28\pm0.7$ & $0.026\pm0.002$ & $0.335\pm0.03$ & $0.139\pm0.03$ & $1.9\pm1$ & $0.56\pm0.3$ & $0.36\pm0.2$    \\ 
J1107$-$4449 &  $0.184\pm0.06$ & $-0.465\pm0.03$ & $-0.236\pm0.02$ & --- & $0.025\pm0.002$ & $0.270\pm0.02$ & $0.321\pm0.04$ & $0.55\pm0.3$ & $0.163\pm0.06$ & $0.26\pm0.1$    \\ 
J1127$-$1857 &  $0.09\pm0.1$ & $-0.583\pm0.007$ & --- & --- & $0.024\pm0.002$ & $0.280\pm0.05$ & --- & $1.64\pm0.8$  & --- & ---         \\
J1146+3958 &  $0.19\pm0.2$ & $-0.287\pm0.04$ & $-0.266\pm0.02$ & --- & $0.022\pm0.002$ & $0.495\pm0.04$ & $0.519\pm0.08$ & $1.08\pm0.5$ & $0.106\pm0.03$ & $0.55\pm0.1$    \\ 
J1229+0203 &  $2.21\pm0.1$ & $-0.472\pm0.03$ & $-0.288\pm0.02$ & $-3.89\pm0.7$ & $0.024\pm0.001$ & $0.393\pm0.03$ & $0.370\pm0.04$ & $1.02\pm0.5$ & $0.174\pm0.06$ & $0.27\pm0.1$    \\ 
J1256$-$0547 &  $2.78\pm0.3$ & $-0.381\pm0.03$ & $-0.189\pm0.02$ & --- & $0.027\pm0.001$ & $0.317\pm0.02$ & $0.307\pm0.08$ & $3.2\pm4$ & $0.139\pm0.07$ & $0.14\pm0.2$    \\ 
J1331+3030 &  $-0.294\pm0.04$ & $-1.079\pm0.04$ & --- & --- & $0.025\pm0.005$ & $0.248\pm0.09$ & $1.08\pm0.9$ & $0.29\pm0.2$ & $0.026\pm0.06$ & $0.27\pm0.6$    \\ 
J1337$-$1257 &  $1.38\pm0.2$ & $-0.350\pm0.03$ & $-0.222\pm0.02$ & $-1.667\pm0.02$ & $0.026\pm0.001$ & $0.267\pm0.02$ & $0.215\pm0.04$ & $2.8\pm2$ & $0.191\pm0.06$ & $0.60\pm0.1$    \\ 
J1427$-$4206 &  $1.34\pm0.1$ & $-0.331\pm0.02$ & $-0.191\pm0.02$ & --- & $0.028\pm0.001$ & $0.428\pm0.03$ & $0.219\pm0.08$ & $0.93\pm0.4$ & $0.146\pm0.08$ & $0.40\pm0.1$    \\ 
J1517$-$2422 &  $0.82\pm0.1$ & $-0.019\pm0.03$ & $-0.240\pm0.02$ & --- & $0.023\pm0.001$ & $0.392\pm0.02$ & $0.226\pm0.04$ & $1.05\pm0.5$ & $0.212\pm0.09$ & $0.34\pm0.1$    \\ 
J1550+0527 &  $0.031\pm0.02$ & $-0.465\pm0.04$ & $-0.235\pm0.02$ & --- & $0.028\pm0.001$ & $0.186\pm0.02$ & $0.137\pm0.03$ & $0.30\pm0.1$ & $0.54\pm0.3$ & $0.26\pm0.2$    \\ 
J1617$-$5848 &  $0.30\pm0.2$ & $-0.599\pm0.03$ & $-0.237\pm0.02$ & --- & $0.027\pm0.001$ & $0.171\pm0.02$ & $0.197\pm0.04$ & $5.7\pm7$ & $0.27\pm0.2$ & $0.00\pm0.3$    \\ 
J1642+3948 &  $1.218\pm0.09$ & $-0.353\pm0.03$ & $-0.343\pm0.02$ & --- & $0.022\pm0.002$ & $0.305\pm0.03$ & $0.455\pm0.08$ & $0.81\pm0.4$ & $0.091\pm0.04$ & $0.46\pm0.1$    \\ 
J1733$-$1304 &  $1.000\pm0.07$ & $-0.417\pm0.03$ & $-0.242\pm0.02$ & --- & $0.031\pm0.001$ & $0.222\pm0.02$ & $0.224\pm0.04$ & $0.83\pm0.4$ & $0.164\pm0.06$ & $0.45\pm0.1$    \\ 
J1751+0939 &  $1.08\pm0.1$ & $-0.155\pm0.03$ & $-0.289\pm0.02$ & $-1.56\pm0.4$ & $0.020\pm0.001$ & $0.685\pm0.04$ & $0.272\pm0.04$ & $0.39\pm0.2$ & $0.136\pm0.05$ & $0.37\pm0.1$    \\ 
J1924$-$2914 &  $1.755\pm0.08$ & $-0.317\pm0.02$ & $-0.289\pm0.01$ & --- & $0.028\pm0.001$ & $0.238\pm0.02$ & $0.281\pm0.05$ & $0.82\pm0.4$ & $0.094\pm0.04$ & $0.42\pm0.1$    \\ 
J2000$-$1748 &  $0.63\pm0.3$ & $-0.335\pm0.04$ & $-0.123\pm0.03$ & --- & $0.018\pm0.002$ & $0.495\pm0.04$ & $0.324\pm0.08$ & $2.0\pm1$ & $0.137\pm0.07$ & $0.44\pm0.2$    \\ 
J2025+3343 &  $0.52\pm0.2$ & $-0.412\pm0.03$ & $-0.330\pm0.02$ & --- & $0.023\pm0.001$ & $0.416\pm0.03$ & $0.369\pm0.05$ & $1.00\pm0.4$ & $0.111\pm0.04$ & $0.33\pm0.1$    \\ 
J2056$-$4714 &  $0.238\pm0.08$ & $-0.329\pm0.02$ & $-0.263\pm0.01$ & $1.36\pm0.7$ & $0.017\pm0.001$ & $0.312\pm0.02$ & $0.392\pm0.04$ & $0.73\pm0.3$ & $0.055\pm0.01$ & $0.44\pm0.1$    \\ 
J2148+0657 &  $0.73\pm0.1$ & $-0.539\pm0.02$ & $-0.383\pm0.02$ & $5.99\pm0.9$ & $0.021\pm0.001$ & $0.191\pm0.01$ & $0.603\pm0.08$ & $2.0\pm1$ & $0.051\pm0.01$ & $0.633\pm0.09$    \\ 
J2232+1143 &  $1.31\pm0.2$ & $0.086\pm0.07$ & $-0.421\pm0.01$ & --- & $0.018\pm0.001$ & $0.598\pm0.04$ & $0.525\pm0.06$ & $0.86\pm0.7$ & $0.25\pm0.1$ & $0.21\pm0.1$    \\ 
J2253+1608 &  $2.64\pm0.1$ & $-0.176\pm0.03$ & $-0.303\pm0.02$ & $-0.3\pm1$ & $0.018\pm0.001$ & $0.301\pm0.02$ & $0.331\pm0.04$ & $1.01\pm0.5$ & $0.198\pm0.07$ & $0.56\pm0.1$    \\ 
J2258$-$2758 &  $0.96\pm0.3$ & $-0.388\pm0.03$ & $-0.246\pm0.01$ & $0.14\pm0.2$ & $0.021\pm0.0009$ & $0.349\pm0.02$ & $0.258\pm0.03$ & $2.28\pm0.7$ & $0.286\pm0.08$ & $0.703\pm0.07$    \\ 
J2357$-$5311 &  $0.052\pm0.08$ & $-0.524\pm0.03$ & $-0.211\pm0.01$ & $0.18\pm0.7$ & $0.023\pm0.0008$ & $0.179\pm0.02$ & $0.107\pm0.02$ & $1.26\pm0.5$ & $0.65\pm0.4$ & $0.53\pm0.1$    \\ 
   \hline
     \end{longtable}
  \end{ThreePartTable}
\end{landscape}

\subsection{Forecasts  and uncertainty intervals.}\label{sec-fore}

Using  the MLE parameters and the model, we can derive  flux forecasts and its stochastic uncertainties 
for each source using the Kalman recursions (\ref{sec-ssr}) and the 
 parametric uncertainty (see \Sref{sec-uncs}). 
In the characteristic monitoring time scale of the grid calibrators, the main source of uncertainty is given by the 
intrinsic time variability. In the long-term ($t>\tau,T$), the parametric uncertainty of the deterministic model usually dominates.

\begin{figure}
\includegraphics[width=0.5\textwidth]{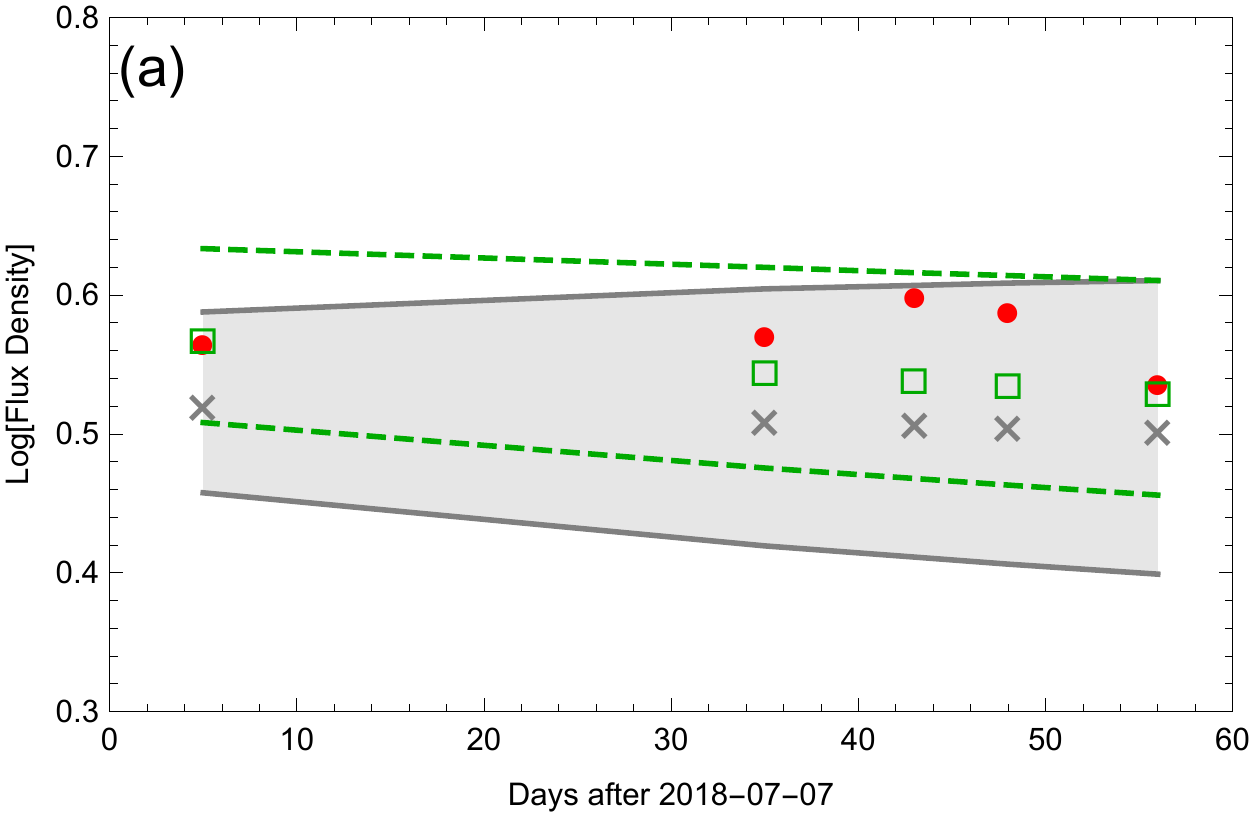}%
\includegraphics[width=0.5\textwidth]{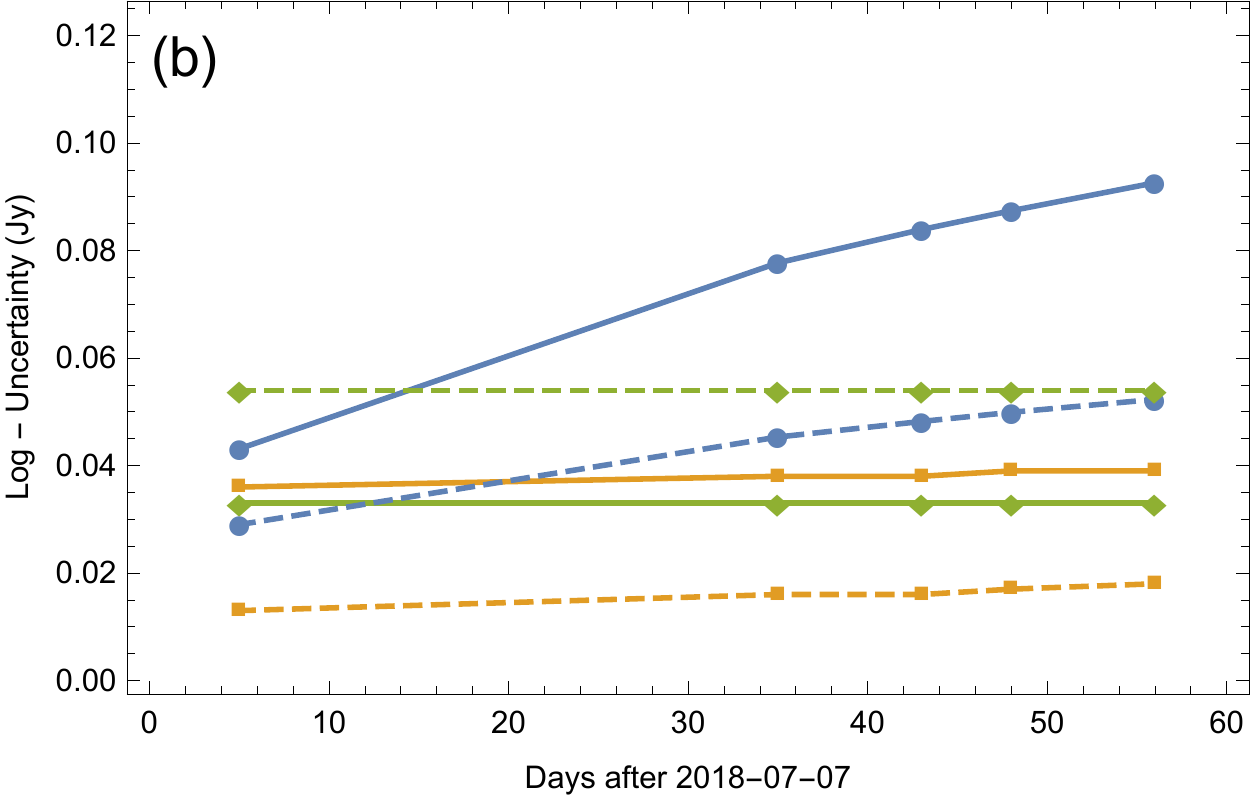}\\
\includegraphics[width=0.51\textwidth]{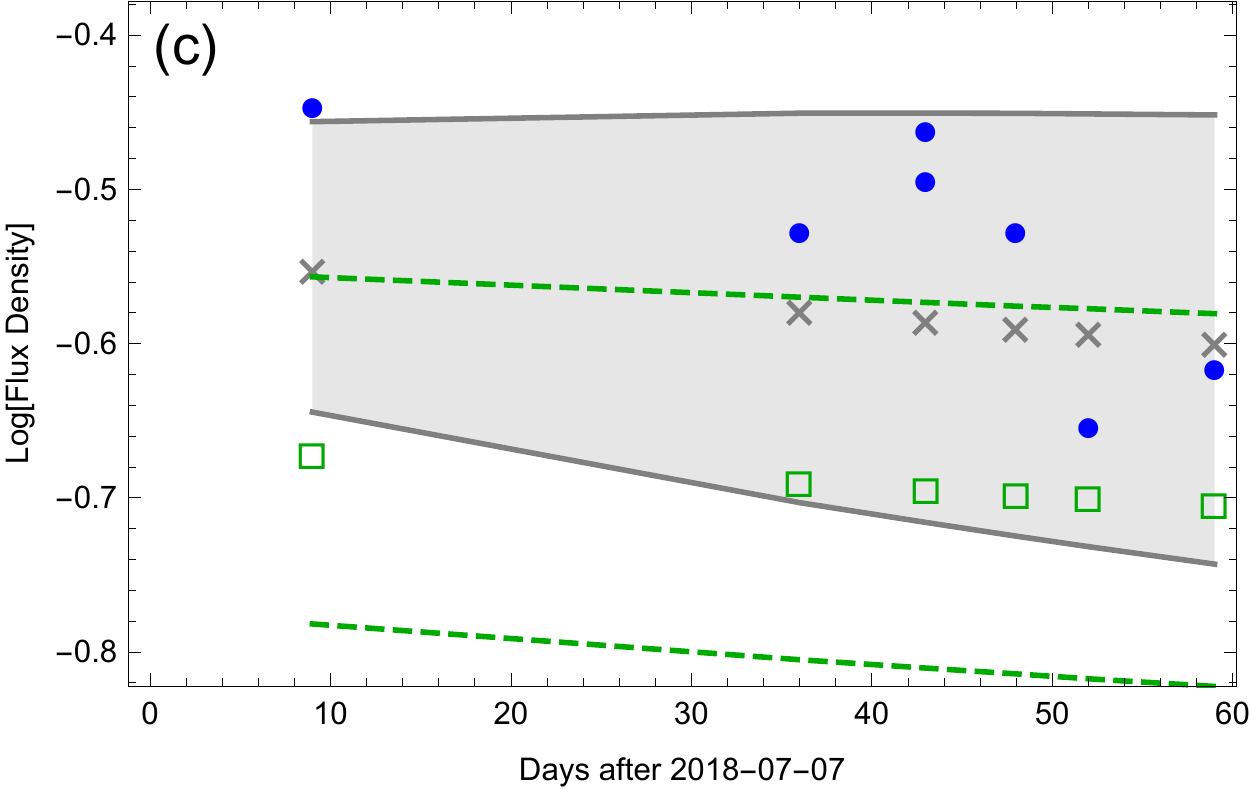}%
\includegraphics[width=0.5\textwidth]{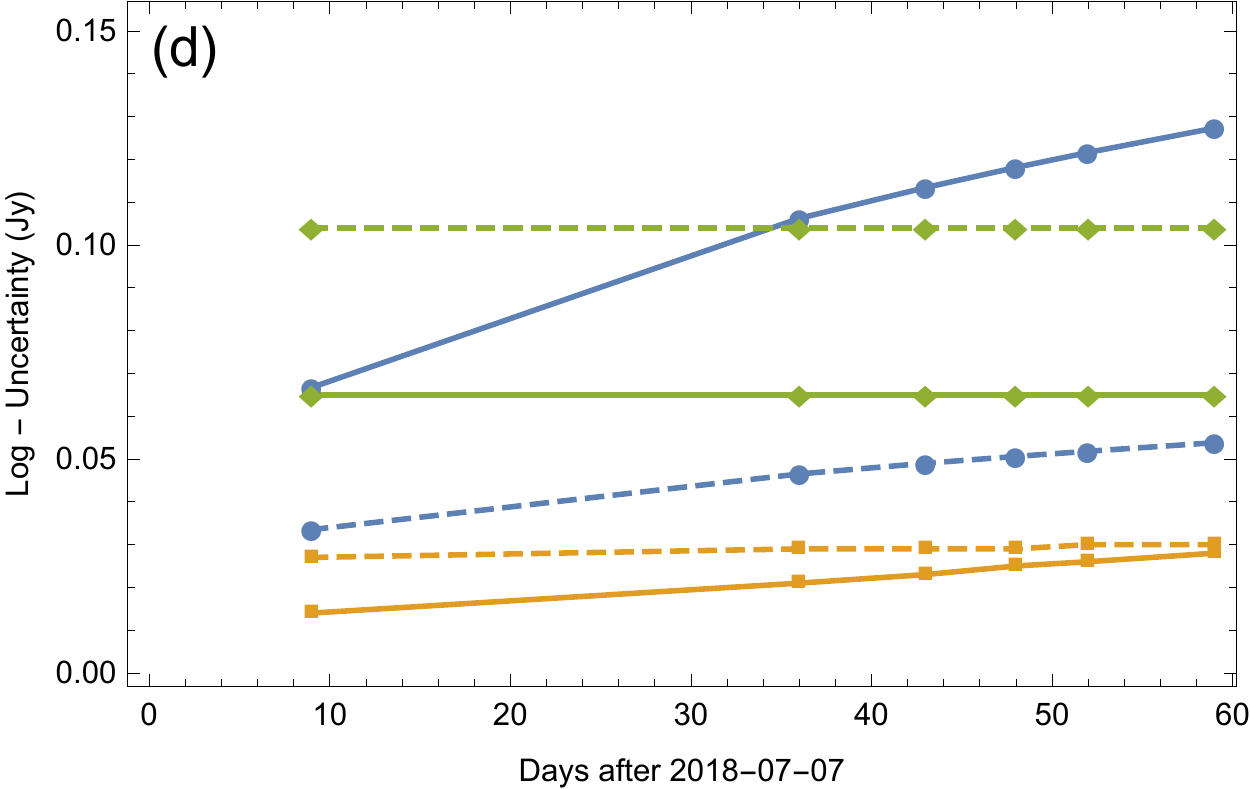}
\caption{Stochastic model forecast of J0635$-$7516 during the $60$ days ensuing the last measurement  used for the  analysis of this work. {(a).} Red dots: 91.5 GHz fluxes. Crosses and squares show the forecasts of the OU-1F1a (\nulag\ + curvature) and the OU-1F0a model, respectively. Forecasts are bracketed by the $\pm1\sigma$ total uncertainty interval (parametric+stochastic) shown in grey continuous and dashed green lines for the OU-1F1a (lag+curvature) and the OU-1F0a model, respectively. {(b).} Uncertainties of the 91.5 GHz flux. In blue (dots), orange (squares), and green (diamonds) the stochastic, parametric uncertainty, and \chgg{measurement} uncertainty levels, respectively. Continuous and dashed lines  indicate the results of the OU-1F1a (lag+curvature) and the OU-1F0a model, respectively. {(c).} Same as panel (a), but with blue dots showing fluxes at 343.5 GHz.  {(d).}  Same as panel (b) but at 343.5 GHz. \label{fig-pred}}
\end{figure}

 \Fref{fig-pred} shows the five following measurements of J0635$-$7516 in Bands 3 and 7, which occur during a time interval of $\sim60$ days after the  
 last measurement considered in this work, taken in 2018-07-07.
 Panels (a) and (c) also shows the forecasts  from the OU-1F0a and the selected OU-1F1a   model  during this time, together with the 
evolving $\pm1\sigma$ uncertainty intervals. These intervals  include the parametric and stochastic variability, added in quadrature.
 Panels (b) and (d) display the breakdown of the uncertainty in its components (\Sref{sec-uncs}). 
\Fref{fig-pred} shows  predictions based only on  the data taken up to 2018-07-07 for reference, 
but in practice it would be desirable to update the model at each new measurement.
 \chg{\Fref{fig-pred} shows that most measurements fall within the $\pm1\sigma$ uncertainty 
interval around the forecasts in  both models, with all measurements falling  within  $\pm2\sigma$. Note that 
the selected model OU-1F1a performs better than the OU-1F0a model,  specially in Band 7.}
  
 We compare the 91.5 and 343.5 GHz flux forecasts for the rest of the grid calibrators
 with their fluxes  measured   immediately after the last observed time given in \Tref{tab-data}. 
 \chg{For the full grid sample, there are 56 and 41 of these measurements in total for Bands 3 and 7, respectively.  
 According to the hypotheses, the standardized differences between the fluxes and forecasts  should be distributed as a standard Gaussian.  
An  Anderson-Darling test on both Band 3 and 7 standardized differences  indicate that we cannot reject the Gaussian hypothesis 
at the 95\% confidence level in either case.  We  test the null hypothesis further by using two more tests.
We calculate the mean of the standardized differences and the sum of its squares. According to the null hypotheses, they 
should distribute as Gaussian with standard deviation $n^{-1/2}$ and a $\chi^2_n$ distribution, respectively, where $n$ is the number of measurements. 
For Band 3, we find that the  standardized differences are consistent with the null hypotheses in both tests. This is also the case for the means of the  Band 7 means standardized differences. 
However, the sum of its  squares is $20.6$, while the 95\%  (two-sided) interval of the $\chi_{41}^2$ distribution starts at $25.22$.
Therefore, it is possible that the uncertainties of the Band 7 forecasts are overestimated by a factor $\sqrt{25.22/20.6}$, or an 11\%.
An overestimation of this type --- meaning that the forecasts are actually closer to the measured values than expected --- would affect  
the reported  flux uncertainties, and it  may decrease the sensitivity of the model to detect outliers. It would also 
overestimate the interpolated flux calibration error, either in a science project  or during the quality evaluation of new data for ingestion 
to the source catalog.}
    

Finally, we evaluate the accuracy of the parametric uncertainty estimation given by \Eref{eq-pu} using Monte Carlo simulations. 
We use a bayesian model to estimate the parametric uncertainty in the general case when the likelihood function is not well 
approximated by a normal near the maximum --- as it happens for some parameters in J0635$-$7516.  This source 
is representative of those calibrators in which the monitoring time is not sufficiently long in order to obtain an accurate estimation
of the decorrelation time.
According to the bayesian prescription, the posterior probability distribution of the parameters is given by the likelihood times the prior. 
We choose the latter to be minimally informative, that is, a constant greater than zero in the  allowed parameter space. The allowed 
parameter space includes only  the positive values for those parameters expected to be positive 
(e.g., decorrelation times and variance rates) and it is unconstrained otherwise. 
We sample the posterior using a Metropolis-Hastings Markov chain  \cite{gregory2005bayesian} starting from the MLE, and calculate the 
forecasts for all the simulated parameters.  This sampling should take into account all the non-linearities involved in the calculation
of the deterministic model and in  deriving a forecast.

We find that the estimations of  the parametric uncertainty derived from \eref{eq-pu} are comparable, although usually larger by a factor between 1--3 times 
 the standard deviation of the forecasts sampled from the posterior distribution. That is, the formulae derived in \Sref{sec-uncs} gives a conservative
 estimation of the parametric uncertainty. Because usually this parametric uncertainty does not dominate the forecast uncertainties in the short term, 
 we  use  \eref{eq-pu} to estimate it for simplicity.

\subsubsection{\chg{Flux interpolation: comparison between two methods.\label{sec-compInterp}}}

 \chg{Because the  stochastic model includes a description of the time variability and the spectral characteristics of the source, we can produce forecasts and interpolations 
 of the flux at any time and frequency. These interpolations and forecasts have uncertainties  derived from the variability of the source and from the 
 measurement error.}
 
 \chg{
 Another way (similar to the one used, for example,  in the ALMA pipeline up to Cycle 5
 \footnote{\chgg{Pipeline documentation: \texttt{https://almascience.org/documents-and-tools/processing/science-pipeline}. 
The specific task used to obtain interpolated fluxes from the source catalog is 
\texttt{getALMAFlux}, 
described in 
\texttt{https://safe.nrao.edu/wiki/bin/view/ALMA/GetALMAFlux}}}) to interpolate (and extrapolate) the flux  from the source catalog uses the 
closest (in time) Band 3 measurement and 
a spectral 
 index derived from the closest pair of measurements --- Bands 3 \& 6 or 3 \& 7 --- taken \chgg{on the same day.} 
Using this Band 3 flux and spectral index, the expected flux density at other frequencies are calculated.
We will refer to this method as P0. The error of this flux is derived from the uncertainty of each flux measurement, 
but there is no estimation of the uncertainty derived from the variability of the source.}

\begin{figure}
\centering\includegraphics[width=0.5\textwidth]{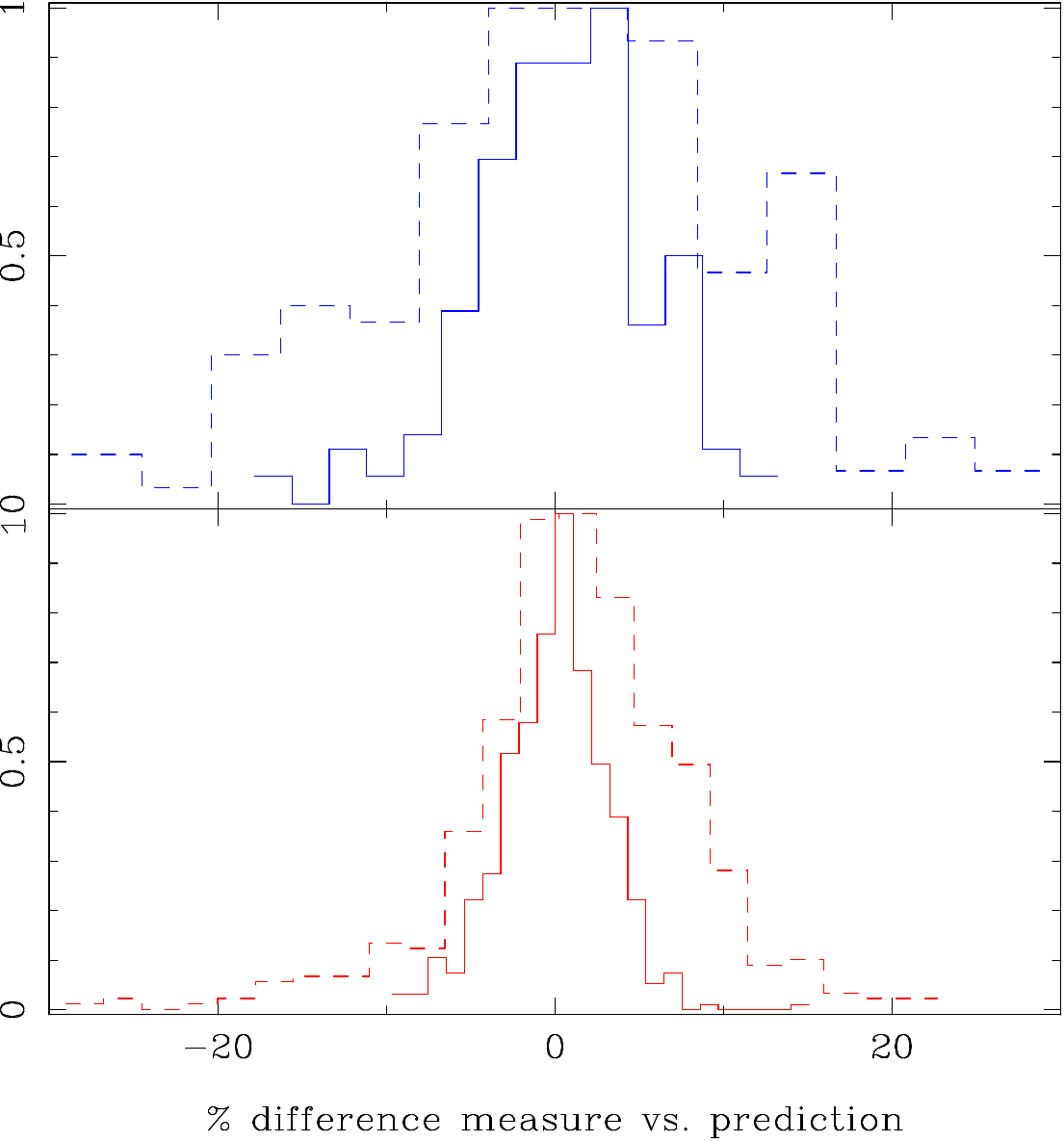}%
\caption{\chg{Bottom and top panels show, respectively, the Band 3 and 7 normalized histograms 
of the percentage difference between the measured flux density and the P0 interpolation (dashed lines) 
and stochastic model interpolation (continuous line).}\label{fig-compInterp}}
\end{figure}
\chg{We compare both methods of interpolating the flux on J0635$-$7516 by ignoring one day of observation and obtaining the interpolated values 
for this day using the stochastic model fitted to the source (OU-1F1a) and P0. We repeat this process for each day, and compare these interpolations 
with the fluxes actually measured. Figure \ref{fig-compInterp} shows  histograms of the percentage difference between the measured data and the P0 and stochastic model interpolations. The histograms make clear that the stochastic model differences respect to the measured data are smaller than those provided by P0. 
For P0, the standard deviation of the differences at Bands 3 and 7 are 7\% and 11\%, respectively,\footnote{These numbers are not to be confused with the typical flux uncertainty of ALMA.} 
while for the OU-1F1a interpolations, these  are 3\% and 5\%, respectively.}


\newpage\section{Discussion}\label{sec-dis}

\chg{In this section we show in more detail some additional applications and extensions of the stochastic modeling (Sections \ref{sec-phase}. and \ref{sec-high})
and evaluate some of its drawbacks and future improvements (\Sref{sec-drawbacks}).}

Stochastic variability studies of extragalactic sources have commonly focused on accurately determining 
 the parameters of the process in order  
to link them with the underlying physics. The use of the stochastic process to make forecasts and to interpolate the flux at 
unobserved times has received less attention in astronomy. 
\chg{Areas of the ALMA  operations  where the work presented in this paper is of practical use are: (i) the monitoring 
of grid calibrators and (ii) science data reduction and  the quality assurance process.}

\chg{During the  calibrator survey monitoring of the grid sources, it is important  to compare each new measurement 
with the previous history of the calibrator. The stochastic model provides updated forecasts of the measurement 
at the new observing times and associated uncertainties based on the observed variability of the source (as shown in \Sref{sec-fore}). 
Both these quantities allow the calibration JAO group to evaluate precisely how consistent is the new measurement with the historical variations of the 
quasar \chgg{(and identify outliers and flaring behavior)}.}

\chg{Regarding possible applications to data reduction, 
note that  the flux of the secondary flux calibrator for each science project is  determined from an 
interpolation (in time and frequency) based on previous measurements obtained by the calibrator survey.
The stochastic model  allow us to calculate this  interpolated flux and the associated (variability induced) uncertainty. 
Furthermore, in a similar fashion compared with the calibrator survey,  the 
stochastic modeling of the amplitude/phase calibrator allows us to evaluate quantitatively whether its derived flux is consistent with the 
previous catalogue values. This assessment  of the reliability and uncertainty of the flux calibration is commonly done during the quality assurance process.
\Sref{sec-phase} shows examples of stochastic modeling on amplitude/phase (non-grid) calibrators. }

\subsection{\chg{Application to less frequently sampled data: phase calibrators}\label{sec-phase}}

\chg{In principle, the methods presented in this study can be applied to the  light-curve of any quasar in mm/sub-mm wavelengths, including the rest of the 
calibrators in the ALMA source catalog. 
Although the present study is focused on the grid sources, this is mainly because they have more measurements compared to 
the rest of the calibrators. 
The scant data on some calibrators  increases greatly the  parameter  degeneracy and uncertainty.}

\chg{To illustrate the applicability  of these methods to phase calibrators not in the grid source list, we select the subset of ten best  sampled sources 
in the ALMA source catalog (\Tref{tab-phaseFore}). Because of the smaller number of points, in order to remove some parameter degeneracy,  
 we used a OU-1F1a model without curvature and with two additional constrains: $\sigma_{N,0}=0.06$ 
fixed and we took equal decorrelation timescales for the flux and spectral indices ($T=\tau$). 
Figures showing the results of the fitting  for all these sources are presented in the supplementary material.}

\chg{In a typical ALMA science project, the flux determined for the phase calibrator is examined in order to 
determine the  plausibility of the flux calibration. If the flux is not consistent with the observed variability, this 
may indicate problems with the data reduction or calibration. 
This is one of the basic tests performed during the quality assurance stage of each ALMA project, and a 
common practice in the radio-astronomy community. However, determining what is 
a flux consistent with the variability of the phase calibrator is not usually well defined. The dispersion of the 
values of the light-curve may serve as a first approximation. However, intuitively, this dispersion is certainly too large  if the phase calibrator 
flux was measured (that is, absolutely calibrated against a primary flux calibrator) not  long before or after  the science project in question. 
The present work  defines quantitative uncertainties on these fluxes taking into consideration the historical variability of the source, and 
therefore defines precisely how consistent is the new measurement with those in the catalog.}

\chg{In \Tref{tab-phaseFore} we compare the interpolations derived from the P0 method (see \Sref{sec-compInterp}) 
and those derived from the stochastic modeling for the ten best sampled phase calibrators. 
We calculate the median (per source and Band) of the absolute value of the relative differences (in percentage) between the  interpolated fluxes  
calculated using both methods and the actual measurements \citeaffixed{Bonato2018MNRAS}{given by}.
We assume the fluxes given by \citeasnoun{Bonato2018MNRAS}  are independent from the measurements of the phase calibrators given in the ALMA source catalog.
We can see that in the majority of the cases (26 of 40, among all Bands and sources) the median of the absolute differences is lower for the stochastic estimations. 
We confirm the significance of this result  using a binomial distribution test, which reject the  hypothesis that 
both methods are equivalent (and favors the stochastic interpolation residuals being lower) 
at a 95\% confidence level. Furthermore, we cannot confirm that the P0 flux interpolation is 
significantly (p-value lower than 0.05) better than the stochastic method in any band.}

\begin{figure}
\centering\includegraphics[height=0.5\textwidth,angle=-90]{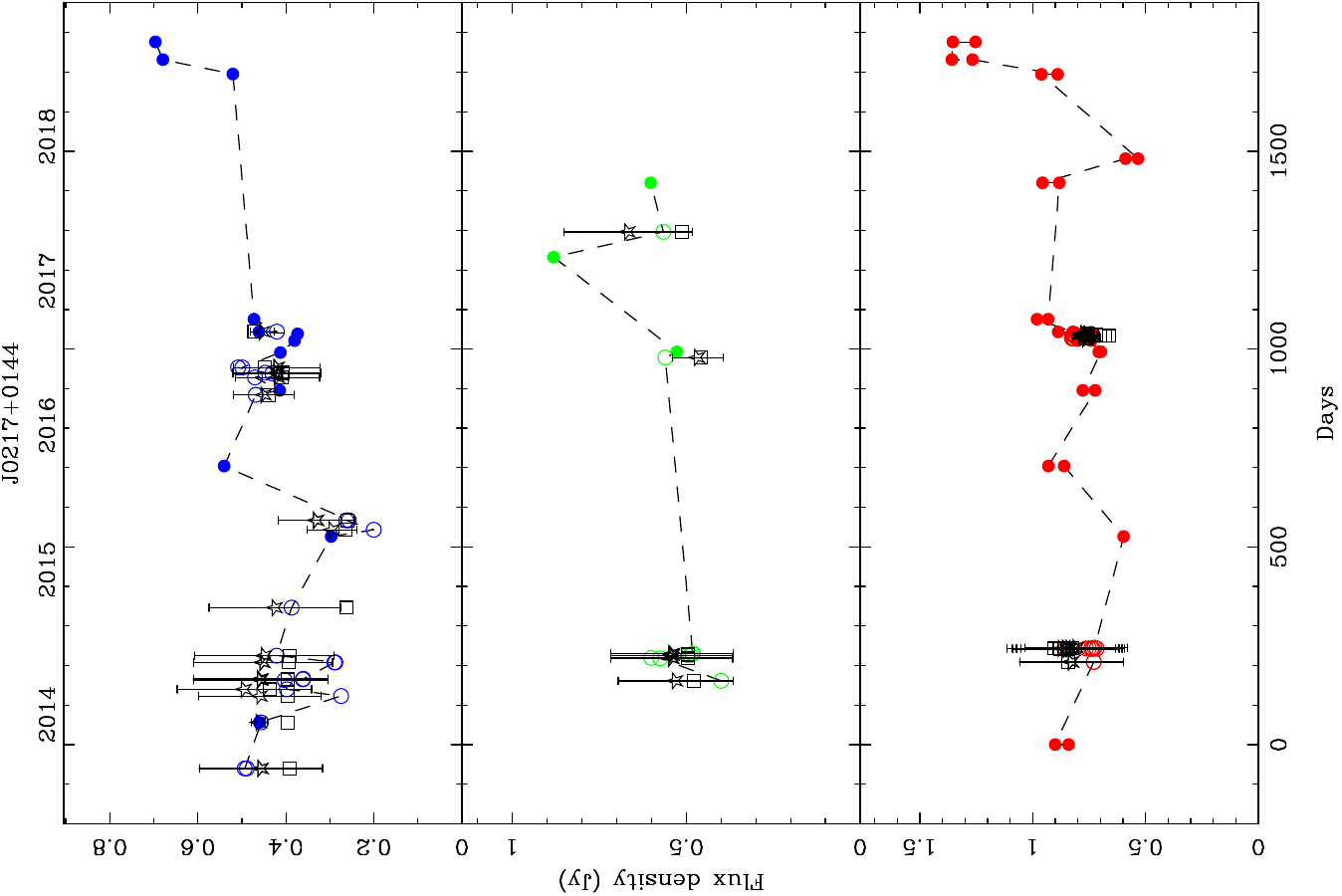}%
\vspace*{-1ex}\caption{\chg{Top panel: Band 7 data of J0217+0144. Blue filled and open circles show the  
source catalog and  \citeasnoun{Bonato2018MNRAS} fluxes, respectively. Inter- and extrapolations from the stochastic model are marked with stars, 
and those from P0 are marked by squares. Error bars centered on the stochastic interpolations show their expected uncertainty. \chgg{Middle panel: same as above but in Band 6  (green). Bottom panel: same as top panel but in Band 3 data (red).}} 
 \label{fig-phaseFore}}
\end{figure}

\chg{\Fref{fig-phaseFore} shows, for J0217+0144 as a representative example,  the data from the source catalog (filled circles), 
from \citeasnoun{Bonato2018MNRAS} (open circles), the interpolation and extrapolations of the 
stochastic modeling (stars) and those of the method described above (squares). The error bars are centered in the stochastic model values, and display the 
uncertainty of the flux including the variability of the source. Note additionally that the first ``prediction" of the model 
is rather a retrodiction (or hindcast) because there is no previous data on the source. 
It is  equivalent to a forecast due to the time reversibility of the stochastic process.}
%
%

\begin{table}
\caption{\chg{Median of the absolute differences in percentage  between the fluxes given by \citeasnoun{Bonato2018MNRAS} and those extra- and interpolated from the 
ALMA source catalog. For each entry, the first number is associated with the P0 flux estimation algorithm, and the second uses the prediction 
based on a OU-1F1a model fitted to each source.}    \label{tab-phaseFore} }
\begin{indented}
\item[]\hspace*{-2em}\begin{tabular}{lrcccccc}
\br
Source & \#& \multicolumn{6}{c}{Band} \\
	           &meas. & 3 & 4 & 6 & 7 & 8 & 9\\\mr
J0106$-$4034 & 14 & $3.7,2.9$ & \ldots & $72.9,60.1$ & $52.6,1.9$ & \ldots & \ldots \\
J0132$-$1654 & 28 & \ldots & \ldots & $25.0,66.0$ & $26.7,6.4$ & \ldots & \ldots \\
J0217+0144 & 42 & $13.4,7.4$ & \ldots & $9.2,12.2$ & $11.0,14.7$ & $30.5,28.5$ & $15.0,18.4$ \\
J0601$-$7036 & 42 & $8.1,4.9$ & \ldots & $11.1,8.8$ & $8.4,1.4$ & $25.8,32.6$ & $66.9,69.0$ \\
J0607$-$0834 & 48 & $7.5,5.4$ & $8.1,11.0$ & $8.3,7.0$ & $7.8,7.3$ & $14.5,12.8$ & $14.6,15.3$ \\
J1626$-$2951 & 33 & $22.1,7.8$ & $20.5,6.6$ & $23.9,25.0$ & $17.3,18.6$ & \ldots & $42.4,31.6$ \\
J1744$-$3116 & 42 & $15.0,10.7$ & $10.2,3.8$ & $9.5,12.3$ & $10.2,18.5$ & $41.6,8.7$ & \ldots \\
J2134$-$0153 & 186 & $10.8,10.0$ & \ldots & $6.3,6.6$ & $4.2,3.6$ & \ldots & $61.8,60.8$ \\
J2157$-$6941 & 16 & $21.5,38.2$ & \ldots & \ldots & $48.6,13.3$ & \ldots & \ldots \\
J2225$-$0457 & 29 & $11.7,5.1$ & \ldots & \ldots & $26.0,22.0$ & \ldots & $57.7,43.8$ \\
\br
\end{tabular}
\end{indented}
\end{table}

\newpage\subsection{\chg{Combining calibrator survey with additional data.}\label{sec-high}}

The flux densities used in the light-curves analyzed in this work come from data taken by the  calibration survey and reported in the source catalog. 
Additional measurements for many calibrators are provided by the ALMACAL survey \cite{Oteo2016ApJ,Bonato2018MNRAS}. In principle, any other sub-mm 
measurements can be used in conjunction with the source catalog  to model the light-curves, as long as  the  new data: (i) have comparable 
measurement errors as the source catalog; and (ii) the new data's calibration is independent from the source catalog. 
Regarding (i), because the main source of measurement error comes from calibration, 
there are no large differences between the ALMACAL flux uncertainties and those of the source catalog. 
\chgg{It is also desirable to combine datasets with similar characteristics and comparable 
cadency and time coverage  in order to  obtain a robust stochastic fitting. 
Requirement (i) may be  circumvented in the future using a more sophisticated heteroscedasticity prescription. 
Regarding  point (ii), while most  ALMACAL data is calibrated independently from the grid source catalog, not all of it is. 
Therefore, we emphasize this section having the purpose of showing how the stochastic model can be extended to data taken
with a larger frequency spread, rather than proposing a more complete model or a procedure applicable 
 to all  the grid sources. In the following and for the rest of this section, therefore, we treat requirement (ii) as an assumption.}
 
\begin{figure}
\includegraphics[width=1.\textwidth]{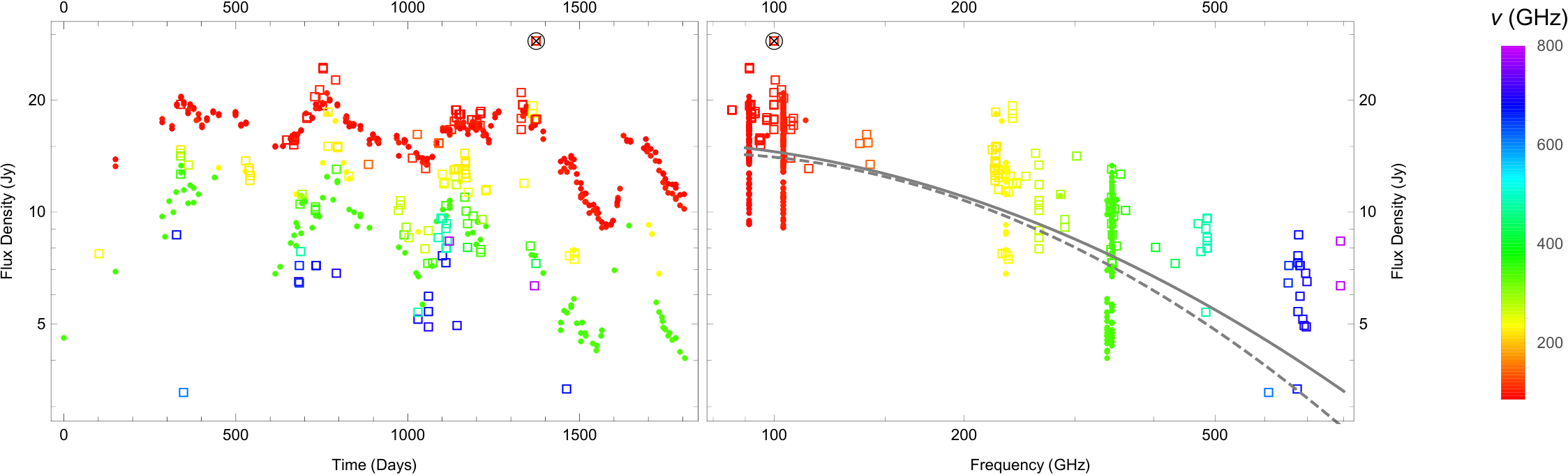}%
\caption{\chg{Combined data from the source catalog (filled circles) and from \citeasnoun{Bonato2018MNRAS} (hollow squares) from J2253+1608.  
The $\otimes$ marks one masked outlier.  Left panel: light-curve. Right panel: Spectrum. Grey continuous and dashed curves 
show the long-term mean spectrum fitted to  the combined dataset and to only the source catalog data, respectively.
 \label{fig-comp}}}
\end{figure}

\chgg{Because the additional ALMACAL data on source J0635$-$7516
 do not include  Bands above 7, for this section testing we choose 
J2253+1608 (3C 454.3), which has been used as a high-frequency calibrator and  has been observed much more  frequently in Bands 8 to 10.}
Figure \ref{fig-comp} shows the light-curve and spectrum of this source including the additional data. We eliminate one Band 3 measurement from the combined dataset 
as an outlier because it is 60\% larger than the source catalog flux. Again, we try to qualify data as outlier very sparingly. 

\begin{figure}
\includegraphics[width=\textwidth]{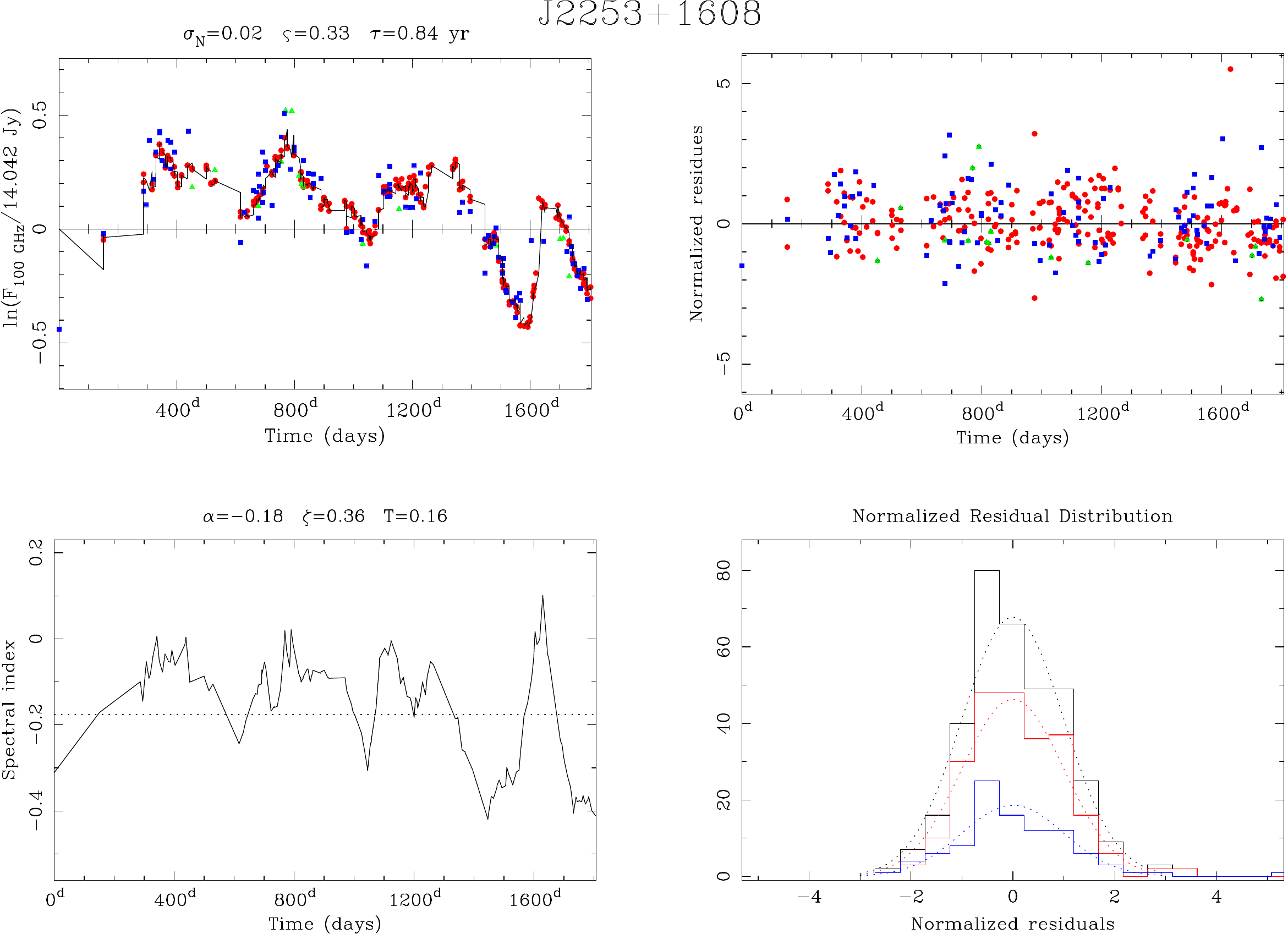}%
\caption{\chg{OU1F1a model with lag fitted to the source catalog data for 
J2253+1608. Panel arrangements, plotted quantities, and line colors follow  the same convention as of \Fref{fig-sel}.  \label{fig-selJ22}}}
\end{figure}

\begin{figure}
\includegraphics[width=\textwidth]{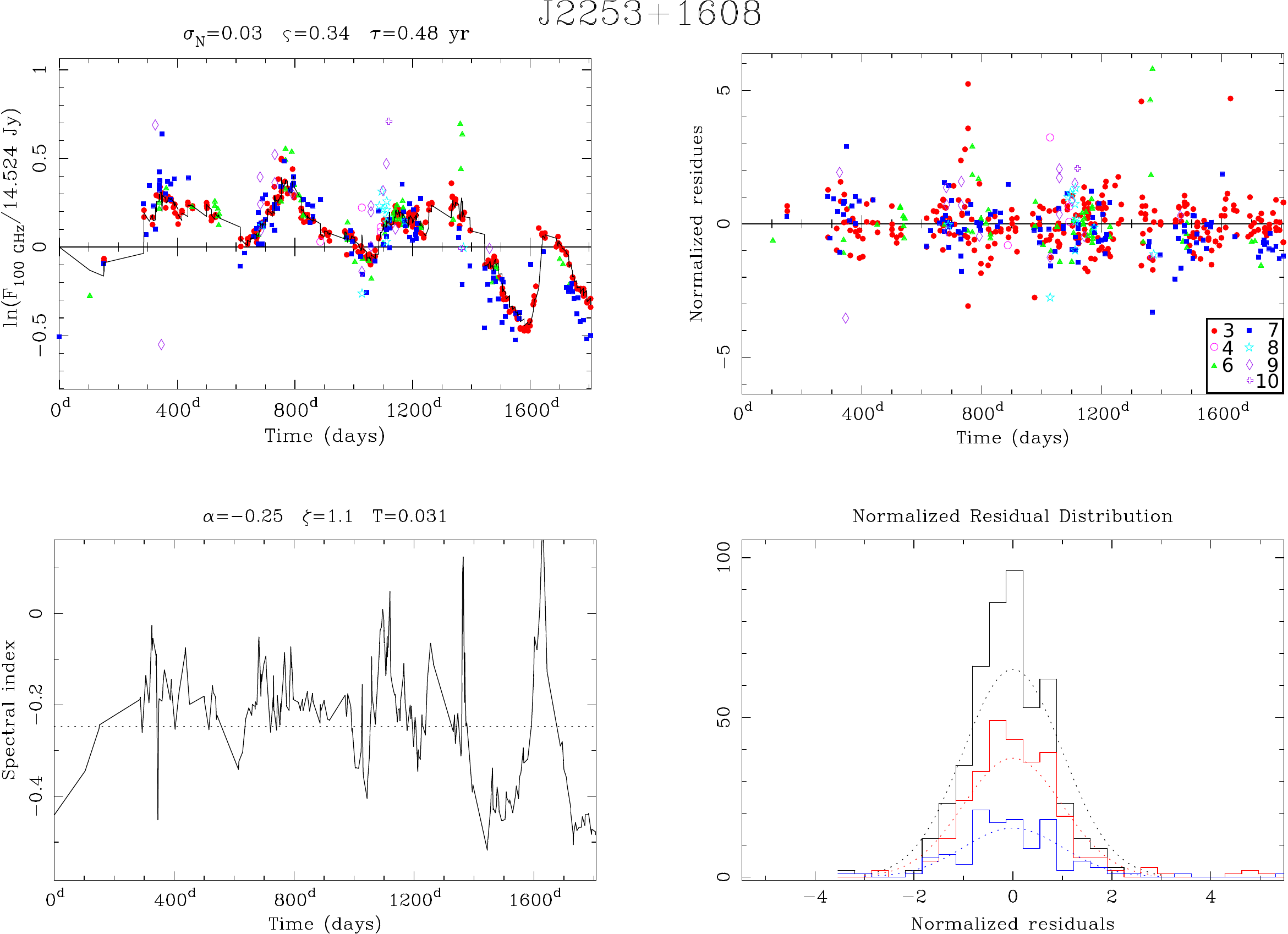}%
\caption{\chg{OU1F1a model with lag fitted to the combined data set  (source catalog and \citeasnoun{Bonato2018MNRAS}) for 
J2253+1608. Panel arrangements, plotted quantities, and line colors follow  the same convention as of \Fref{fig-sel}. 
Top panels include  measurements at Bands 3, 4, 6, 7, 8, 9, and 10, 
whose symbols are indicated in the legend of the top-right panel. \label{fig-selB18}}}
\end{figure}

\chg{Figures  \ref{fig-selJ22} and \ref{fig-selB18} shows the results of the OU-1F1a model fitted to J2253+1608, to the source 
catalog data only and the combined dataset, respectively. 
The models shown in Figures  \ref{fig-selJ22} and \ref{fig-selB18} are similar in general, but there are some noticeable differences in
 the long-term spectrum. This is  illustrated in 
the right panel of \Fref{fig-comp}: Band 8 and 9 data indicate that the model fitted to the data between Bands 3 and 7 has  too a pronounced
concavity and that the actual spectrum is slightly flatter.
Other  differences between both models is that the spectral index curve of the combined data 
(\chgg{bottom-left} panel of Figure \ref{fig-selB18}) seem to be more variable compared with the smoother spectral index curve in Figure \ref{fig-selJ22}, which is a  
consequence of the smaller derived decorrelation time. The residuals  shown in the \chgg{top-right} panel of Figure \ref{fig-selB18} do not
show evident trends, although there are several points  above $5\sigma$ which may indicate  outliers.  
Additionally, the histogram of the residuals (\chgg{bottom-right} panel) of the combined data seems to be less consistent with a Gaussian distribution than that of \Fref{fig-selJ22}. It is possible that 
a different model, either with a more sophisticated  spectral description of the data, 
or a better adapted stochastic process --- like a CARMA or an infinite mixture model \cite{Kelly2011ApJ,Takata2018ApJ} --- could provide a better fit to the light-curves at all frequencies between Bands 3 and 10. }

{\subsection{Drawbacks and future refinements.}\label{sec-drawbacks}}

To address the current limitations of the proposed modeling, 
first  we must stress that the perspective of the current study is admittedly phenomenological. That is, it emphasizes the 
capability of the models to infer the behavior of the calibrators'  fluxes rather than to understand the nature of their variability.
\chg{The criterion which we use to evaluate whether a mixture of Ornstein-Uhlenbeck processes is an adequate model for the blazar variability 
is if it is capable of describing  the light-curve variations leaving residuals consistent with Gaussian white noise at all frequencies.}

On this vein,  a more pertinent criticism may be our  reliance on Gaussian distributions as generators of the stochasticity and noise. 
Indeed, the general theory of time series  uses more general distributions \cite{Brockwell2002ITS&F}, but in astronomy, by far
the majority of past and current development is in Gaussian time series modeling   \citeaffixed{2017AJ....154..220F}{e.g.,}.
Albeit it  is common when stochastic methods are  presented to  remark that Gaussian hypotheses are not crucial, 
to explicitly present and explore a non-Gaussian model is rare \citeaffixed{Eyheramendy2018MNRAS}{e.g., as done briefly in}. 
In any case, the present work is thought as a first step toward a more adequate  calibrator modeling, which may  not be based on 
Gaussian hypotheses.  

Another limitations of our modeling are related with the current cadency of the calibrator survey and  the limited time span of the monitoring.
One direct consequence of the former is that the short-term (intraday) variability likely 
forms part of what we call white noise residuals  ($\sigma_N$), together with the instrumental and atmospheric noise. 
A dedicated study of this time-frequency domain (for example, by doing a high-cadency monitoring for a limited period of time)
could break down which fraction of this noise is actually intrinsic to each source.
 Limitations  stemming from short time span of the monitoring are of two types: 
unaccounted transient variability and 
 systematic uncertainties and bias in the estimated parameters.
 The first one, more than a limitation, is the risk that
  the source just may unexpectedly change qualitatively  the  behavior it has characterized it during the last years.
 Longer studies of blazar variability may illustrate what the grid calibrators could do in the future  in  decade-long timescales. 
For example, monitoring of four grid calibrators in a $\approx14$ yr \chg{timescale} was   performed by the IRAM 30m telescope 
 in the mm/sub-mm frequencies \cite{Trippe2011AA}. These are J1229+0203, J1642+3948, J2253+1608, and J0319+4130; respectively,
 3C 273, 3C 345, 3C 454.3, and 3C 84. The fluxes and spectral indices  presented in \Sref{sec-data} are consistent with those observations 
 and the variations observed for the most part. However,  the stochastic processes presented here will likely not be able to  recover or forecast all variability features. 
Indeed, the change experienced by J2253+1608 in $\sim$2005 CE \cite[Fig.~1]{Trippe2011AA} is unlikely to be well 
 described by a stationary stochastic process. More complete models for the future monitoring  will likely need to include some non-stationarity,
 either in the deterministic model or in its stochastic parameters. 
 
 Alternatively, for some applications it may be desirable to select only a fraction (the most recent, for example) of the history of the light-curve. 
 For example, in \Fref{fig-sel} the \chgg{top-right} panel shows a somewhat  larger scattering of residuals before $t=800 $ d. 
Including the data from this more unstable phase of J0635$-$7516 in the modeling produces an increase  of the uncertainty 
due to variability in the current forecasts.  
Indeed, removing these data from the modeling decreases \chg{the uncertainties associated to the forecasts}.
While this type of censoring of the data may be justified in some cases, we must bear in mind that a new phase of \chgg{increased} instability is essentially
  unpredictable. In a way, a larger uncertainty in the forecasts is  the way the model accounts for the possible occurrence of another
  unstable phase in the future. Alternative models which include variation of more parameters with frequency (like the variance rates)
   or which include non-stationary flaring activity could provide more flexibility and allow for tighter uncertainty bounds for the forecast through the entire spectral coverage. 
  
The second  drawback derived from a relatively short monitoring time baseline is  bias and  systematic uncertainty of the fitted parameters. 
Specifically, the decorrelation time --- which is defining of the OU-process --- cannot be recovered. 
This is well illustrated by  \citeasnoun{2017A&A...597A.128K} and by \Eref{eq-uncAR1tau}  in the case of an AR(1) process. 
From the latter equation  is not difficult to see, for example,  that the MLE of $\phi=\exp(-\delta t/\tau)$  is not located farther 
from 1 (where $\tau$ becomes unconstrained) than $\sqrt{\Delta T/2\tau}$ times the standard deviation of the estimator, regardless of 
the cadency or the number of samples $N$. Since the MLE is the asymptotically most efficient estimator, there 
is no really a way to improve the estimation of $\tau$ without increasing $\Delta T$, the total monitoring time.
\citeasnoun{2017A&A...597A.128K} illustrates this limitation by noting that a short time baseline
 cannot probe the short frequency section of the power spectrum of the process. 
Other non-parametric estimators of the decorrelation timescale  \cite{2012MNRAS.419.1197K,2017ApJ...835..250K}
 which may perform better  for finite samples than the MLE are based 
 on determining an accurate power spectra of the process, for which we require a more regular sampling.
 Note also that  when $\Delta T/\tau$ is not sufficiently large, 
 the long-term mean $\mu$  also becomes unconstrained. Roughly,  $\Delta T/\tau$ gives the number of independent 
  samples of the process. These are useful to  determine $\mu$, whose  MLE uncertainty goes like  
  $\sqrt{\tau/2\Delta T}$ \eref{eq-muhat}. 

 Fortunately, as shown in Sections \ref{sec-uncs} and \ref{sec-fore}, 
 the large parametric uncertainty  derived from a long decorrelation time does not
 produce  a large uncertainty in the forecasts of the process, at least in the relatively short  times characterizing the 
 current cadency of the calibrator survey. In the long-term, the largest 
 effect is produced by the uncertainty of $\mu$.
When $\Delta T/\tau<10$, a procedure to determine  the 
 maximum parametric error we may incur due to the unconstrained $\tau$
 may be to model the stochastic process
 with a regular (i.e., undamped) Brownian motion. Alternatively, resampling the likelihood with 
 the  bayesian procedure presented in \Sref{sec-fore} can provide in principle 
 a better estimation for the parametric uncertainty \chgg{when $\tau$ is less constrained}.
 
 \section{Conclusions}\label{sec-con}

We present a stochastic modeling of the time series measurements made by ALMA of a list of \totalsamplesize\ 
blazar sources used as flux calibrators. The main results, conclusions, and advantages of the 
procedures presented in this  work are summarized as follows:
\begin{enumerate}
\item{Mixtures of  Ornstein-Uhlenbeck (OU) process in flux and spectral index are  able to model reasonably well the
multi-frequency light-curves of the calibrators in the 90--350 GHz range, on timescales of $\sim5$ yr \chgg{and cadences of one measurement every 1-2 weeks}. 
A single OU-process and a single spectral index provides a first model approximation. Further features of the modeling include additional stochastic process for the spectral index variations, additional curvature of the spectrum, and 
a frequency dependent time lag.}
\item{Using maximum likelihood estimation and based on a statistical assessment of the residuals, we determine that most (38) sources are better modeled
with an additional OU-process describing  the spectral index variations. In addition, 37 of the 39 sources evidence a decreasing spectral index with frequency. Five sources are apparently characterized by a significant frequency dependent lag between their Band 3 and 7 light-curves. Four of them 
are better described by a  $\sim2$ d lag with high-frequency leading, and one source with a 7 d low-frequency leading lag.}
\item{The modeling provides forecasts and flux interpolations based on the history 
of the time-variable source, together with their associated  uncertainties. 
These can be use to perform diagnostics on 
 new measurements of the calibrators,  \chg{determine flux uncertainties due to the variability of the secondary flux calibrator,} and to evaluate the confidence of the flux calibration in the ensuing  data quality assurance.}
\item{Ill-constrained decorrelation times of the OU-process due to monitoring limitations should not 
greatly affect the determination of forecasts and interpolations. These are  the most relevant quantities for 
the practical use  in the ALMA calibration routine.}
\item{Additional advantages of the formalism presented in this work are the flexibility they accommodate
 inhomogeneous sampling, updated measurements, \chg{and potentially higher frequency data}. Future improvements  could include further refinements 
 in the noise terms (e.g., separating explicitly atmospheric and instrumental) and inclusion of non-Gaussian hypotheses. } 
\end{enumerate}

\ack
Authors thank M.\ Bonato for helping and answering our questions about ALMACAL. Authors also thank an anonymous referee for detailed comments which have improved this paper. This paper makes use of ALMA data: ADS/\-JAO.\-ALMA\-\#2011.0.00001.\-CAL. ALMA is a partnership of ESO (representing its member states), NSF (USA) and NINS (Japan), together with NRC (Canada), MOST and ASIAA (Taiwan), and KASI (Republic of Korea), in cooperation with the Republic of Chile. The Joint ALMA Observatory is operated by ESO, AUI/NRAO and NAOJ. In addition to already \texttt{R} cited packages, this publication made use of \texttt{stringr} \cite{stringr} and  \texttt{forecast} \cite{forecast}. Analysis presented here made use of the Perl Data Language (PDL, \texttt{http://pdl.perl.org}) developed by K. Glazebrook, J. Brinchmann, J. Cerney, C. DeForest, D. Hunt, T. Jenness, T. Lukka, R. Schwebel, and C. Soeller.

\appendix

\section{Homogeneous sampling without noise: the AR (1) model.}\label{sec-ar1}

The AR(1) process is one of the simplest processes which fulfills the features described in \sref{sec-data}: is characterized by a 
long-term mean ($\mu$) with a correlation between terms which decreases monotonically with time-lag.  
The Gaussian AR(1) process is defined by the following recurrence 
\begin{equation}
x_{n}-\mu=\phi (x_{n-1}-\mu)+\sigma^2  Z_{n}~~,\label{eq-AR1}
\end{equation}
where the $Z_n$ is a  standard GWN and $0<\phi<1$. This process is Markovian, that is, the distribution of $x_{n}$ given the entire series until the $(n-1)$-th term ($\ldots, x_0, x_1, \dots, x_{n-1}$) depends solely on $x_{n-1}$.
Sometimes the process receives the name ``mean-reverting" because $x_{n}$ is distributed around a value closer to $\mu$ than $x_{n-1}$. Defined in this way,  the covariance between two terms is given by 
\begin{equation}\cov\left(x_{n+h},x_{n}\right)=\frac{\sigma^2\phi^h}{1-\phi^2}~~,\end{equation} \label{eq-ar1cov}
that is, it depends only on the time difference (stationary process). 

A simple interpretation may assign the $x_{n}$ to the log-flux of a calibrator measured at a time $n$. In this case, the time series correspond to equally spaced observations. As shown in \eref{eq-ar1cov}, $\phi$ modulates the degree of correlation between consecutive measurements and thus it must depend on the time between observations. For $\phi\ll1$,  the $x_n$ are almost uncorrelated between each other, characteristic of largely spaced observations. Analogously, $\phi\approx1$ could represent tightly correlated, very frequent measurements.
Indeed, the AR(1) series can be interpreted as the discrete, homogeneous sampling of an OU continuous time process.  This process also receives the name continuous AR process (CAR) and damped random walk. Under this interpretation, 
\begin{equation}
\phi=e^{-\delta t/\tau}~~,\label{eq-phi}
\end{equation} 
where $\tau$ is known as the decorrelation timescale, 
and $\delta t$ is the time between consecutive samplings.  It is worth noticing that because the time-sampling is homogeneous, the discrete series is stationary. If the time sampling where inhomogeneous, then $\delta t$ and $\phi$ would depend on the index of the series. 
Thus, the process would no longer be stationary,  but only due to the sampling: the continuous process is  still stationary. 

The parameters of an AR(1) process $(\mu, \phi,\sigma^2)$ can be estimated from a specific sample (or realization) of the process, say$\{x_n\}_{n=1}^N$ .
Under reasonably  general circumstances \cite{Eyheramendy2018MNRAS}, the irregular process is ergodic for the mean, and  $\mu$ can be approximated efficiently by the average of the $x_n$. In the AR(1) case, the average distributes approximately normal  with
\begin{equation}
\hat{\mu}=\frac{1}{N}\sum_{n=1}^Nx_n \approx \mathcal{N}\left(\mu,\frac{\sigma^2}{N(1-\phi)^2}\right)~~.\label{eq-muhat}
\end{equation}
We see immediately that for $\phi\approx1$ --- or following \eref{eq-phi},  $\delta t\ll\tau$ --- the estimations of the long term mean becomes more uncertain.

 We define the centered process  $\tilde{x}_n=x_n-\mu$, which is an AR(1) process with zero long-term mean. In practice, we approximate   
 $\tilde{x}_n$ using  $x_n-\hat{\mu}$.  Parameters $\phi$ and $\sigma^2$ can be estimated using maximum likelihood on 
 $\tilde{x}_n$, assuming it is an AR(1) process. The log-likelihood function is given by
\begin{equation}
\mathnormal{l}(\phi,\sigma^2)=-\frac{N}{2}\ln(2\pi\sigma^2)-
\frac{1}{2\sigma^2}\sum_{n=1}^N\left(\tilde{x}_n-\phi\tilde{x}_{n-1}\right)^2~~,
\end{equation}
assuming $\tilde{x}_0=0$. This last requirement is artificial, but not relevant for large $N$ \cite{Hamilton1994TSA}. The maximum likelihood estimators (MLE) and its  normal asymptotic distributions are given by
\begin{eqnarray}
\hat{\phi}&=&\frac{\sum_{n=1}^N \tilde{x}_{n-1}\tilde{x}_{n}}{\sum_{n=1}^N\tilde{x}_{n-1}^2}\approx \mathcal{N}\left(\phi,\frac{1-\phi^2}{N}\right) ~~,\label{eq-uncAR1tau}\\
\hat{\sigma^2}&=&\frac{1}{N}\sum_{n=1}^N\left(\tilde{x}_{n}-\hat{\phi}\tilde{x}_{n-1}\right)^2\approx \mathcal{N}\left(\sigma^2,\frac{2\sigma^2}{N}\right)~~.
\end{eqnarray}

It is also possible to give simple formulae for the best --- in the sense of minimum mean square error --- predictions and interpolations. The  best prediction (or forecast) of the $(n+h)$-th term of a centered AR(1) process given the $(n-1)$ previous observations ($\tilde{x}_{n+h|n-1}$, $h\ge 0$) is given by \cite{Brockwell2002ITS&F}
\begin{equation}
\tilde{x}_{n+h|n-1}=\phi^{h+1} \tilde{x}_{n-1}~,\quad\mathrm{E}\left(\tilde{x}_{n+h}-\tilde{x}_{n+h|n-1}\right)^2=\frac{1-\phi^{2h+2}}{1-\phi^2}\sigma^2~~,\label{eq-predAR1}
\end{equation}
where the second equation gives the mean square error of the prediction.
Similarly, based on observations of $\tilde{x}_i$, $i=1,\ldots(n-1)$ and $\tilde{x}_{n+k}$, the best interpolation for $\tilde{x}_{n+h}$ ($0\le h<k$) is 
\begin{equation}
 \hspace*{-5em}\mathrm{E}\left(\tilde{x}_{n+h}|\tilde{x}_1,\ldots\tilde{x}_{n-1},\tilde{x}_{n+k}\right)=%
\frac{\tilde{x}_{n-1} \phi^{h+1}  \left(1-\phi ^{2 (k-h)}\right)+\tilde{x}_{n+k}\phi ^{k-h} \left(1-\phi^{2(h+1)}\right)  }{1-\phi ^{2 (k+1)}}~~.\label{eq-ARint}
\end{equation}
Note that \eref{eq-ARint} depends only on the measurements taken immediately after and before  time $n$. When $\phi\rightarrow1$, \eref{eq-ARint} becomes the linear 
interpolation between $\tilde{x}_{n-1}$ and $\tilde{x}_{n+k}$. The mean squared error is 
\[\mathrm{E}\left(\tilde{x}_{n+h}-\mathrm{E}\left(\tilde{x}_{n+h}|\tilde{x}_{n-1},\tilde{x}_{n+k}\right)\right)^2 =\sigma^2\frac{1-\phi^{2(h+1)}}{1-\phi^2}\frac{1-\phi^{2(k-h)}}{1-\phi^{2(k+1)}}~~.\] 
This last expression tends  to the prediction error \eref{eq-predAR1} when $k$ is large. 
As can be expected, observations in the distant past or future are only useful 
to determine the  global parameters of the process, but their specific values do not constrain estimations or forecasts at present time. 

We can estimate the influence of the uncertainty of the typically unknown parameter $\mu$ by assuming
we replaced it with a random variable 
$\hat{\mu}$ with mean $\mu$ and variance $\sigma_\mu^2$. According to \eref{eq-predAR1},
the best predictor of $x_{n+h}$ given $n-1$ observations ($x_{n+h|n-1}$) would be
 $\hat{\mu}+\phi^{h+1} (x_{n-1}-\hat{\mu})$. 
Therefore, the variance  of $x_{n+h|n-1}$ due to the $\mu$-uncertainty 
--- that is, the parametric uncertainty --- and to the stochastic
variability are, respectively, 
\begin{equation}
(1-\phi^{h+1})^2\sigma_\mu^2~~,\qquad \sigma^2\frac{1-\phi^{2h+2}}{1-\phi^2}\le\sigma^2(h+1)~~.\label{eq-parUncAR1}
\end{equation}
A short time between the last observation and the prediction is characterized by low values of 
$h$ and $\phi\approx1$, which decreases the uncertainty due to the unknown $\mu$. 
Note that we expect a low influence of $\mu$ in the prediction uncertainty 
 also if $\tau$ is very large. In this latter case, the process becomes closer to a Brownian motion
 and the concept of  long-term mean is meaningless.

\section{State space representation and Kalman recursions}\label{sec-ssr}

We follow the notation and treatment of  \citeasnoun{Kitagawa1996SPATS}.
A time series $Y_n$ allows a Gaussian state space representation if it can be expressed in the following way
\begin{eqnarray}
Y_n=H_n X_n+\epsilon_n~~,\label{eq-ssrmeas}\\
X_{n+1}=F_n X_n+ G_n w_n~~,\label{eq-ssrdyn}\\
\left(\begin{array}{c} \epsilon_n\\ w_n\end{array}\right)\sim\mathcal{N}\left( \bm{0}, \left[ \begin{array}{cc} R_n & 0 \\0 & Q_n \end{array}\right] \right)~~,\label{eq-ssrnoi}
\end{eqnarray}
where \eref{eq-ssrmeas} and \eref{eq-ssrdyn} are called the measurement and state equations, respectively. The random variables $\epsilon_n$ and $w_n$
are uncorrelated. In the case of an OU-rFsa process,  \eref{eq-ssrmeas} corresponds  to \eref{eq-vsi} with 
\begin{equation}
\hspace*{-1em}H_n=(\underbrace{1,\ldots,1}_{r},\underbrace{\ln(\nu_n/\nu_0),\ldots,\ln(\nu_n/\nu_0)}_{s})~,\quad R_n=\sigma_N^2(\nu_n) ~~.\label{eq-H}
\end{equation}
\Eref{eq-ssrdyn} corresponds to \eref{eq-srgral}  with $F_n=\exp\left(-\delta t_n \bm{\tau}^{-1}\right) $ and  $X_{n}$ 
being the vector in the left hand of \eref{eq-srgral}.
Unless one intends to simulate the process,  it is not necessary to define $G_n$ and $Q_n$ explicitly but only up to the requirement that 
 $G_nQ_nG_n^\mathrm{T}=\cov(\eta_n)$ (see \eref{eq-aleamat}).
 We denote in this Appendix a general process as $Y_n$. A lowercase $y_n$ refers specifically to the  stationary model adjusted to a calibrator, as in
 \eref{eq-cd}.

Let us denote $X_{n|n-1}$ the conditional expectation of $X_n$ given all the previous information, which is 
represented by $(Y_1,\ldots,Y_{n-1})$, $i=1,\ldots,n-1$. We define $V_{n|n-1}=\mathrm{E}\left(X_n-X_{n|n-1}\right)\left(X_n-X_{n|n-1}\right)^\mathrm{T}$,
that is, the variance of the predictor. 
The $Y_i$ represent either an observation of $Y(t)$ at time $t_i$, or a filler value in case no observation was taken at time $t_i$. This flexibility will be useful 
later to introduce best predictions at time $t_i$ given all the other observations (the interpolation or smoothing problem). 
Evidently, this filler value should have no influence in the calculated best estimators.
 We define  respectively $X_{1|0}$ and $V_{1|0} $ as the predictor and variance of $X_1$, assuming no previous information.
It is useful sometime  consider that  a ``previous"  observation  of the process $Y_0$ took place in the infinite past 
($\delta t_1=t_1-t_0 \rightarrow +\infty$), and is therefore uninformative.
 Analogously, let $X_{n|n}$ and $V_{n|n}$ denote  the conditional expectation of  $X_n$  and its variance, respectively, 
  given information $Y_i$ up to $i=n$.  
  
  The first set of Kalman recursions are
\begin{eqnarray}
K_n=\left\lbrace\begin{array}{ll} V_{n|n-1}H_n^\mathrm{T} (H_nV_{n|n-1}H_n^\mathrm{T}+R_n)^{-1} & ; Y_n ~\mathrm{observed}\\ 0 & ;Y_n ~\mathrm{not~observed}\end{array}\right.~~,\label{eq-Kgain}\\
X_{n|n}=X_{n|n-1}+K_n(Y_n-H_nX_{n|n-1})~~,\label{eq-Kfilter}\\
V_{n|n}=(I-K_nH_n)V_{n|n-1}~~,\label{eq-KfilterV}\\
X_{n+1|n}=F_{n+1}X_{n|n}~~,\label{eq-Kpred}\\
V_{n+1|n}=F_{n+1}V_{n|n}F_{n+1}^\mathrm{T}+G_{n+1}Q_{n+1}G_{n+1}^\mathrm{T}~~,\label{eq-KpredV}
\end{eqnarray}
for $n\ge1$.   
 Equations \eref{eq-Kgain} to \eref{eq-KfilterV} define the Kalman filter, and 
Equations \eref{eq-Kpred} and \eref{eq-KpredV} define the Kalman predictions. $K_n$ is known as the Kalman gain.
The least mean square prediction for $Y_n$ and its variance are
\begin{equation}
\fl Y_{n|n-1}=H_nX_{n|n-1}~~,\qquad \mathrm{E}\left(Y_n-Y_{n|n-1}\right)^2=:r_n=H_nV_{n|n-1}H_n^\mathrm{T}+R_n~~.\label{eq-predY}
\end{equation} 

The Kalman predictions are useful not only as forecast, but also to calculate the log-likelihood, which is given by
\begin{equation}
-2\ln \mathcal{L}=\sum_{n~\mathrm{observed}} \ln(2\pi r_n)+\frac{\left(Y_n-Y_{n|n-1}\right)^2}{r_n}~~.\label{eq-ll}
\end{equation}

The second set of Kalman recursions are 
\begin{eqnarray}
A_n=V_{n|n}F_{n+1}^\mathrm{T}V_{n+1|n}^{-1}~~,\label{eq-KsmooA}\\
X_{n|N}=X_{n|n}+A_n(X_{n+1|N}-X_{n+1|N})~~,\label{eq-Ksmoo}\\
V_{n|N}=V_{n|n}+A_n(V_{n+1|N}-V_{n+1|n})A_n^\mathrm{T}~~,\label{eq-KsmooV}
\end{eqnarray}
where $N>n$. In \eref{eq-KsmooA}, $V_{n+1|n}^{-1}$ can be replaced by a pseudo-inverse in case is not invertible \cite{anderson2012optimal}. 
Equations \eref{eq-KsmooA}--\eref{eq-KsmooV} define the  Kalman smoothing, and give the best prediction of $X_n$ 
given all the $Y_i$ (past and future) until $N$.  The interpolated expected  value of $Y(t)$ at $t_n$ (assuming it was not observed), given all the 
data, is  $Y_{n|N}=H_n X_{n|N}$.

\section{Supplementary material}\label{sec-supp}

In the supplementary material we provide additional results, plots, and \texttt{R} (v.~3.5.0) computer programs to generate the simulated 
datasets presented through the 
paper. The supplementary material includes  plots like those of Figures \ref{fig-data},
\ref{fig-OU10ex},  \ref{fig-acfComp}  and \ref{fig-sel} for each source, and for all models 
specified in \Tref{tab-comp}. A log file for each model gives details about the MLE, including the
 final error and cross-correlation matrices, AIC, \pad, and \plb\ values. A text file also indicates the 20 measurements masked 
 from the catalog as possible outliers. We also include data and details of the modeling for the non-grid calibrators analyzed in \Sref{sec-phase}.
 As indicated in the text, we provide the scripts used to generate the synthetic light-curves 
 used in \Sref{sec-sim}. 
 PDL code implementing the Kalman recursions (Appendix A)
  is given   in the supplementary material
  and can also be obtained from the Astrophysics Source Code Library \cite{ascl}.

\section*{References}

\end{document}